\documentclass[prd,12pt]{article}

\usepackage{amsmath,amssymb,graphicx,multirow,xspace,slashed}
\usepackage[colorlinks=true,urlcolor=blue,anchorcolor=blue,citecolor=blue,filecolor=blue,linkcolor=blue,menucolor=blue,pagecolor=blue]{hyperref}
\usepackage[compress,numbers]{natbib}
\usepackage{subfigure, placeins}
\usepackage{booktabs}
\usepackage{blindtext, rotating}
\usepackage{afterpage}
\usepackage{enumitem}
\usepackage{ marvosym }
\usepackage{authblk} 
\usepackage{verbatim}
\usepackage{soul} 
\usepackage[normalem]{ulem}
\usepackage{pifont}
\newcommand{\cmark}{\text{\textcolor{darkgreen}{\ding{51}}}}%
\newcommand{\xmark}{\text{\textcolor{red}{\ding{55}}}}%
\newcommand{\xmarkk}{\text{\textcolor{purple}{\ding{54}}}}%

\usepackage{floatrow}
\newfloatcommand{capbtabbox}{table}[][\FBwidth]

\usepackage[font=footnotesize,labelfont=bf]{caption}

\usepackage{lineno}

\allowdisplaybreaks

\long\def\symbolfootnote[#1]#2{\begingroup%
\def\thefootnote{\fnsymbol{footnote}}\footnote[#1]{#2}\endgroup}

\newcommand{\newc}{\newcommand}
\newc{\gsim}{\lower.7ex\hbox{$\;\stackrel{\textstyle>}{\sim}\;$}}
\newc{\lsim}{\lower.7ex\hbox{$\;\stackrel{\textstyle<}{\sim}\;$}}
\newc{\gev}{\,{\rm GeV}}
\newc{\mev}{\,{\rm MeV}}
\newc{\ev}{\,{\rm eV}}
\newc{\kev}{\,{\rm keV}}
\newc{\tev}{\,{\rm TeV}}
\newc{\MHT}{$H_T^{\text{miss}}$}
\newc{\MET}{$\slashed{E}_T$}
\newc{\MTT}{$M_{T2}$}

\def\ln{\mathop{\rm ln}}

\newc{\mz}{M_Z}
\newc{\mpl}{M_*}
\newc{\mw}{m_{\rm weak}}
\newc{\nr}[1]{N^c_R{}_{#1}}

\renewcommand{\dag}{\dagger}

\def\beq{\begin{equation}}
\def\eeq{\end{equation}}
\newcommand{\bea}{\begin{eqnarray}\begin{aligned}}
\newcommand{\eea}{\end{aligned}\end{eqnarray}}
\def\bitem{\begin{itemize}}
\def\eitem{\end{itemize}}
\newcommand{\CO}{O}

 \numberwithin{equation}{section}

\newcommand\fverb{\setbox\fverbbox=\hbox\bgroup\verb}

\newbox\fverbbox

\usepackage[usenames,dvipsnames]{xcolor}

\definecolor{darkgreen}{rgb}{0,0.5,0}
\definecolor{goodorange}{rgb}{0.9,0.4,0}

\begin{document}

\baselineskip 0.6cm

\begin{titlepage}

\thispagestyle{empty}

\begin{center}

\vskip 1cm

{\Large \bf Digging Deeper for New Physics in the LHC Data}

\vskip0.2cm{}

\vskip 1.0cm
{\large Pouya Asadi, Matthew~R.~Buckley, Anthony DiFranzo,\\\vskip0.2cm Angelo Monteux and David Shih }
\vskip 1.0cm
{\it NHETC, Dept.~of Physics and Astronomy\\ Rutgers, The State University of NJ \\ Piscataway, NJ 08854 USA} \\
\vskip 2.0cm

\end{center}

\begin{abstract}

In this paper, we describe a novel, model-independent technique of ``rectangular aggregations" for mining the LHC data for hints of new physics. A typical (CMS) search now has hundreds of signal regions, which can obscure potentially interesting anomalies. Applying our technique to the two CMS jets+MET SUSY searches, we identify a set of previously overlooked $\sim 3\sigma$ excesses. Among these, four excesses survive tests of inter- and intra-search compatibility, and two are especially interesting: they are largely overlapping between the jets+MET searches and are characterized by low jet multiplicity, zero $b$-jets, and low MET and $H_T$. We find that resonant color-triplet production decaying to a quark plus an invisible particle provides an excellent fit to these two excesses and all other data -- including the ATLAS jets+MET search, which actually sees a correlated excess. We discuss the additional constraints coming from dijet resonance searches, monojet searches and pair production. 
Based on these results, we believe the wide-spread view that the LHC data contains no interesting excesses is greatly exaggerated.

\end{abstract}

\end{titlepage}

\setcounter{page}{1}

\tableofcontents

\vfill\eject

\section{Introduction}

The Large Hadron Collider (LHC) has recently achieved major milestones. At the ICHEP 2016 \cite{ICHEP} and Moriond 2017 \cite{MoriondEW,MoriondQCD} conferences, the ATLAS and CMS collaborations presented the results of many searches for new physics using $\sim10$/fb and $\sim 35$/fb of data, respectively, at 13 TeV. At these integrated luminosities, the sensitivity to new physics begins to truly outstrip what was previously achieved in the $\sim 20$/fb of 8 TeV data collected in Run I. When analyzed by the experimental collaborations, no evidence for new physics has emerged from this data. 

It would be fair to say that the lack of new physics in the ATLAS and CMS data has reinforced a growing sense of unease among particle physicists. While new physics around the weak scale remains as theoretically well-motivated as ever, there is a general feeling that it should have showed up by now. For example, although natural regions of supersymmetric parameter space remain, the simplest versions of supersymmetry (which were the pre-LHC expectations) are excluded by the null results \cite{Feng:2013pwa,Evans:2013jna,Evans:2013uwa,Kowalska:2016ent,Han:2016xet,Buckley:2016kvr,Buckley:2016tbs,Ross:2017kjc}.
If solutions to dark matter or the hierarchy problem have failed to manifest in 35/fb of 13 TeV data, why should we expect them to appear in the next 35 or 100/fb?

In this paper, we wish to push back on the characterization of the LHC data as {\em clearly} devoid of interesting signals with the potential to be new physics. We believe this is overly pessimistic. It is entirely possible that signatures of new physics are present in the existing data. At the very least, this possibility cannot be excluded without significantly more work by both theorists and experimentalists.

The issue is that the LHC searches, especially those from CMS, now typically  contain hundreds of signal regions (SRs) categorized by various bins in topology (number of leptons, jets, $b$-jets, etc.) and kinematics ($H_T$, $M_{\rm eff}$, $\slashed{E}_T$, etc.). Slicing the data this finely is a potentially powerful, model-independent approach, and it allows the searches to be sensitive to a much wider variety of models than the small set of benchmarks that have been studied so far. Indeed, we are very grateful to the experimental collaborations for providing such a wealth of information. However, having so many SRs also results in a noisier dataset. Adding in the presence of non-trivial correlations in the background predictions, it can be very challenging to get a sense of the presence or absence of statistically significant anomalies in the data.

The conclusion that there is no evidence of new physics in the data is primarily based on the study of a handful of benchmark ``simplified" models by the experimental collaborations. These typically consist of a few particles, with pre-determined cross sections and simple branching ratios -- pair produced gluinos decaying 100\% to $q\bar q\chi$, for example. These models only ever populate a small subset of the many SRs in the LHC searches. 
 Interesting excesses could exist in the data and yet be completely overlooked by these analyses of simplified models. Even within a given simplified model topology, one could have an excess at lower masses, where the simplified model is naively excluded, provided the cross section were somehow reduced.\footnote{To this end, significance plots provide more useful and important information. CMS has started to include these plots in their supplementary information~(see e.g.~Additional Figures 2-10 for \cite{CMS36}),  and we would like to encourage them to continue.}
Showing limit plots for just a narrow set of simplified models paints a potentially misleading picture of the data -- they are simply not an adequate basis set to cover data as complex as what the LHC is now providing.

More generally, while it is straightforward to use the full set of SRs (and their correlations) to test for the presence of a specific new physics scenario, this presupposes knowledge of the new physics model to be tested.  Imagine that new physics is present in the LHC data, but with a drastically different set of signatures and kinematics than realized in the set of popular simplified models. The new physics would result in statistical excesses in some subset of SRs, but with so many SRs in the analysis (and so many more chances for random fluctuations), this would not be immediately apparent. Likelihood calculations using signal templates derived from the current set of simplified models would likewise miss the new physics, as they populate very different SRs. 

What we need then is some method which provides a more comprehensive ``basis set'' of signal templates than the existing simplified models. This will allow us to take a more data-driven approach towards discovering new physics at the LHC: rather than asking ``is this particular model of new physics present in the LHC data?'' we can instead ask ``what potentially interesting excesses are present in the data? And what type of new physics models are compatible with them?" In our opinion, this approach is better suited to the situation particle physics currently finds itself in.

In this work, we will attempt to provide such a method by scanning over all possible ``rectangular aggregations" (RAs) of individual SRs. We are motivated by the fact that a true signal (as opposed to a statistical fluctuation) would tend to populate a set of kinematically and topologically neighboring SRs. Moreover, given how finely the SRs are sliced now, the combination of detector resolution and underlying physics (such as angular distributions and ISR/FSR) would tend to spread the signal over  multiple SRs. Not knowing more about the distribution of events within neighboring SRs, we choose to simply aggregate together signal and background counts in rectangular regions in the (multi-dimensional) space of cut variables (for example, one rectangle might be $2\le N_j\le 3$, $N_b=0$, $500\le H_T\le 1000\gev$, $300\le \slashed{E}_T \le 500\gev$). By performing a full profile-likelihood analysis including correlations within and outside of each RA, we can generate a model-independent list of possible excesses over the background, which can then be more carefully examined to determine if they are possibly consistent with a new-physics interpretation.

Our approach should be contrasted with the alternative approach taken by some analyses (see e.g.~\cite{CMS33,CMS36}) to define a small set (typically $\CO(10)$) of ``aggregate signal regions,"\footnote{Sometimes also called ``super signal regions'' or ``combined regions.''} coarser selections than the individual SRs that are motivated by various signal topologies. While these aggregate SRs can be potentially useful, they are still too signature-dependent and too few in number to adequately assess whether there are any interesting excesses in the data. 
Also, any choice of aggregate SRs is prone to wash out underlying excesses, which we refer to as \textit{over-aggregation}: this will be the case any time a bin (or a set thereof) with a significant excess is combined with bins consistent with the background.

Clearly, our  method is only possible if the SRs are non-overlapping (i.e.~exclusive) in the space of kinematic variables. While this is the case for the CMS searches, most of the ATLAS searches released to date have a small set of overlapping signal regions\footnote{The ATLAS searches also tend to have much higher kinematic thresholds, for unclear reasons. This makes them much less powerful than their CMS counterparts.} (see, for example, \cite{ATLAS22}). We concentrate therefore on the CMS results, and will consider the corresponding ATLAS data only after potentially interesting anomalies are identified in particular CMS channels. We will further focus on the two CMS jets+$\slashed{E}_T$ searches \cite{CMS33,CMS36} which have 174 and 213 SRs, respectively. These searches provide full covariance matrices,\footnote{As we will discuss in detail later, the background estimates can be highly correlated in these searches, so the covariance matrices play a critical role in our rectangular aggregation technique, and in performing accurate statistical calculations more generally. Without them, our list of interesting excesses would have been completely different. We thank CMS for providing the full covariance matrices and encourage them to continue.} are well documented, and are thus recastable. We consider them to be a good testing ground for our approach. Obviously, it would be interesting to continue in this vein with all of the other viable CMS searches.

Our rectangular aggregation technique results in $\sim $~7,000 and $\sim$~33,000 possible aggregations in the jets+$\slashed{E}_T$ searches \cite{CMS33,CMS36}, respectively. Within these, we find 10 and 14 minimal rectangles that contain statistical excesses with a local SM-only $p$-value below 1\% ($N_\sigma > 2.57$); these rectangles are minimal in the sense that they cannot be decreased in size without significantly lowering the size of the excess. These rectangles can be further grouped into three (five) clusters for \cite{CMS33} (\cite{CMS36}) which share similar kinematic and topological features, which we will refer to as regions of interest (ROI).  Finally, we dig deeper into these potentially interesting rectangles to determine whether they are more likely to be statistical fluctuations, or whether they are compatible with a new physics interpretation. We examine distributions of $N_j$, $N_b$, $H_T$, etc.\ to make sure they look sensible, and we also test for compatibility between the two searches. Of the eight total statistically-significant ROIs in the two analyses that our aggregation technique revealed, we found only four that appeared consistent with a new physics interpretation (two in \cite{CMS33} and two in \cite{CMS36}), all with local statistical preference at approximately the $3\sigma$-level. (As we will discuss in more detail later, we estimate a global significance of $\sim 2 \sigma$.) We consider these excesses to be very interesting and deserving of further study. 

In particular, one of the two excesses in \cite{CMS36} is broadly consistent with one of the excesses in \cite{CMS33}. In Section~\ref{sec:monojet} we consider this anomaly in more detail. It is characterized by low jet multiplicity, no $b$-tagged jets, and relatively low $\slashed{E}_T$ or $M_{T2}$, and so we refer to this as a ``monojet'' excess. Having identified this excess through our rectangular aggregation technique, we can now attempt to analyze it using a more conventional approach: constructing simplified models and performing full fits to their parameter spaces using all of the SRs. We can also use the CMS exotica ``monojet''  search \cite{CMS48} and the ATLAS jets+$\slashed{E}_T$ search \cite{ATLAS22} to further refine our calculations. 

After considering several simplified models, we find that the following provides an excellent fit to the data: resonant production of a colored mediator that decays promptly to a quark plus an invisible particle. Combining the CMS and ATLAS jets+$\slashed{E}_T$ searches, we find a local significance of $3.5\sigma$ for this model. 
(Interestingly, the  ATLAS jets+$\slashed{E}_T$ search also shows an excess in its most relevant SR, so including it actually {\em increases} the preference for signal.) On the other hand, the best fit cross section is in some tension with the results of the CMS monojet search \cite{CMS48}; including the constraints from this reduces the local significance to $3\sigma$.  This model is also consistent with the limits from dijet searches\footnote{In fact, there appears to be a $\sim 2\sigma$ upward fluctuation in the data around $\sim 1.1\tev$ which could increase the significance of the excess, but unfortunately, not enough information is provided by the CMS collaboration to be able to precisely calculate the significance.} \cite{cms-dijet},
which would be implied by the associated decay of the mediator back to a pair of colored particles. 

The stated $3\sigma$ excess is, of course, only the local significance. While one might expect that the global significance to drop significantly after the application of the look-elsewhere-effect (after all, our rectangular aggregation technique covers some 33,000 rectangles), this is in fact not what occurs. The look-elsewhere-effect (LEE) (for a nice discussion, see e.g.~\cite{Gross:2010}) is rigorously defined only in terms of a specific model, and as we will demonstrate in Section~\ref{sec:lookelsewhere}, for the resonant colored particle model we find the global significance is $\sim 2\sigma$ after the LEE is applied. In essence, even though there are a very large number of rectangular aggregations which we scan over, they are highly correlated (rectangles overlap and nest inside one another), so there are not 33,000 independent chances for our technique to ``look-elsewhere'' and find a random fluctuation. Also, any particular model will only populate a small subset of the RAs, further reducing the impact of the LEE; for instance, the monojet model described above will only populate SRs with $N_j\lesssim 3$ and $N_b=0$.

Regardless of the LEE, we believe that the method of rectangular aggregations developed in this paper has value in identifying potential ``hot-spots'' in the existing LHC data.  It is important to proactively identify these hot-spots now: if new physics is accessible at the LHC, there will be a time when its statistical evidence is at the (somewhat marginal) $\sim 3\sigma$ level considered in this paper. These hot-spots are clearly worthy of further study by both theorists and experimentalists. For theorists, they are a useful starting point for model building, which can in turn focus attention on additional correlated channels (such as the dijet resonances in the worked jets+$\slashed{E}_T$ example), and lead to more optimized search strategies. For experimentalists, these hot-spots are regions which should be continually monitored with more data to see which grow and  which fade away.
Ideally, the event selections for these analysis regions should be frozen to the extent possible, allowing the evolution of their statistical significance to be tracked as more data is collected. Without this proactive approach, increases in triggers and selection cuts could blind the LHC to nascent excesses. 

Having identified these potentially interesting regions in the existing data, any future statistical significance does not pay a price from the LEE, and anomalies that grow with time would of course be immensely interesting. In this sense, the excesses we identify are postdictions of the current dataset; moving forward, they become predictions, and since the ultimate dataset will be 100$\times$ larger, these predictions still have great value.\footnote{We would like to thank Marumi Kado for emphasizing this point to us.} Constructing and tracking a model-independent list of anomalies is a program that will span the life of the LHC, and our method of rectangular aggregations is only the first step in this effort.

The outline of our paper is as follows. Section~\ref{sec:asr} provides a general description of our technique and an application to the jets+$\slashed{E}_T$ searches. 
In Section~\ref{sec:monojet}, we investigate in more detail the most promising excess identified by the RA technique, attempting to fit it to models and studying the correlated signatures.
The look-elsewhere-effect is quantified in Section~\ref{sec:lookelsewhere}. We conclude in Section~\ref{sec:conclusion}. Our statistical technique is described in detail in Appendix~\ref{app:stat}, while Appendix~\ref{app:val} describes our recasting of the CMS and ATLAS searches, and Appendix~\ref{app:badexcess} describes RAs with statistically significant deviations from background which we believe are likely {\em not} consistent with new physics.

\section{Aggregating Signal Regions}\label{sec:asr}
\subsection{Technique of rectangular aggregations\label{sec:technique}}

In this section, we describe a new, model-independent method to mine the increasingly complex and numerous SRs of the LHC searches for statistically significant excesses. As explained in the Introduction, our method is motivated by the simple idea that any new physics scenario will tend to populate some set of multiple SRs which are ``close'' to each other in both topology and the kinematic variables. Lacking a more sophisticated-yet-model-independent template for how events are distributed across SRs, we choose to simply aggregate together SRs in a ``rectangular'' fashion.\footnote{Clearly, our reliance on rectangular aggregation regions could be sub-optimal for new physics scenarios that populate SRs in more complicated patterns, for example when two variables are highly correlated or when multiple particles with different masses or decay topologies are produced. This is an interesting possibility, which deserves future study.}  By considering all possible rectangular aggregations (RAs) of any size in a given search, we ensure sensitivity to a wide range of possible signals.

\begin{figure}[t]
\begin{center}
\includegraphics[scale=0.8]{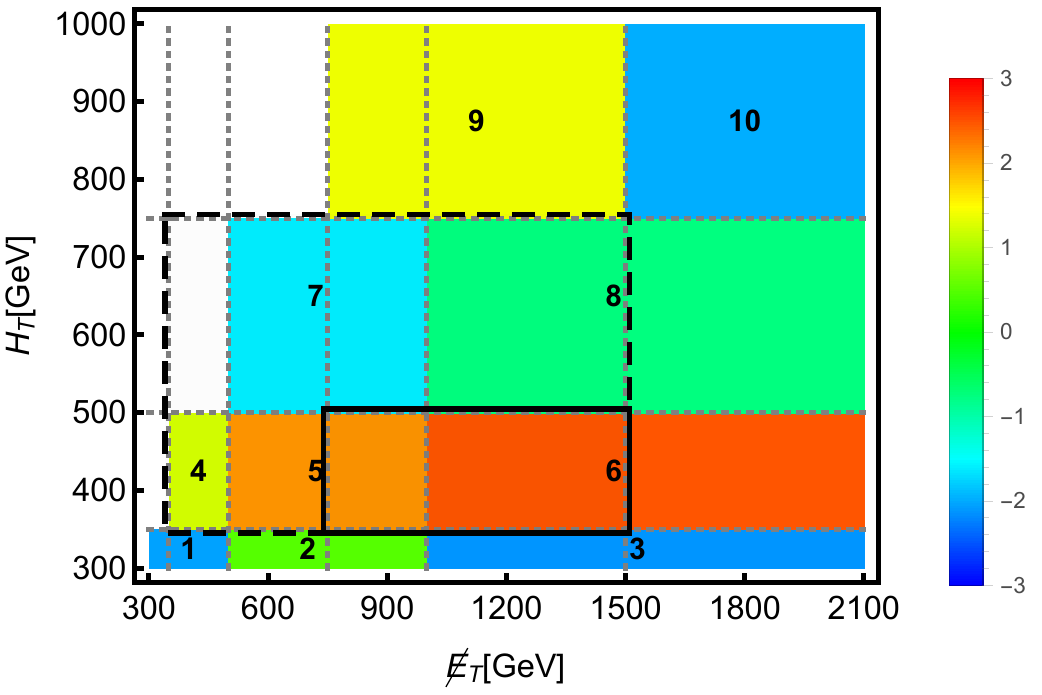}
\caption{ The rectangular aggregation method with two kinematic variables on a particular set of 10 SRs from \cite{CMS33}. The colored rectangles represent individual signal regions, while the dotted gray lines illustrate the narrowest binning in each kinematic variable. The color-coding of each SR shows the pulls (in units of standard deviations) of a mock dataset which includes statistical fluctuations and an injected signal, with red (blue) regions having upward (downward) fluctuations with respect to the background. The black rectangles (solid and dashed) delimit two of the many possible choices of RAs. As described in the text, the SRs overlapping with a rectangle will be aggregated to form an RA.
}
\label{fig:rectangle}
\end{center}
\end{figure}

We illustrate the general idea behind this method in Figure~\ref{fig:rectangle}, where two kinematic variables ($H_T$ and $\slashed{E}_T$) are used in the SR definitions. (For searches that have more than two kinematic variables -- as is generally the case -- our rectangular aggregation scheme is extended in the obvious way.)
 Each bin is color-coded according to the statistical pull with respect to the background in a mock dataset, while the black rectangles show a few possible RAs. 
 In particular, the solid rectangle shows a small aggregation (SRs 5-6) resulting in large local significances, while the dashed rectangle exemplifies an over-aggregated RA (SRs 4-8) which washes out the excesses in the underlying bins.  Figure~\ref{fig:rectangle} also illustrates an important complication: the SRs are typically not uniformly spaced in the kinematic variables. In this case we make all possible rectangles given by the finest possible binning (denoted by dotted gray lines in Figure~\ref{fig:rectangle}) in each kinematic variable. For each rectangle, we then add the SRs that {\it overlap} with that rectangle due to the non-uniform binning. 

To compute the local significance of a given RA containing an excess, we make use of the (asymptotic) profile likelihood method described in  \cite{Cranmer1007.1727}. Specifically, we compute the test statistic $q_0$, which is the difference of the profile log-likelihoods of the background-only hypothesis and the background plus best-fit signal-strength hypothesis. As shown in \cite{Cranmer1007.1727},  in SRs with large event counts, $N_\sigma=\sqrt{q_0}$ can be translated to a $p$-value  through the Gaussian distribution, e.g. $N_\sigma=2$ corresponds to a $p$-value of 0.05. 
(The full details of our statistical technique are discussed in Appendix~\ref{app:stat}.) Here our key assumption about the signal hypothesis is that the signal {\it populates only the given RA} and nowhere else.\footnote{We ignore the possibility of control region contamination. We discuss this issue further in the context of specific models in Section \ref{sec:monojet}.   
} 

In the calculation of the profile likelihood, correlated uncertainties in the background estimates of individual SRs play an essential role. 
These correlations are often sizable, especially between nearby bins whose background expectations are inferred from correlated control samples. If these correlations are not included in the calculation of $q_0$, the resulting significances can be wildly off. 

Fortunately, the CMS collaboration has recently started releasing full correlation and covariance matrices for their searches. To incorporate them into our RA procedure, we should sum the entries from the aggregate bins in the full covariance matrix $V$ to form a reduced matrix $V_R$. Explicitly, we construct a new vector of observed and expected events, as well as a new covariance matrix, given by:
\begin{align}
\vec{O}_R & =\left(	\sum_{i \in R} O_i ,~ O_I \right),\qquad 
\vec{E}_R  = \left(	\sum_{i \in R} E_i ,~ E_I \right),\qquad
\label{eq:eventR}
V_{R} &=\left(\begin{matrix}
\sum_{i,j \in R} V_{ij} & \sum_{i\in R} V_{iJ}\\
\sum_{j\in R} V_{Ij} & V_{IJ}
\end{matrix} \right) ,
\end{align}
where $\vec{E}\ (\vec{O})$ is the vector of expected background (observed data) for all bins, $(i,j)$ refer to bins that are being aggregated in the rectangle $R$, and $(I,J)$ refer to bins that are not included in the rectangle $R$. The signal hypothesis (with arbitrary normalization) is then the vector $\vec S_R=(1,0,\dots, 0)$ in this basis. 

As an example of the importance of correlations, in ROI 1b constructed from the SRs of  \cite{CMS36} (to be described below), the full calculation including correlations yields 2.95$\sigma$, while neglecting correlations reduces the significance to just 1.8$\sigma$. The reason can ultimately be traced back to the fact that the correct measure of the error is $1/(V_R^{-1})_{11}$ (see Eq.~\eqref{eq:simplL}), and this is very different (and in this case much smaller) than $(V_R)_{11}$, due to the presence of correlations.

Using our method of rectangular aggregations, we can generate in any search a model-independent list of RAs with locally high statistical significance.  We can then follow this up with a more detailed study of the excesses in these RAs, and whether they are compatible with any actual new physics models. In the next subsection, we will apply this method to the CMS jets+$\slashed{E}_T$ searches.

\subsection{Application: jets plus missing energy searches \label{sec:jetsMET}}

Currently, CMS has two jets+$\slashed{E}_T$ searches using the full 36/fb Run II dataset: \cite{CMS33} and \cite{CMS36}, which we will also refer to via their PAS identifiers: CMS033 and CMS036, respectively. The kinematic variables used in CMS033 are the number of jets $N_j$ with $p_T>30\gev$, number of $b$-tagged jets $N_b$, the scalar sum of jet $p_T$ ($H_T$), and the missing transverse momentum \MHT$=\slashed{E}_T$.\footnote{\MHT~ is defined as the missing energy formed from jets only. However, given the lepton veto in these searches, in this case the distinction from $\slashed{E}_T$ is negligible. Thus, we will refer to it as $\slashed{E}_T$.}  CMS036  uses $N_j$, $N_b$, $H_T$, and the stransverse mass variable $M_{T2}$ \cite{Lester:1999tx,Barr:2003rg,Lester:2014yga}. Apart from small triggering differences and the use of $\slashed{E}_T$~vs.~\MTT, the selections are very similar, so that events in both searches are largely overlapping. 

In fact, it is even possible to directly and rigorously map SRs of CMS036 onto a set of SRs of CMS033, using a simple inequality between $M_{T2}$ and  $\slashed{E}_T$. We will make use of this fact to test the compatibility of any excesses in the former with the latter. $M_{T2}$ is calculated in CMS036 by iteratively grouping all jets into two pseudo-jets, and then computing the transverse mass using each pseudojet and the missing momentum. The key observation is that for two (pseudo)jets, the stranverse mass can actually be calculated analytically as $M_{T2}^2=2p_{Tj1}p_{Tj2}(1+\cos\theta_{12})$ \cite{Lester:2011nj} (which is the same expression as the contransverse mass $M_{CT}$  \cite{MCT1,MCT2}). Meanwhile, $\slashed{E}_T^2 = p_{Tj1}^2+p_{Tj2}^2 + 2p_{Tj1}p_{Tj2}\cos\theta_{12}$, making it clear that $M_{T2} \le \slashed{E}_T$ in every event.

There are 174 individual SRs in CMS033  and 213 in CMS036, which when combined in four-dimensional rectangles result in roughly 7,000 and 33,000 possible aggregations, respectively.  After scanning over all rectangular collections of signal regions, we found several deviations from the background-only hypothesis; we summarize the most statistically significant (with a local $p$-value below 1\%, or equivalently, $N_\sigma>2.57$) discrepancies  in Table \ref{tab:excess36} for CMS036~\cite{CMS36} and Table \ref{tab:excess33} for CMS033~\cite{CMS33}. In order to avoid double-counting aggregations where adding nearby SRs does not appreciably increase the significance, we only list aggregations which do not contain a smaller RA with $N_\sigma$ greater than $0.9$
of the larger region's significance.

We find 14 and 10 aggregations in CMS036 and CMS033, respectively, that are above our threshold of 1\% local $p$-value. However, they are not all independent from each other. Rather, they form distinct clusters or ``hot-spots" in parameter space, with nested and overlapping aggregations. We will refer to these clusters as ``regions of interest" (ROIs) in what follows.  
Altogether, we find five ROIs in Table~\ref{tab:excess36} for CMS036 and three in Table~\ref{tab:excess33} for CMS033.

\begin{table}
\resizebox{\columnwidth}{!}{%
\begin{tabular}{cc|c|c|c|c|c|c|c}
\hline 
\multicolumn{2}{c|}{ROI} & bins & $N_j$ & $N_b$ & $H_T$ (GeV) & \MTT (GeV) & $N_\sigma$ & compatible?\\
\hline 
\multirow{5}{*}{1} & a & 126-130, 132-136 & $2-3$ & $0-1$ & $1000-1500$ & $\geq 400$ & 3.5   & \multirow{5}{*}{\xmarkk \MET}\\ 
 & b & 126-127, 132-133 & $2-3$ & $0-1$ & $1000-1500$ & $400-800$ & 3.36  &\\
 & c & 126-127 & $2-3$ & $0$ & $1000-1500$ & $400-800$ & 3.09  \\
 & d & 127-130, 133-136 & $2-3$ & $0-1$ & $1000-1500$ & $\geq 600$ & 2.68   \\ 
 & e & 126, 132 & $2-3$ & $0-1$ & $1000-1500$ & $400-600$ & 2.57   \\
\hline 
\multirow{5}{*}{2}  & a & 1, 2, 8, 9, 13, 16 & $1-3$ & $0-1$ &$250-450$ & $200-300$ & 3.3 & \xmarkk$N_b$\\
 & b & 1, 2, 13 & $1-3$ & $0$ &$250-450$ & $200-300$ & 2.95 & \cmark\\
 & c & 1, 8, 13, 16 & $1-3$ & $0-1$ &$250-450^{*}$ & $200-300$ & 2.93 & \xmarkk$N_b$\\
 & d & 1, 13 & $1-3$ & $0$ &$250-450^{*}$ & $200-300$ & 2.74  & \cmark \\
 & e & 1, 2, 8, 9 & $1$ & $0-1$ &$250-450$ & $-$ & 2.6  & \xmarkk$N_b$\\
\hline
\multirow{2}{*}{3} & a &12, 79 & $1-3$ & 1 & $575^{\dag}-1000$ & $200-300$ & 3.03 &\multirow{2}{*}{\cmark}\\
 & b & 79 & $2-3$ & 1 & $575-1000$ & $200-300$ & 2.84\\
\hline
4 & & 44, 45, 60, 61 & $2-6$ & 2 & $450-575$ & $\geq 400$ & 2.76 & \xmark $H_T$ \\
\hline
5 & & 99 & $4-6$ & 1 & $575-1000$ & $300-400$ & 2.75&\xmark \MTT\\
\hline
\end{tabular} 
}
\caption{The aggregated regions in CMS036 \cite{CMS36} with the highest local discrepancy between the data and the background. We group significant subsets and overlapping aggregations into ROIs in this table. The asterisk (*) in the $H_T$ columns marks a requirement that do not apply to all the aggregated bins, in particular bins 1 and 8 have only $H_T<350\gev$. Similarly, the dagger ($\dag$) denotes that bin 12 has $H_T>700\gev$. Also note that the \MTT~requirement does not apply to  $N_j=1$ bins. We mark the compatibility of each excess by \cmark~(if compatible), \xmark~(if not compatible with nearby SRs in the same search) and \xmarkk~(if incompatible with other searches). In case of incompatibility, we list the kinematic variable responsible.
}
\label{tab:excess36}
\end{table}

\begin{table}
\centering
\resizebox{\columnwidth}{!}{%
\begin{tabular}{cc|c|c|c|c|c|c|c}
\hline 
\multicolumn{2}{c|}{ROI} & bins &$N_j$ & $N_b$ & $H_T$ (GeV) & \MHT (GeV) & $N_\sigma$ & { compatible?}\\
\hline 
\multirow{4}{*}{1} & a & { 13,16, 23,26, 43,46, 53,56, 63,66} & $2-4$ & $\geq1$ & $>1000$ & $300-500$ & 3.11 & \xmark $N_j,N_b$\\
 & b & { 13,16, 23,26, 43,46, 53,56 } & $2-4$ & $1-2$ & $>1000$ & $300-500$ & 2.77 & \cmark \\
 & c & { 13,16, 43,46, 83,86, 120,122}  &$2-8$  & $1$ & $>1000$ & $300-500$ & 2.65 & \xmark $N_j$\\
 & d & { 21-26, 51-56, 61-66}  & $2-4$ & $\geq2$ & $>300$  & $300-500$ & 2.64 & \xmark $N_j,N_b$\\
\hline
\multirow{4}{*}{2} & a & 1, 4, 31, 34, 71, 74 & $2-6$ & 0 & $300^{*}-500$ &$300-500$& 2.96 &\cmark \\
 & b & 71, 74, 81, 84 & $5-6$ & $0-1$ & $300^{*}-500$ &$300-500$ & 2.70 & \cmark \\
 & c & 1, 4, 31, 34& $2-4$ & 0 & $300^{*}-500$ &$300-500$ & 2.64 &\cmark\\
 & d & 31, 34, 71, 74& $3-6$ & 0 & $300^{*}-500$ &$300-500$ & 2.57 &\cmark\\
\hline
\multirow{2}{*}{3} & a & 125-126 & $7-8$ & 1 & $>750$ &$>750$ & 2.81&  \multirow{2}{*}{\xmark $N_j$}\\
 & b & 126 & $7-8$ & 1 & $>1500$ &$>750$ & 2.73 \\
\end{tabular} 
}
\caption{The aggregated regions in CMS033~\cite{CMS33} with the highest local discrepancy between the data and the background. The asterisks in the $H_T$ column mark requirements that do not apply to all the aggregated bins, that is, $H_T>350\gev$ for SRs 4, 34, 74, 84. We mark the compatibility of each excess by \cmark~(if compatible), \xmark~(if not compatible with nearby SRs in the same search) and \xmarkk~(if incompatible with other searches). In case of incompatibility, we list the kinematic variable responsible.
}
\label{tab:excess33}

\end{table}

\subsubsection{Discriminating statistical fluctuations from viable excesses \label{sec:statfluct}}

Obviously, we do not expect all of these excesses to be due to new physics, and at the very least several, if not all, of them should be caused by statistical fluctuations of the SM background. While in some cases the only way to determine whether an excess is a real signal of beyond-the-Standard Model (BSM) physics is to wait for more data, there are two tests we can apply in order to guide our reasoning with the information on hand: 
\bitem
\item Incompatibility with nearby bins in the same search, due to general properties of kinematic variables.

For example, a signal populating an RA with high jet multiplicity and a narrow $H_T$, $M_{T2}$  or $\slashed{E}_T$ range can be disfavored if nearby bins see deficits or no sizable excesses, because it is unlikely for these distributions to be so localized if they come from a realistic many-jet signal. Similarly, we expect distributions of $N_j$ to be smeared around some underlying partonic value due to ISR/FSR. Finally, we expect $N_b$ distributions to be consistent with $b$-tagging rates (or $c$-mistagging rates). 

In Tables~\ref{tab:excess36} and \ref{tab:excess33}, we denote with a \xmark~symbol those statistical excesses we have identified in CMS036 and CMS033 which we believe are not compatible with the signal regions in the same search. 
   
\item Incompatibility with similar bins in other searches. 

As discussed above, CMS033 and CMS036 are highly overlapping -- the kinematic variables defining the SRs in the two searches are largely identical, except that CMS036 uses $M_{T2}$ and CMS033 uses $\slashed{E}_T$. Thus an excess in one search will usually populate analogous bins in the other.  We can make this more precise using the inequality $M_{T2}\le\slashed{E}_T$ derived above: a signal generating an excess in a particular RA of CMS036 will show up in specific SRs of CMS033 (the converse is not always true, in particular CMS036 would not be sensitive to a model with low \MET~and $M_{T2}\ll\slashed{E}_T$, which would only populate CMS033 SRs).

In Table~\ref{tab:excess36}, we denote with a \xmarkk~symbol those statistical excesses we have identified in CMS036 which we believe are not compatible with the signal regions of CMS033.

\eitem

Applying these arguments to the excesses listed in Tables~\ref{tab:excess36} and \ref{tab:excess33} marks roughly half of the anomalies as unlikely to be anything other than statistical fluctuations. As the detailed listing is rather tedious, we point the reader to Appendix~\ref{app:badexcess}, and in particular to Figures~\ref{fig:SRhistos36} and~\ref{fig:SRhistos33} for histograms illustrating the incompatibility. Of course, it is possible that some of these disfavored excesses could be due to a combination of new physics events and an upward fluctuation in background. So while we do not spend time constructing models for them, tracking their evolution with more data will still be useful and important.

\begin{figure}[t]
\begin{center}
\includegraphics[width=0.45\columnwidth]{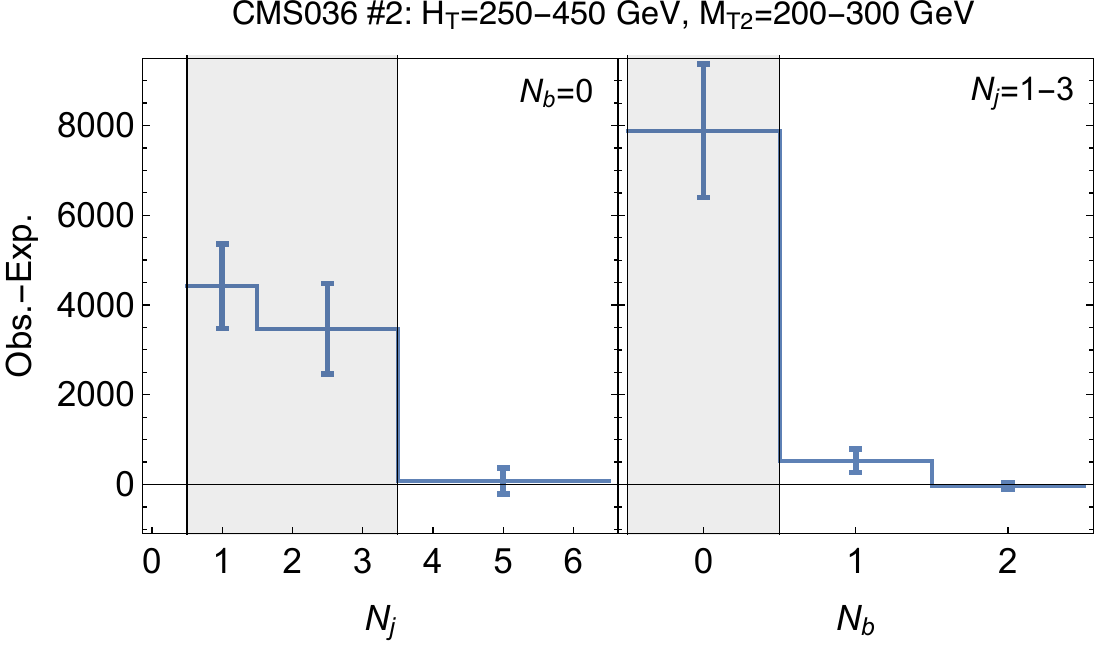}\qquad
\includegraphics[width=0.45\columnwidth]{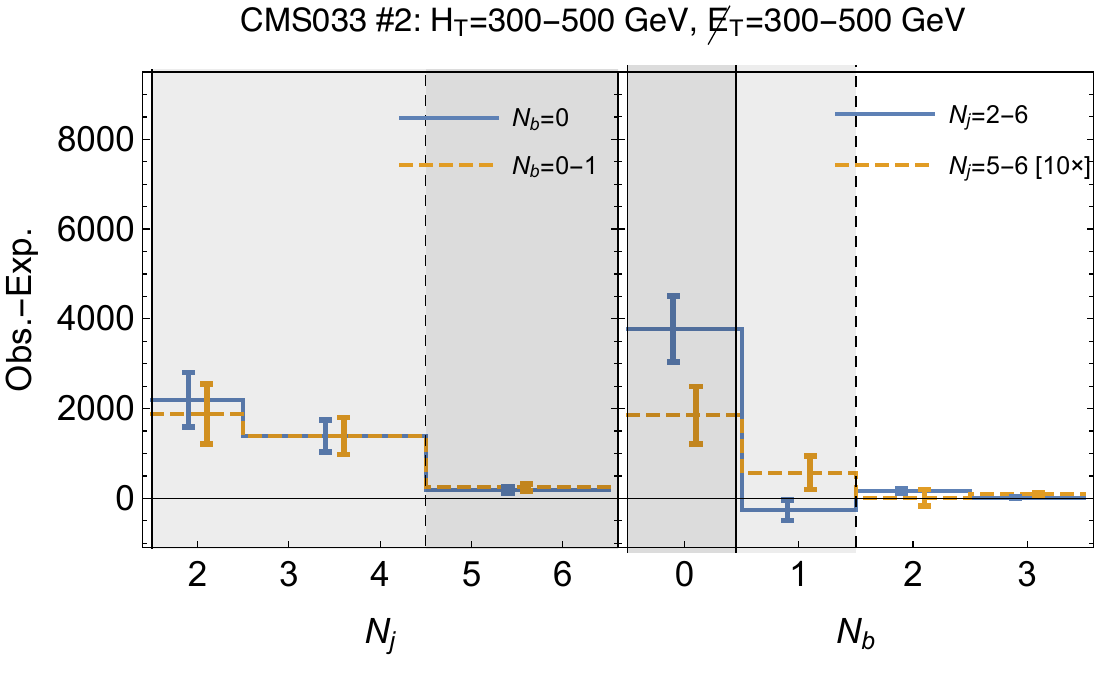}

\includegraphics[width=0.42\columnwidth]{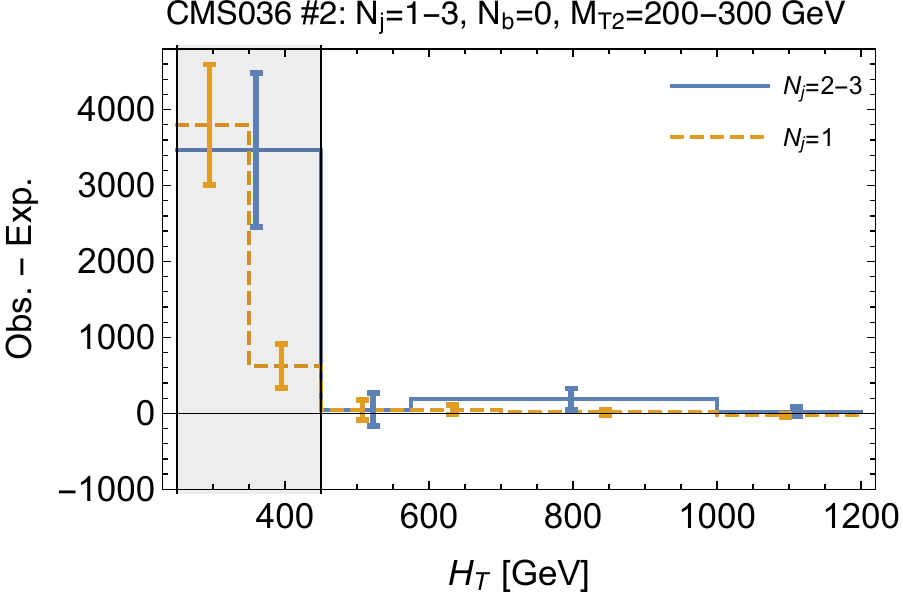}\qquad
\includegraphics[width=0.42\columnwidth]{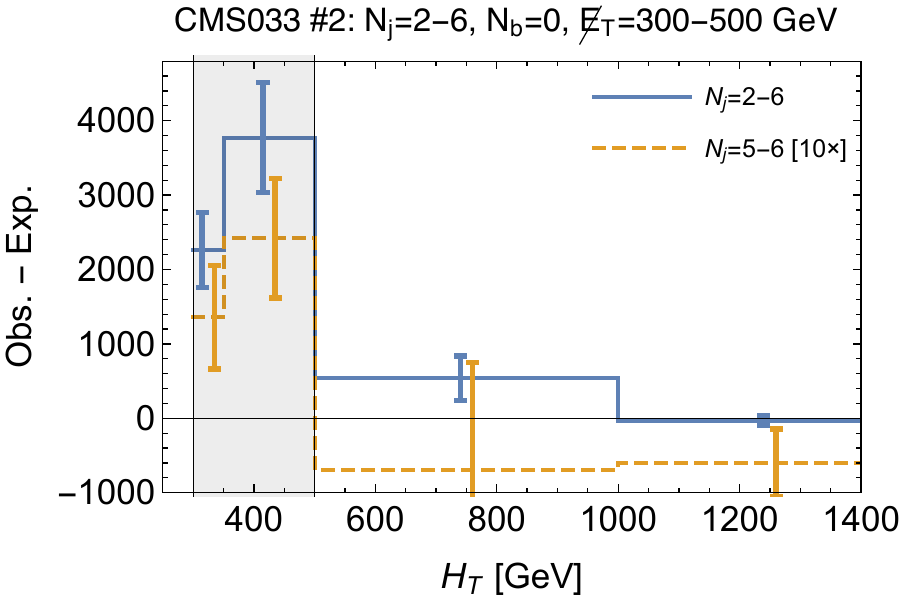}

\includegraphics[width=0.42\columnwidth]{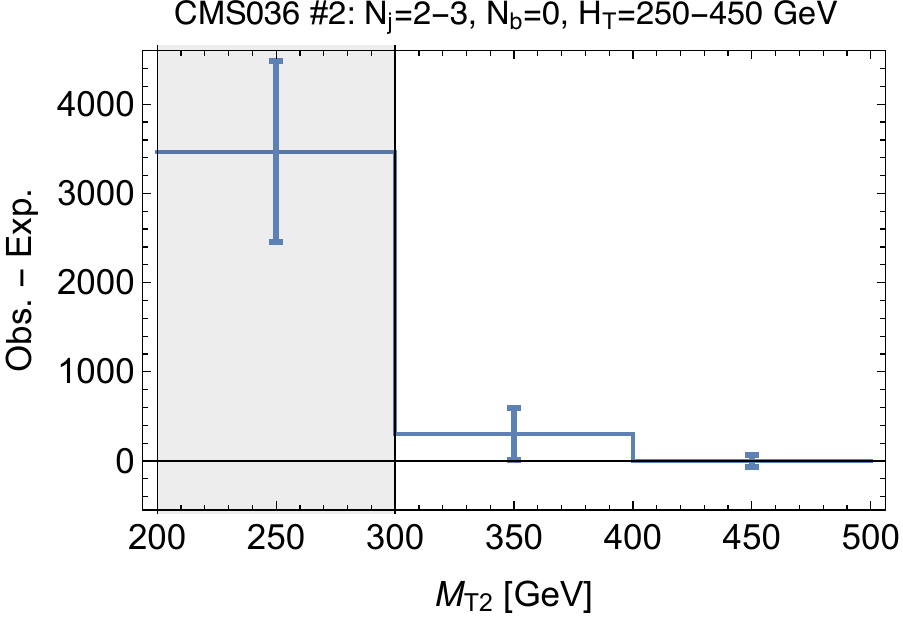}\qquad
\includegraphics[width=0.42\columnwidth]{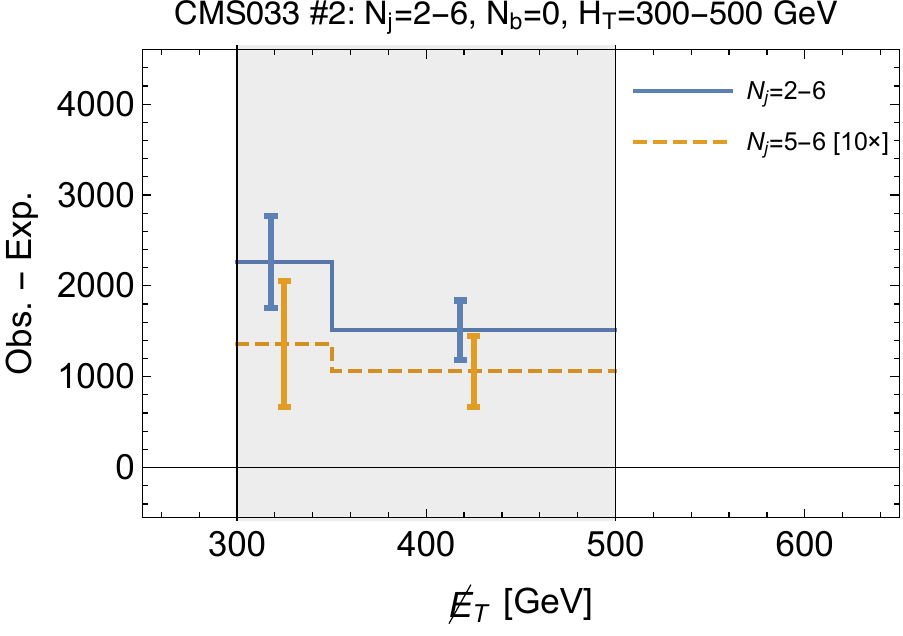}

\caption{Distributions of the residuals (observed minus expected counts) for the broadly compatible excesses of CMS036 and CMS033, ROIs \#2 in Table~\ref{tab:excess36} and Table~\ref{tab:excess33}, with error bars denoting the uncertainty, as explained in the text. The left column shows kinematic distributions for CMS036 ROI \#2 while the right column displays CMS033 ROI \#2. Within each column, from top to bottom we show the $N_j,N_b,H_T$ and \MTT~(\MET~ for CMS033) distributions of the significant aggregation (shaded in gray) and the neighboring bins in that direction in kinematic space. Solid and dashed lines show different components of each aggregation, as labeled in the legends. See text for more details.}
\label{fig:histo_excesses36}
\end{center}
\end{figure}

\begin{figure}[t]
\begin{center}

\includegraphics[width=0.45\columnwidth]{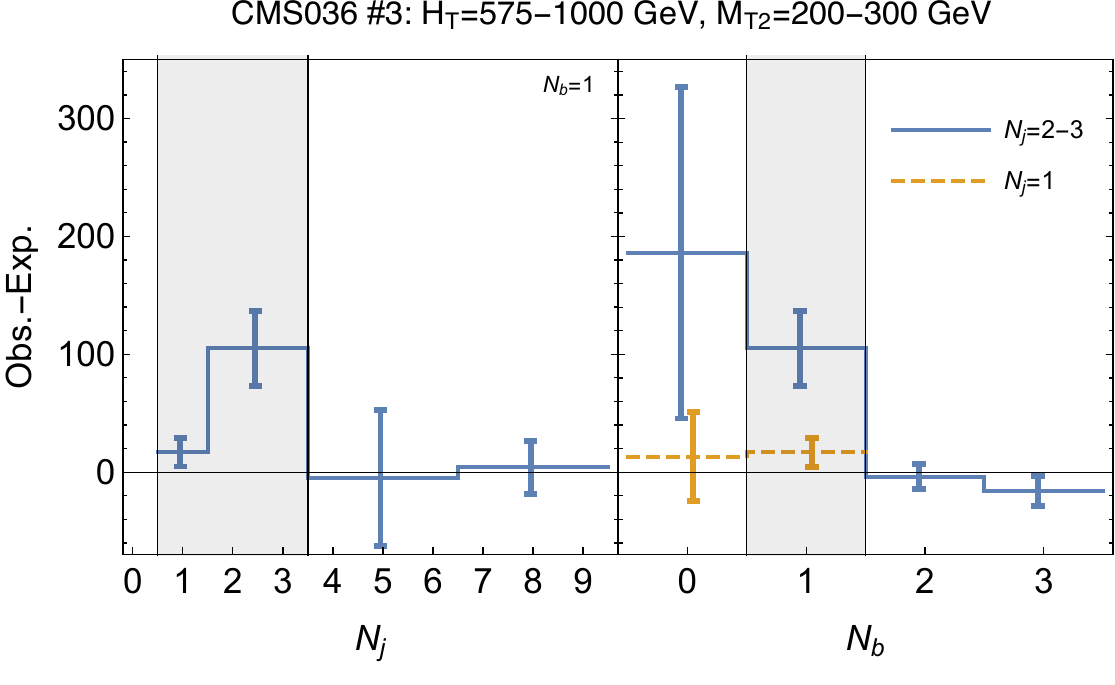}\qquad
\includegraphics[width=0.45\columnwidth]{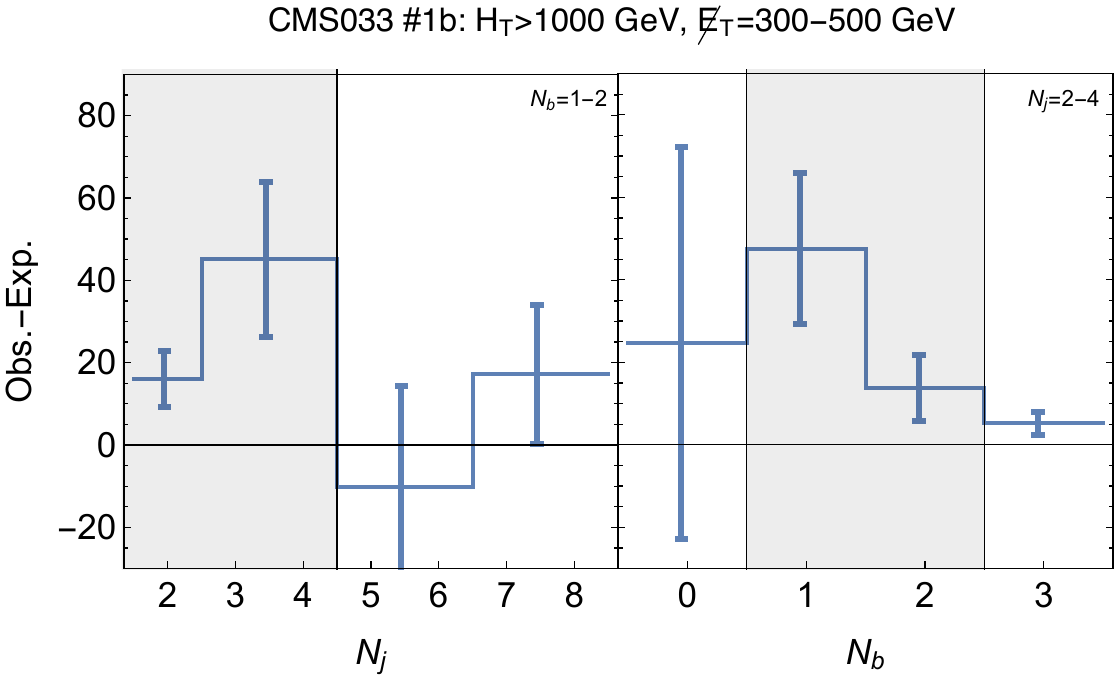}

\includegraphics[width=0.42\columnwidth]{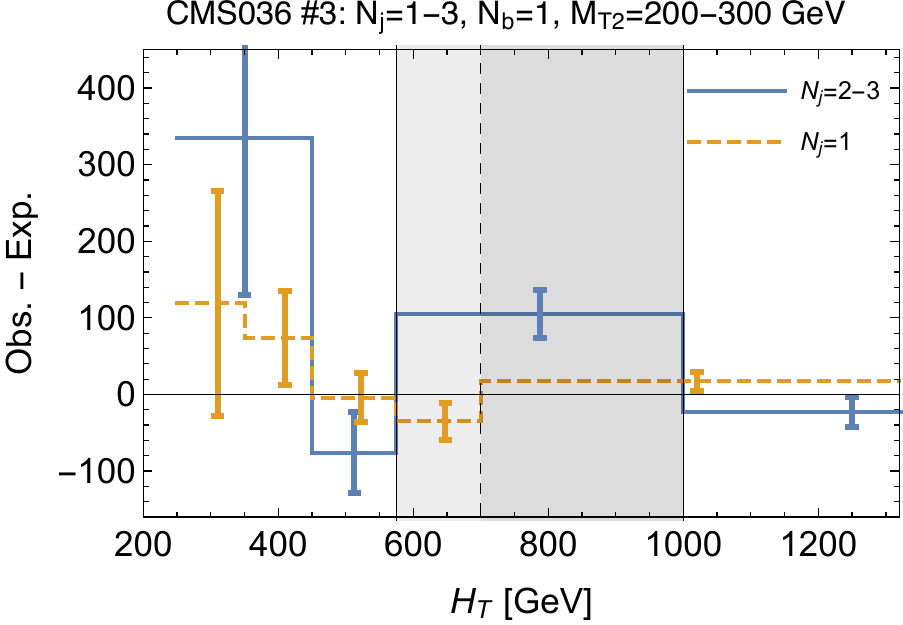}\qquad
\includegraphics[width=0.42\columnwidth]{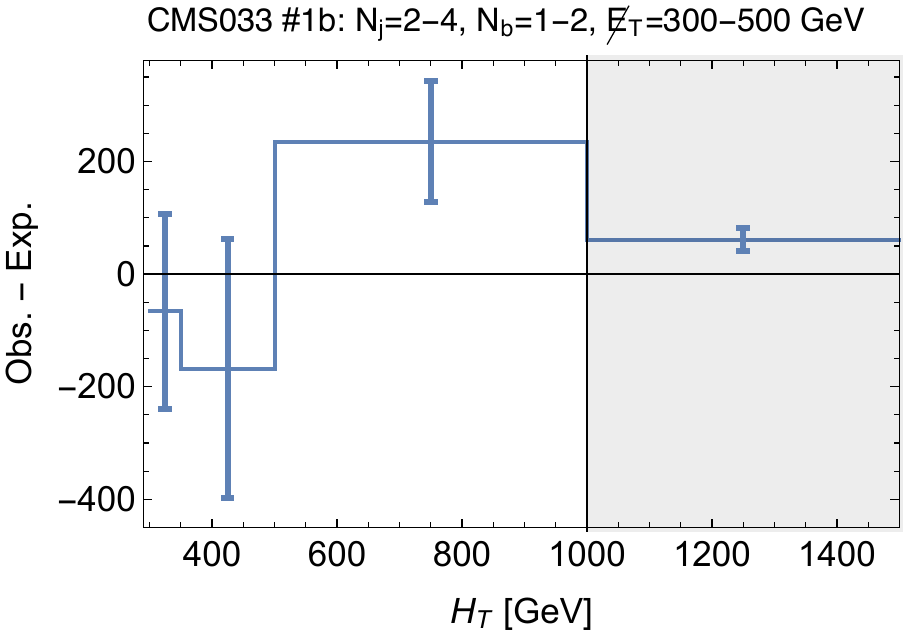}

\includegraphics[width=0.42\columnwidth]{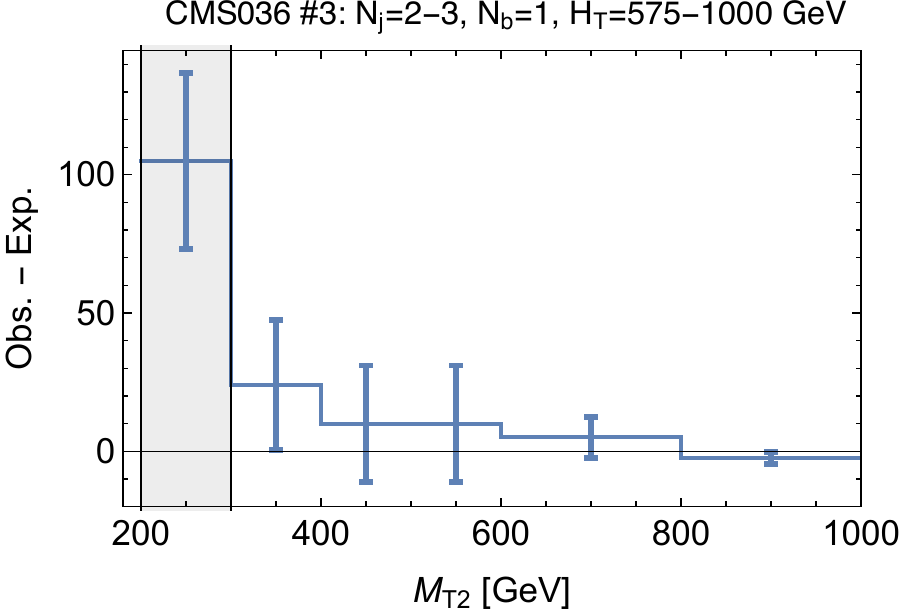}\qquad
\includegraphics[width=0.42\columnwidth]{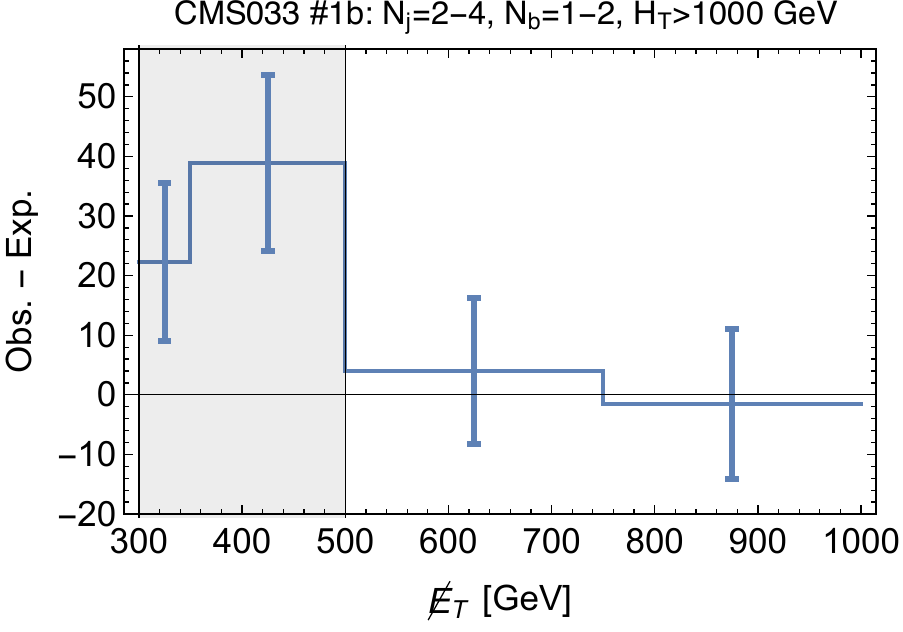}

\caption{Distributions of the residuals  (observed minus expected counts)  for the remaining excesses in Tables~\ref{tab:excess36} and ~\ref{tab:excess33}, with error bars denoting the uncertainty, as explained in the text. The left column shows  kinematic distributions for CMS036 ROI \#3 while the right column displays CMS033 ROI \#1b. The variables plotted and the color-coding are the same as in Fig.~\ref{fig:histo_excesses36}.
}
\label{fig:histo_excesses33}
\end{center}
\end{figure}

\subsubsection{Promising excesses}

We now focus on the anomalies which we believe have the most potential to be new physics. In Figures~\ref{fig:histo_excesses36} and \ref{fig:histo_excesses33}, we show the kinematic distributions of the residuals (difference between observed and expected event counts) for the viable groups of excesses in both searches. We highlight the location of the excess in each kinematic variable ($N_j,N_b,H_T,M_{T2}$ or \MET) with a gray shading. 
 The error bars represent (approximated) uncertainties on the background expectations: the error on the $i$-th bin is taken to be $(E_i+1/(V^{-1})_{ii})^{1/2}$, where $E$ is the vector of expected backgrounds and $V$ is the covariance matrix (possibly reduced after aggregating, as in Eq.~\eqref{eq:eventR}). As discussed in Section~\ref{sec:technique}, 
because of the correlations, $1 / (V^{-1})_{ii}\neq V_{ii}$, and the inverse of the covariance matrix is the correct measure of the uncertainty, as it enters the likelihood calculation, see Eq.~\eqref{eq:simplL}. 
It can be seen that for each excess and kinematic variable, the neighboring regions can accomodate tails of a BSM signal (as opposed to the other non-viable aggregations in Appendix~\ref{app:badexcess}), either due to deviations from the background, or large error bars.

Each column in Figures~\ref{fig:histo_excesses36} and \ref{fig:histo_excesses33} shows a viable cluster of excesses as listed in Tables~\ref{tab:excess36} and~\ref{tab:excess33}.
\bitem
\item We start in the left column of Figure~\ref{fig:histo_excesses36} with CMS036 ROI \#2, which has low jet multiplicity, low $H_T$ and low \MTT. In the top plot, we show the $N_j$ and $N_b$ distributions, in particular only showing the aggregation \#2b; the center plot shows the $H_T$ distribution, where we separately plot the different $N_j$ bins as solid and dashed lines. The finer binning for $N_j=1$ (orange-dashed) demonstrates that a signal would likely be steeply falling with $H_T$. Note that aggregations \#2b and \#2d only differ by the second bin of the $N_j=1$ distribution ($p_T^{j_0}=350-450\gev$) which only marginally increases the significance. Finally,  the bottom plot shows the $M_{T2}$ distribution of the $N_j\geq2$ bins (note that $M_{T2}$ is not defined for $N_j=1$).
\item In the right column, we show the kinematic distributions for the very similar CMS033 ROI \#2, which is formed by signal regions with low $H_T\sim$\MET. Four separate significant aggregations are possible, of which two are shown in solid blue (\#2a), dashed yellow (\#2b, which for presentation purposes is rescaled by a factor of $10$). We do not plot aggregation \#2c and \#2d to avoid cluttering the figure:  \#2c is mostly degenerate with  \#2a, differing only by the $\sim100$ events in the $N_j=5-6$ bins, while \#2d drops the highly populated $N_j=2$ bin, but has similar shapes as \#2a for the other distributions. In the $N_j, N_b$ plots, we separately show the excess location with overlapping gray shading: for example, aggregation \#2b has $N_j=5-6$ and $N_b=0-1$ which is represented by a gray shaded area delimited by a dashed line. It should be noted that in the $H_T=300-500\gev$ range, CMS033 does not have bins at $\slashed E_T>500\gev$, which is why there are no data points above $500\gev$ in the bottom plot.
\item We now turn our attention to the remaining excesses in Figure~\ref{fig:histo_excesses33}. The left column of Figure~\ref{fig:histo_excesses33} illustrates CMS036 ROI \#3, which has low jet multiplicity, moderate $H_T$ and low \MTT: we note that the core of the excess (\#3b) has $N_j=2-3$, with the $N_j=1$ bin increasing the significance only slightly (as seen in the $N_j$ plot on top). We therefore show separately these two bins as solid and dashed. In particular, we note that aggregation \#3a (which includes $N_j=1$) requires a narrower $H_T$ range, shown in darker gray delimited by dashed vertical line.
\item Finally,  in the right column, we show the remaining viable CMS033 excess, \#1b. This excess has relatively low jet multiplicity, one or two $b$ jets, low missing energy and high $H_T$. While this might be hard to reproduce in a specific model, it is not clearly excluded according to our criteria.
\eitem

Of the excesses listed above,  CMS036 \#2b and CMS033 \#2c are particularly interesting, as they both have low jet multiplicity, no $b$-jets, and low $H_T$, $M_{T2}$ and $\slashed{E}_T$. This opens the possibility that both searches are observing the same events due to new physics. In the rest of this work, we will discuss possible BSM explanations of this pair of excesses. While we focus on this excess for the remainder of the paper, we encourage model-building efforts for the other significant aggregations listed above, as they could just as well be due to new physics. In any case, even at this point we think it is interesting and noteworthy that several $\sim 3\sigma$ anomalies can be identified in the experimental data, which is not the commonly received wisdom in the community at this point in time. 

\section{Analysis of the Mono-jet Excess \label{sec:monojet}}

In this section, we try to fit  the $\sim 3\sigma$ anomaly corresponding to CMS033 \#2b  and CMS036 \#2c to a BSM model.
For definiteness, we repeat here the kinematic properties of the two RAs: 
\begin{align}&
\begin{array}{c|cccc}
\text{aggregation (significance)}  & N_j  & N_b & H_T (\gev) & M_{T2}, \slashed{E}_T (\gev) \\\hline
\text{CMS036 \#2b (2.95$\sigma$)}  & 1-3   & 0 & 250-450 & 200-300   \\
\text{CMS033 \#2c (2.64$\sigma$)}  & 2-4   & 0 & 300-500 & 300-500   \\
\end{array}
\label{eq:monojet_excess}
\end{align}
As our calculation of the statistical preference for signal over background relied crucially on the covariance matrix, and this is only an approximation provided by the CMS Collaboration, we confirmed with the experimentalists directly that their full calculation for signal preference in these aggregated rectangles matches our results \cite{CMSpc}.

Given this final state, we also make sure to include any search that is expected to have good sensitivity. In particular, we also reinterpret the ATLAS 2-6 jets + $\slashed{E}_T$ search~\cite{ATLAS22} and the CMS mono-jet search \cite{CMS48}. The ATLAS search defines large overlapping SRs (using the variable $M_{\rm eff}=H_T+\slashed{E}_T$), of which the first one (2j-Meff-1200) has some sensitivity to (the tail of) the $N_j=2$ component of our excess.%
\footnote{However, the $M_{\rm eff}>1200\gev$ cut is too hard and greatly reduces the effectiveness of the ATLAS search. This is a prime example of the difference in the approaches of the CMS and ATLAS SUSY groups to designing their analyses, and how the many-exclusive-SR approach of CMS is much more powerful.} Meanwhile, the CMS monojet+$\slashed{E}_T$ search (denoted as CMS048 in the following) has a significant overlap with the events of CMS033 and CMS036. This search has very loose requirements ($p_T^{j_0}>100\gev$ for the leading jet and $\slashed{E}_T > 250\gev$), with any number of jets allowed, and its SRs are simply $\slashed E_T$ bins. In both cases,  we do not apply our aggregation technique to these searches, as an excess is easily identified by eye; we simply use these additional datasets to constrain the excess found in CMS033 and CMS036. 

For all the BSM models considered, we generate parton-level LHC events with \textsc{MadGraph5v2.5.3} \citep{Alwall:2014hca}, after which initial and final state radiation, as well as hadronization, are handled by \textsc{Pythia8.219} \citep{Sjostrand:2014zea}. We then simulate the detector response with \textsc{Delphes3.4} \citep{deFavereau:2013fsa} tuned to the ATLAS and CMS detectors (depending on the relevant analysis). Each recasted analysis is validated against the simplified models considered by the collaboration, see Appendix~\ref{app:val} for validation plots and more details. 

We then compute efficiencies by taking the fraction of events populating each bin, and quantify the significance of each model with the test statistic $q_0$ (described in Appendix~\ref{app:stat}) as a function of the model parameters (usually masses of the particles in the decay chain). 
It should be noted that the putative signal model for the RA method is in general different than for a defined BSM  model, in two important ways: first, our method aggregated several bins into one RA, while a BSM model can individually populate different bins within each aggregation, and potentially reach higher significance if its differential distributions are shaped like the excess events. Second, a full model typically has non-negligible tails populating nearby bins, which can both lower or increase the significance. 
Therefore, although the aggregations described in Section~\ref{sec:jetsMET} pointed us to this particular final state, we now use the full set of underlying bins (including their full correlations) to test the significance of different models of new physics. 

As noted previously, in this work we neglect the possibility of control region (CR) contamination for our hypothetical signal models. We believe this is unlikely to be an issue for the following reasons. For the jets+$\slashed{E}_T$ searches, the main background sources are $(W\to \ell\nu)$+jets (where the lepton is not reconstructed),  $(Z\to \nu\nu)$+jets and QCD multijet events where the missing energy comes from mismeasurements of the visible jets. In the first two cases, leptonic CRs are defined, while for multijets different methods are used, including inverting the $\Delta\phi$ requirement between the jets and the missing energy. Since we will only consider purely-hadronic signal models, there should be no risk in contaminating the leptonic CRs. The multijet background is typically at most a few percent of the whole background, so CR contamination should not be problematic. In any case, we expect that signal contamination would increase the number of measured events in the control regions, therefore overestimating the backgrounds in the SRs, which means that our significance estimate could be even higher.

\subsection{Possible explanations \label{sec:explanations}}

As the significance of this excess is driven by the $N_j=1$ SRs, we focus here on final states with at most one parton  and missing energy in the hard process, as we expect additional jets from ISR to populate the higher $N_j$ bins. In particular, we compare:
\bitem
\item A squark-neutralino simplified model, where a squark is produced in association with a neutralino LSP. The squark then decays to a quark and the LSP.
\item A simplified model where a particle $\phi$ is resonantly produced and decays to a jet and an invisible fermion $\psi$, resulting in missing energy. Note that $\phi$ needs to be a color triplet, as an octet cannot have a renormalizable operator leading to a two-body decay into a gluon and a color-singlet. We will discuss this model in more detail below, but here we comment that the $\psi$ particle can decay back to three quarks, and so there must be a hidden sector into which it can also decay with significant branching ratios. For the purpose of fitting the kinematics of the observed excesses, we assume the branching ratio of $\psi$ to the hidden sector is 100\%. 

\begin{figure}[t]
\begin{center}
	\subfigure[]{\includegraphics[width=0.31\columnwidth]{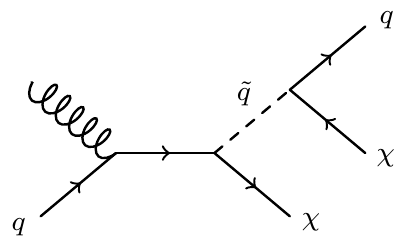}\label{fig:diagramsA}}
	\hfill
	\subfigure[]{\includegraphics[width=0.31\columnwidth]{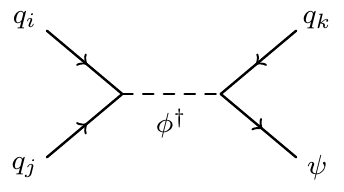}}
	\hfill
	\subfigure[]{\includegraphics[width=0.31\columnwidth]{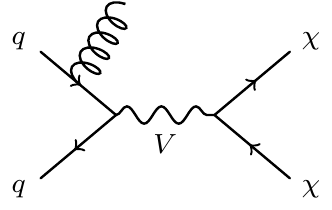}}
\caption{Representative Feynman diagrams of the prospective models: (a) squark-neutralino associated production, (b) resonant colored scalar $\phi$ decaying to a quark and a singlet fermion $\psi$, and (c) a resonant singlet vector $V$ decaying to singlet fermions $\chi$.}
\label{fig:diagrams}
\end{center}
\end{figure}

\item A simplified model with a vector mediator $V$ decaying to dark matter, $\chi$. The only jets in the event are due to ISR, with the mediator and thus the missing energy recoiling against it.
\eitem
Feynman diagrams for the three models are shown in Figure~\ref{fig:diagrams}. These models provide a representative (though not exhaustive) set of possible topologies that could fit the excess. Each model produces quite different kinematics: in the first case the squark and the first neutralino momenta are set by the proton parton distribution functions and there is a continuum choice of initial momenta leading to the production of the pair, resulting in broad distributions for the final states. In the second case, $\phi$ is resonantly produced at rest, therefore the leading jet $p_T$ and the missing energy are set by the $\phi-\psi$ mass difference, with additional jets from ISR. In the last case the mediator is resonantly produced at rest and both the jet momentum and the missing energy are distributed like the ISR, which is just a steeply falling power law dictated by QCD, with no characteristic scale.

\begin{figure}[t]
\begin{center}
\includegraphics[height=0.25\columnwidth]{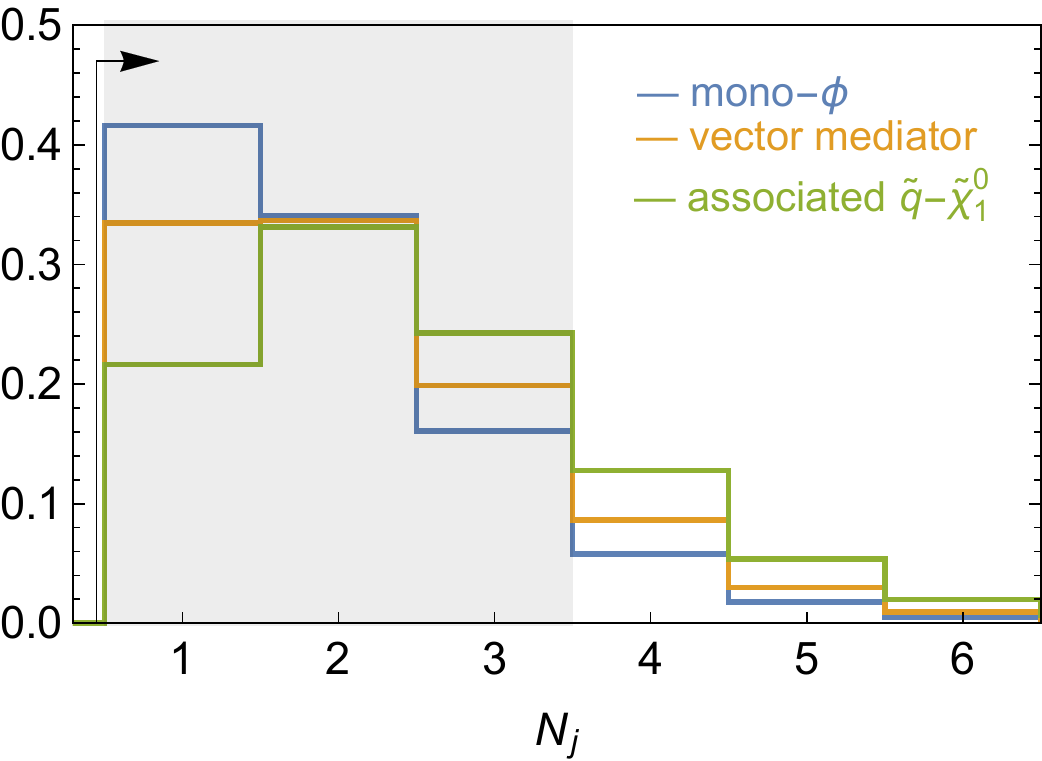}\qquad
\includegraphics[height=0.25\columnwidth]{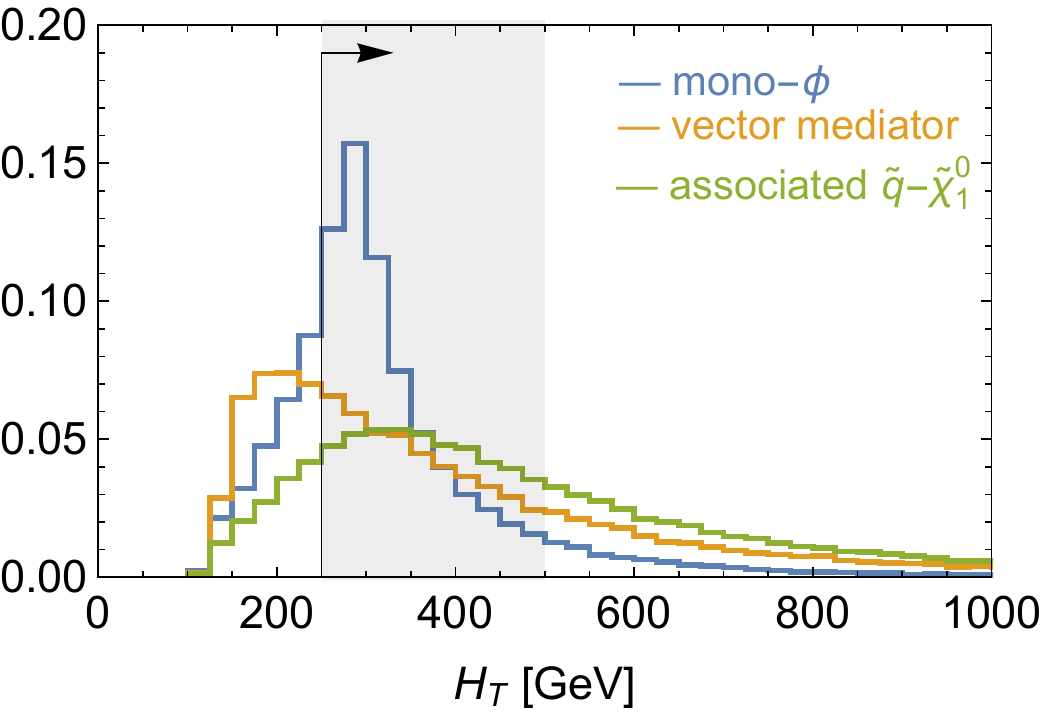}
\includegraphics[height=0.25\columnwidth]{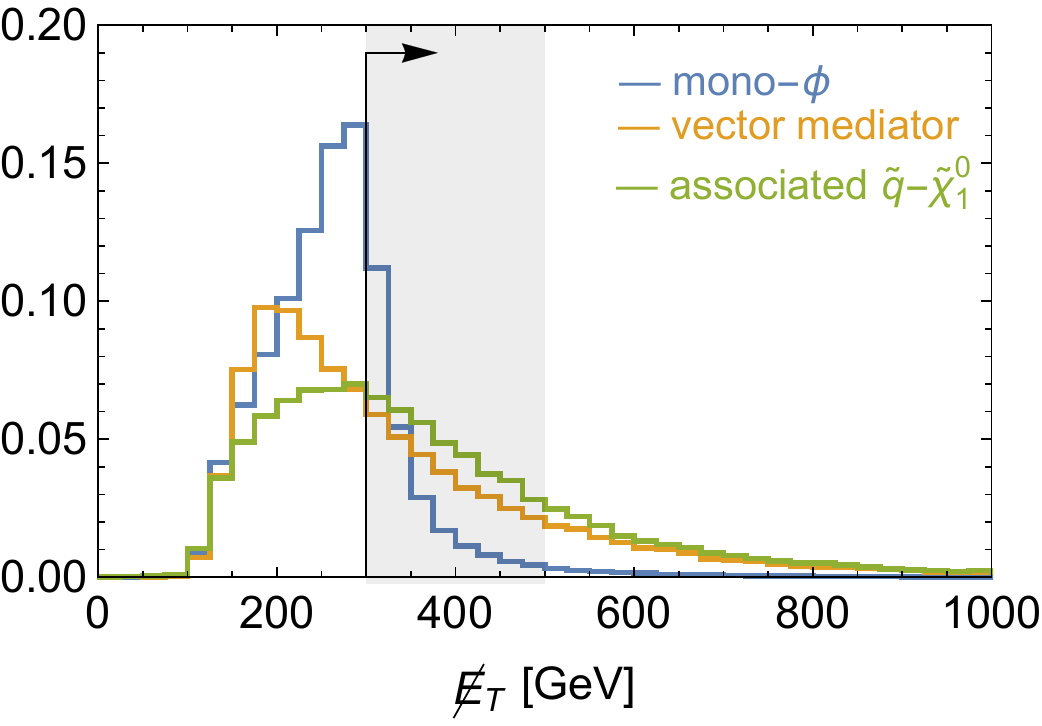}\qquad
\includegraphics[height=0.25\columnwidth]{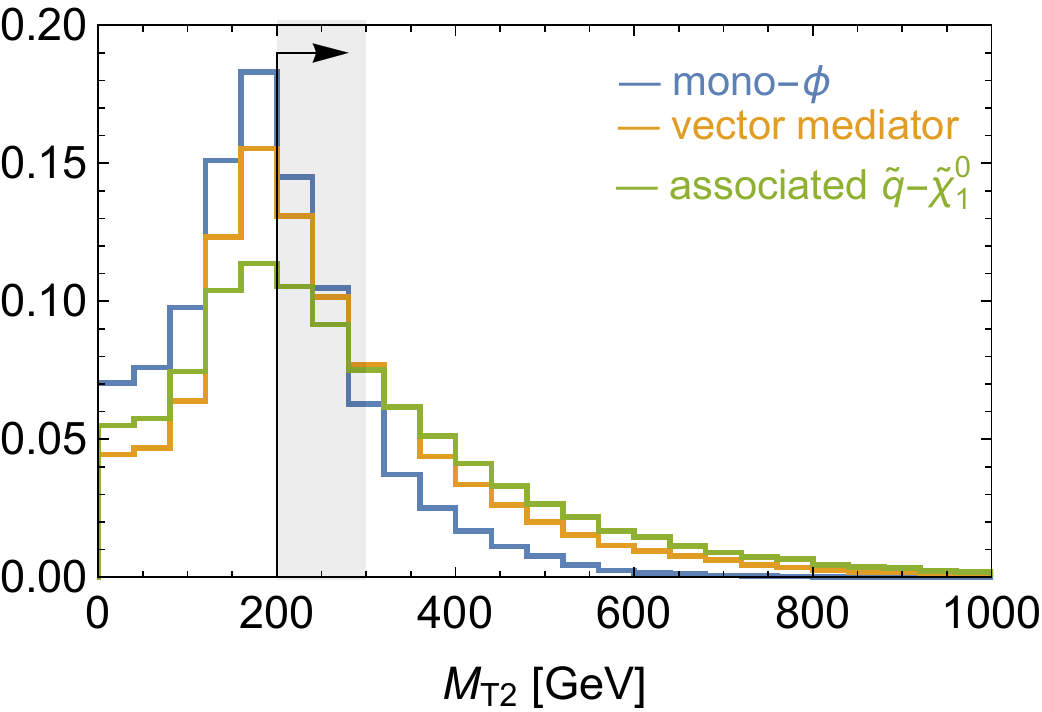}
\caption{Distributions of $N_j$, $H_T$, $\slashed{E}_T$, \MTT~ for the three models described in the main text. In grey, we show the parameter range where the excess is located. The black lines with arrows indicate the lowest SR boundary for each kinematic variable.
}
\label{fig:1jhisto}
\end{center}
\end{figure}

We illustrate the difference between these models in Figure~\ref{fig:1jhisto}, which shows the distributions of $N_j$, $H_T$, $\slashed{E}_T$, and \MTT~(for events passing the CMS036 triggers) at a benchmark point in the mass plane. To fit the excess, the hardest jet should have $p_T^{j_0}\sim300-400\gev$ with comparable missing momentum. We choose the masses accordingly, with $\tilde q, \phi$ and $V$ at 1.2~TeV while the invisible particle is at $850\gev$ for $\tilde{q}$ and $\phi$ and $600\gev$ for $V$ 
(for the vector mediator case the distributions are largely insensitive to the invisible particle mass).
While the distributions peak at the excess, it is clear that the squark-neutralino model has no chance in populating only the excess in Eq.~\eqref{eq:monojet_excess}, in particular because the $H_T$ tails at $H_T>500\gev$ are much wider than for the other models. The difference between the vector mediator and mono-$\phi$ models is also evident, the former having fatter tails and the latter sharply peaked: in particular, as the dark matter model relies on ISR to generate both the jet momentum and the missing energy, it is distributed like the background (mostly $Z\to\nu\bar\nu$ with a ISR jets). If it were to populate the $N_j=1$ bins in CMS036, this model would also generate a consistent excess across a large fraction of bins where no deviation was observed in the data.  Hence, neither the squark-neutralino or the dark matter model reach a significance  above 1.5$\sigma$ across their mass planes. On the other hand, the resonant $\phi$ model seems to fit well.\footnote{As the number of jets and their $p_T$ distributions are among the primary features characterizing this excess, it is important to be confident in their modeling. In particular, a more correct procedure would be to generate the hard events in \textsc{MadGraph5} and \textsc{Pythia8} matched to extra jets using the MLM scheme \cite{Mangano:2006rw}. For the squark-neutralino and vector mediator models, this can be done without issue, and the differences in the jet distributions were slight. However, due to a bug in the \textsc{Pythia8} color-connection algorithm, this was not possible for our resonantly produced color-triplet. As the specific issue was with the triplet-triplet-triplet vertex, we generated a resonantly produced color-octet decaying to a gluon and an invisible particle at both the matched and unmatched level. This set of color-assignments is extremely difficult to justify in any reasonable new physics model, which is why we do not use it as our benchmark scenario. However, no significant difference was seen in the experimental acceptances due to matching, which we believe allows us to ignore (for now) matching in the color-triplet model. 
}

\begin{figure}[t]
\includegraphics[width=0.33\columnwidth]{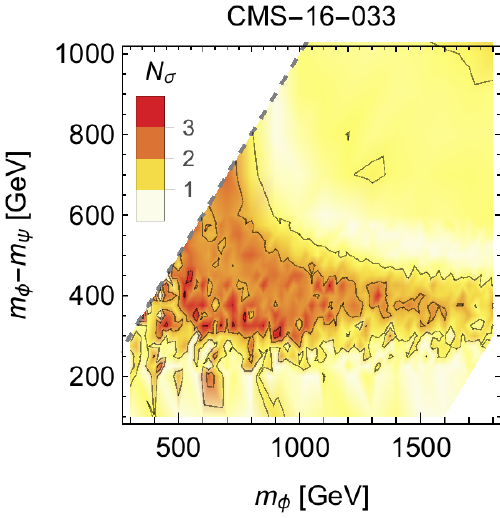}%
\includegraphics[width=0.33\columnwidth]{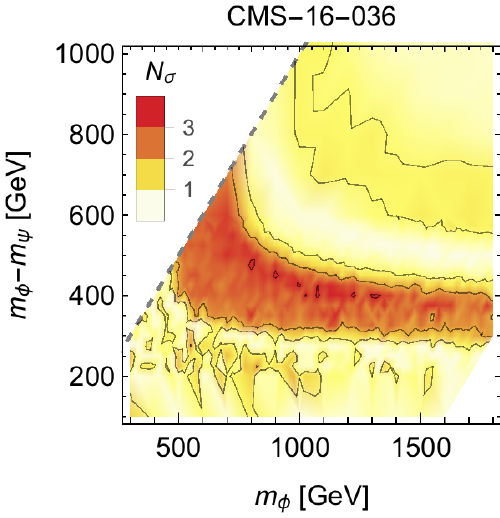}%
\includegraphics[width=0.33\columnwidth]{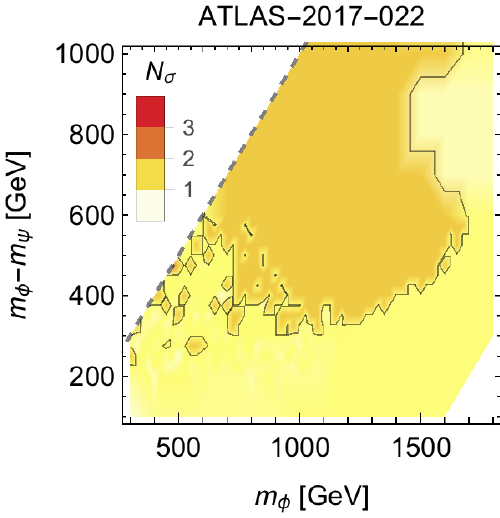}
\includegraphics[width=0.5\columnwidth]{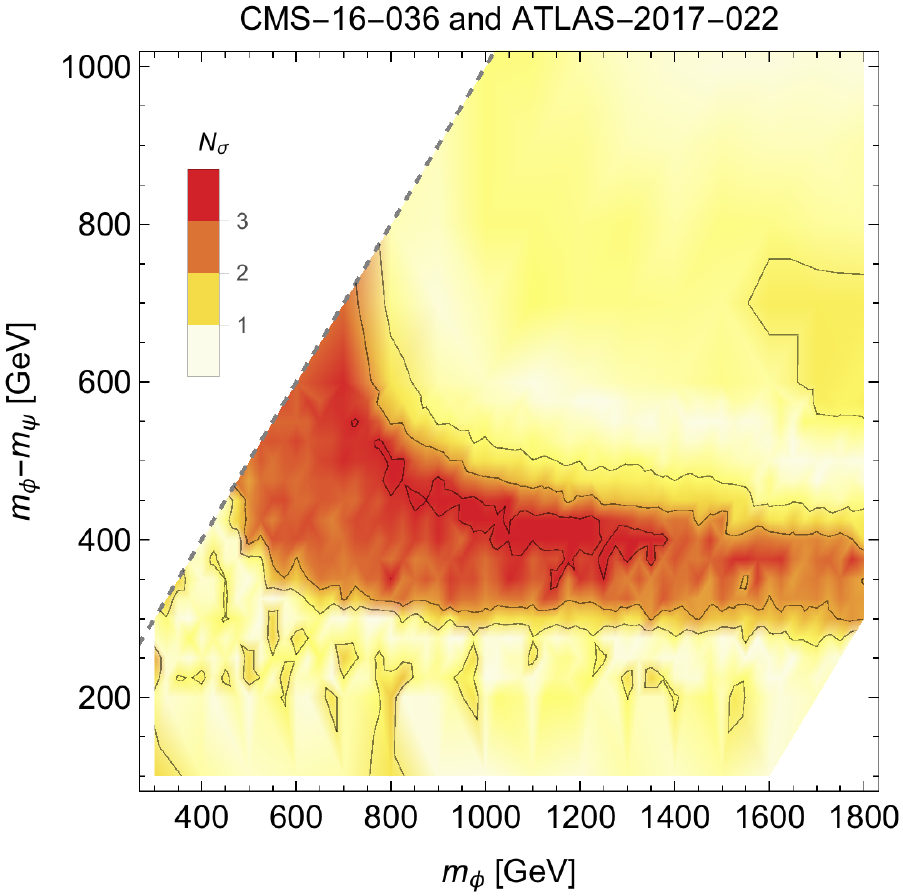}%
\includegraphics[width=0.5\columnwidth]{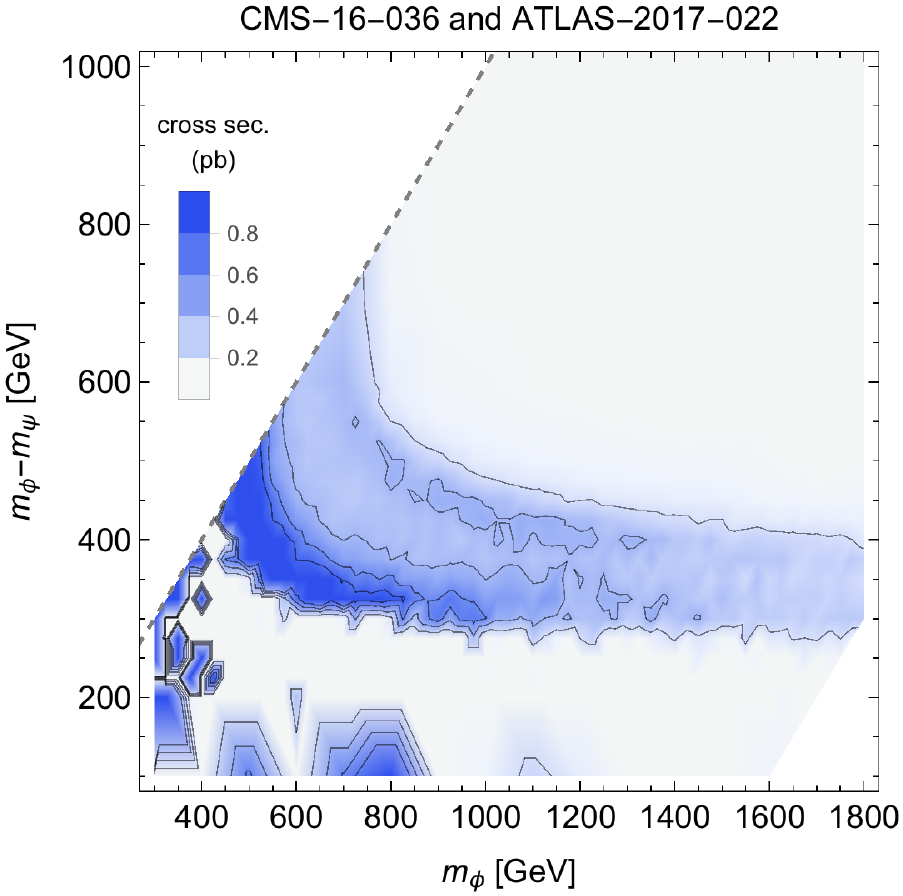}
\caption{Top row: significance for resonant production of $\phi$, decaying to a jet and an invisible particle $\psi$, as a function of $m_\phi$ and $m_\phi-m_\psi$, for the CMS-16-033 (left), CMS-16-036 (center), and ATLAS 2017-022 (right) analyses. The lower row shows the combined significance for ATLAS 2017-022 and  CMS-16-036 (left), and the cross section corresponding to that significance (right). 
\label{fig:monosquark_LL}}
\end{figure}

In the top row of Figure~\ref{fig:monosquark_LL}, we show the significance for the mono-$\phi$ model in the $\phi-\psi$ mass plane (because the jet momentum is set by the $\phi-\psi$ mass difference, we set the vertical axis to $m_\phi-m_\psi$) for each individual search. In particular, it can be seen that  the same region of parameter space generates a significant excess in both CMS036 \cite{CMS36} and CMS033 \cite{CMS33}, which is exactly how a first glimpse of new physics would appear. In the bottom row, we show the significance achieved combining the independent ATLAS and CMS datasets (left), which brings a slight increase to the likelihood (due to a $\sim1.5\sigma$ excess in the ATLAS search), and the cross section necessary to achieve that significance (right). Note that the best-fit value of the cross section varies between different searches, so that the same model with given parameter values  cannot achieve 3$\sigma$ in both CMS036 and CMS033. In particular, the latter would prefer a higher cross section (by a factor of two), but its significance still reaches $N_\sigma=2.5$ if the signal cross section  is set to the best-fit value of CMS036. Given the overlapping datasets, we cannot combine the significance of the two CMS searches.

We see that the mono-$\phi$ model is preferred with respect to the Standard Model {\it at more than 3$\sigma$ (local significance)} in the broad range $m_\phi\sim800-1400\gev$, $m_\phi-m_\psi\sim 400\gev$ and a cross section of order $0.2-0.4$~pb. At the best fit point $(m_\phi,m_\psi)=(1300,900)\gev$, we get $N_\sigma=3.5$ with a best-fit cross section of $\sigma=0.4$~pb.

\begin{figure}[t]
\includegraphics[width=0.5\columnwidth]{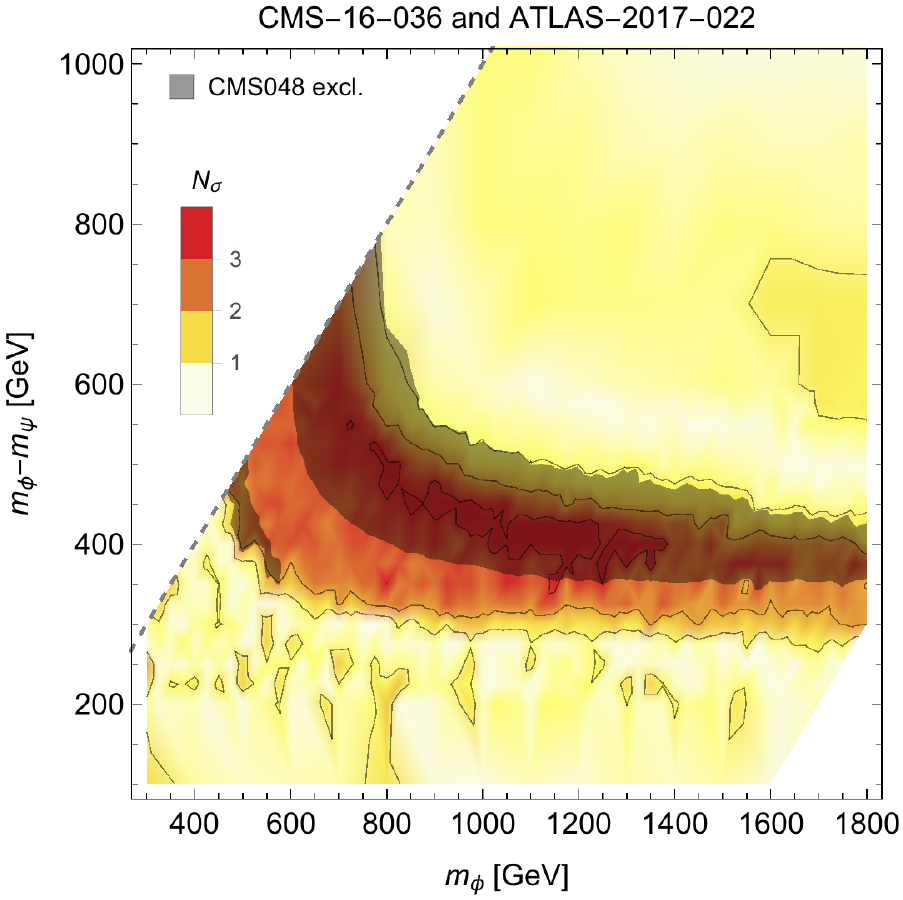}%
\includegraphics[width=0.5\columnwidth]{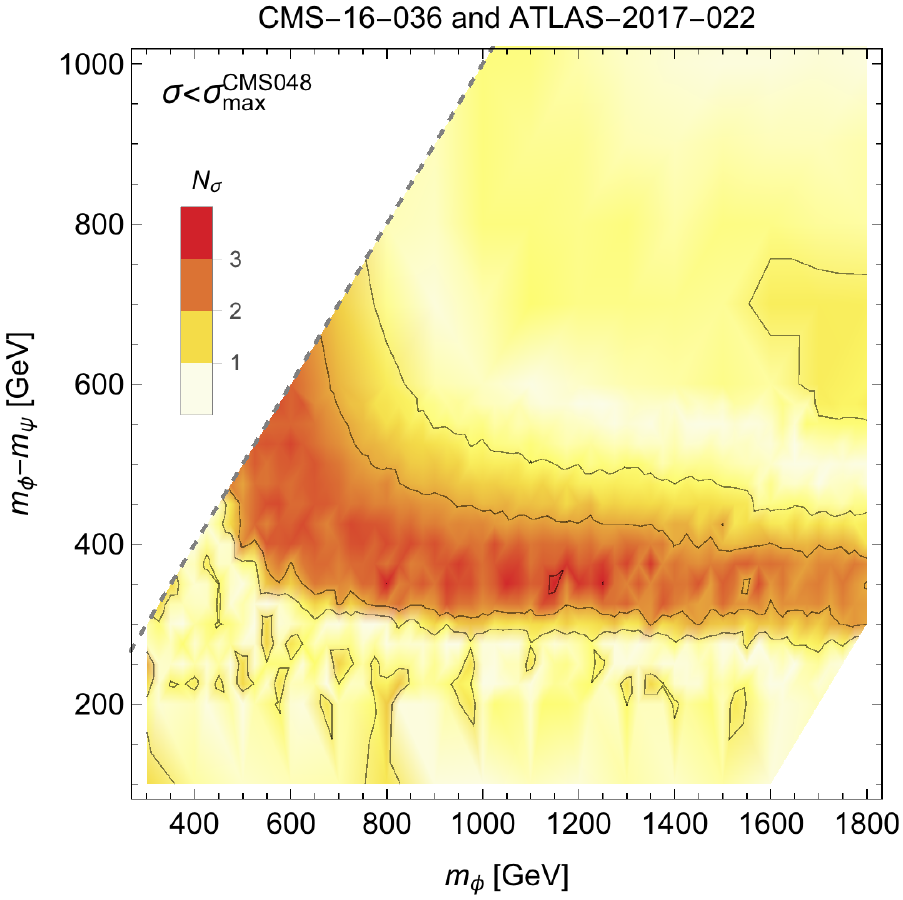}%
\caption{Limits on the parameter space of the mono-$\phi$ simplified model from CMS monojets (CMS048). In the plot on the left, the cross section favored by the combined CMS036 and ATLAS022 analyses is excluded by CMS048 at the 95\% C.L. in the dark gray region. On the right, we show the maximum significance of the combined CMS036 and ATLAS022 analyses allowed by CMS048. 
\label{fig:monosquark_048}}
\end{figure}

Finally, we include limits from the CMS048 monojet search \cite{CMS48}: this search also shows a modest excess of events in the low $\slashed{E}_T$ bins, but because the bin width is much finer than for the $N_j=1$ bins in CMS036, it gives a strong discriminatory power for this model, whose $\slashed{E}_T$ peaks near 300\gev~if it is to explain the excess in CMS036. As the datasets are overlapping between CMS036 and CMS048, we cannot compute a joint likelihood, as we did with ATLAS022. We show the effect of the limits in two ways in Figure~\ref{fig:monosquark_048}: on the left, for each mass point we set the cross section to be the best-fit cross section for the combined ATLAS and CMS excesses (as in Figure~\ref{fig:monosquark_LL}) and then see if that signal is excluded by the monojet search at the 95\% C.L: the resulting exclusion is shaded in gray. On the right, we set the cross section to its best-fit value unless it is excluded by the monojet search, in which case we set it to the 95\% C.L. upper limit given by that search.  We see that, while the best fit value of the mono-$\phi$ model is ruled out, {\it a local significance of nearly $3\sigma$ is still allowed by all the present data}.

Additional signatures of this simplified model are: 
\bitem
\item Dijet resonance: as $\phi$ is resonantly produced, it will also decay back to jets. The cross section shown in Figure~\ref{fig:monosquark_LL} is then $\sigma(pp\to \phi)\times BR(\phi\to j+\slashed{E}_T)$, accompanied by a model-dependent dijet cross section $\sigma(pp\to \phi)\times BR(\phi\to jj)$ (depending on the coupling to $\psi$). 
\item Two jets and $\slashed{E}_T$: Given that $\phi$ must be color-charged, it is also pair-produced, so that the signal must be accompanied by a $2j+\slashed{E}_T$ signature. In addition, depending on the branching ratio to dijets, there will be final states with $3j+\slashed{E}_T$ as well as  $4j$.
\item If the branching ratio of $\psi$ to three jets is non-negligible (a possibility we do not consider in this paper for simplicity), 
resonant $\phi$ production will form a four-jet resonance with a nested three-jet subresonance, which is currently unconstrained at the LHC. In addition, pair production can generate eight-jet final states ($4j+4j$), six-jet final states ($4j+2j$), or five jets +\MET.
\eitem

\begin{figure}[t]
\begin{center}
\includegraphics[width=0.5\columnwidth]{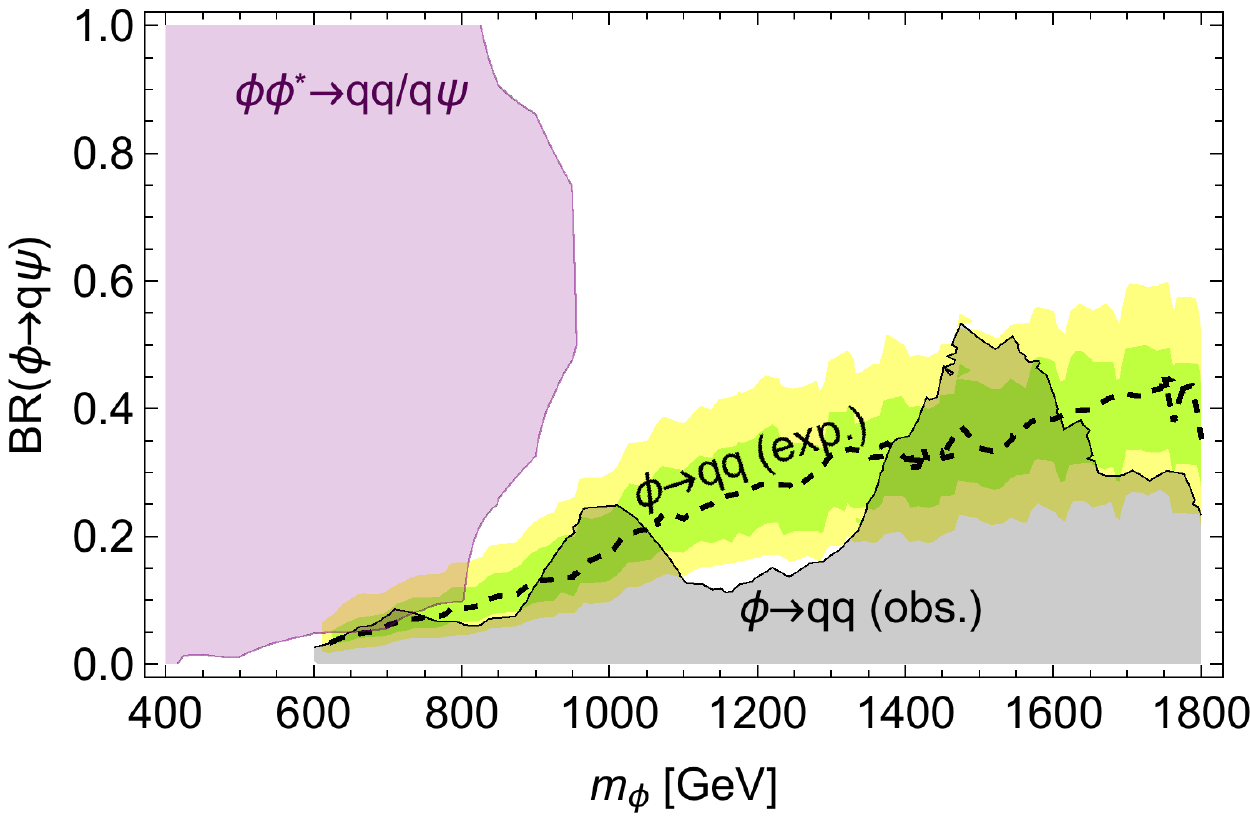}
\caption{Additional limits on the mono-$\phi$ model given by single- and pair-produced dijets. We here set $m_\phi-m_\chi=400\gev$ and show the dependence on $m_\phi$ and the branching ratio, while keeping constant the signal cross section to the best fit value as in Figure~\ref{fig:monosquark_LL} (see the text for more details). The combined ATLAS+CMS036 significance is above $3\sigma$ in most of the plane. The shaded gray area is excluded by the observed dijet limits \cite{cms-dijet}, with the dashed line showing the dijet expected limits. In purple, we show exclusions from $\phi$ pair production followed by mixed decays according to the branching ratio on the vertical axis, set by CMS033 \cite{CMS33}.}
\label{fig:lim_dijets}
\end{center}
\end{figure}

While the branching ratio of $\phi$ into dijets depends on the relative size of the couplings (which will be discussed in the context of a full model in the next section), we can still show model-independent limits as in Figure~\ref{fig:lim_dijets}. Here, we set $m_\phi-m_\psi=400 \gev$  (which maximizes the significance of the excess), set the cross section to the best-fit cross section $\hat \sigma$ (as given in Figure~\ref{fig:monosquark_LL}), and then vary the scalar mass and the branching ratios. In the presence of only two decay channels, we have $Br(\phi\to jj)=1-Br(\phi\to q \psi)$, and having fixed $\sigma(pp\to\phi) Br(\phi\to q\psi) = \hat \sigma$, we find the observable dijet cross section as
\beq\label{eq:sigmajj}
\sigma_{jj}=  \frac{1-Br(\phi\to q\psi)}{Br(\phi\to q\psi)}\hat \sigma A,
\eeq
where $A$ is the acceptance of the CMS dijet search \cite{cms-dijet}. We compute it at the parton level as recommended in \cite{cms-dijet}, requiring $|\Delta\eta_{jj}|<1.3$, and $H_T>250\gev$, $m_{jj}>0.49\tev$ for the low mass range considered there, or  $H_T>900\gev$, $m_{jj}>1.25$\,TeV in the high-mass range. We find acceptances between $0.5$ and $0.6$, in agreement with the quoted value of $0.6$. Finally, we compare  $\sigma_{jj}$ in Eq.~\eqref{eq:sigmajj}  to the dijet limits on narrow quark-quark resonances from \cite{cms-dijet}, as a function of the mass $m_\phi$ and the branching ratio $Br(\phi\to q\psi)$, and show the excluded region in gray in Figure~\ref{fig:lim_dijets}.%
\footnote{We also show the expected limits as a dashed line, with the 1$\sigma$ (2$\sigma$) bands in  green (yellow). As the dijet and other CMS data samples are independent, it would be possible to compute the combined log-likelihood of the two searches and possibly reach even higher significances. For example, note that near $m_{\phi}=1.2\tev$, there is a $\sim2\sigma$ deviation from the expected limits, which if naively added in quadrature to our  significance could reach a combined $4\sigma$. Unfortunately, the event counts in the dijet mass distribution from the preliminary results are not public (!!), so it is not possible to reinterpret the data for assessing the significance of a particular model, and we can only use the quoted limits.}

We also set limits from pair-production of $\phi$ decaying to two or three jets and $\slashed{E}_T$, purple shading in Figure~\ref{fig:lim_dijets} (the strongest limits are set by CMS033 \cite{CMS33}). The limits are more constraining when the decays are mixed, as on average there are more jets in the final state, while still retaining some missing energy. As the branching ratio into dijets increases, the limits are weaker as fewer events have missing energy at all: for $Br(\phi\to q\psi)=0$, values of $m_\phi<400\gev$ are excluded from the ATLAS paired dijet resonance search \cite{ATLAS:2017gsy}.

\subsection{Full model}
\label{sec:model}

The simplified model described is so far incomplete: in the absence of other fields, the decay $\psi\to 3j$ would happen on collider timescales (due to the large coupling needed for resonant production), resulting in a $4j$ final state. While currently there are no direct limits on this final state, in order to describe the excess the (dominant) $\psi$ decay channel should be to invisible particles, suggesting a rich hidden sector.
In addition, to avoid potentially dangerous baryon-number-violating processes at low energy such as dinucleon decay (as present in RPV SUSY, see e.g. \cite{Barbieri:1985ty}), $\psi$ cannot be a Majorana fermion, and a Dirac mass term is needed. The Dirac partner of $\psi$ could couple exclusively to the hidden sector which would easily explain the missing energy signature of the excess. 

The minimal Lagrangian for the mono-$\phi$ model is the following:
\beq
\mathcal{L} \supseteq   g\phi^* q_i^c \psi + \lambda\phi q^c_i q^c_j + m_\psi \psi  \psi' + m_\phi^2 |\phi|^2+ g'\psi' N \tilde N  ,
\label{eq:model}
\eeq
where $q_i^c$ are right-handed quarks, $N,\tilde N$ are neutral, hidden sector fields. The scalar $\phi$ is a right-handed color-triplet and its electric charge can be either $+\frac23$ (up-like) or $-\frac13$ (down-like), for which we respectively have couplings to quarks of the form $\phi_u d^c_i d^c_{j\neq i}$ and $\phi_d u_i^c d_j^c$. A conserved global baryon number can be defined, with $Q_B(q^c)=\frac13$, $Q_B(\phi)=-\frac23$, $Q_B(\psi)=-Q_B(\psi')=-1$. Constraints from (baryon-number conserving) flavor-changing neutral currents can be satisfied if only one $\phi q_i^cq_j^c$ combination is dominant.

The requirement that $\psi$ decays mostly to the hidden sector can easily be achieved either kinematically (e.g. if $m_\psi>m_{\tilde N}+m_N$ the hidden decay is two-body, while the SM decay mode is three-body), and/or if $g' > g$. For the model to fit the excess, see Eq.~\eqref{eq:monojet_excess}, it is imperative that the final state jet is not tagged as a $b$-jet. We have checked that even a $c$ quark with a roughly 20\% mistagging rate would generate too many events in signal regions with $N_b=1$, and  would not reach $3\sigma$ significance as in Figure~\ref{fig:monosquark_LL}.

\begin{figure}[t]
\begin{center}
	\subfigure[]{\includegraphics[width=0.43\textwidth]{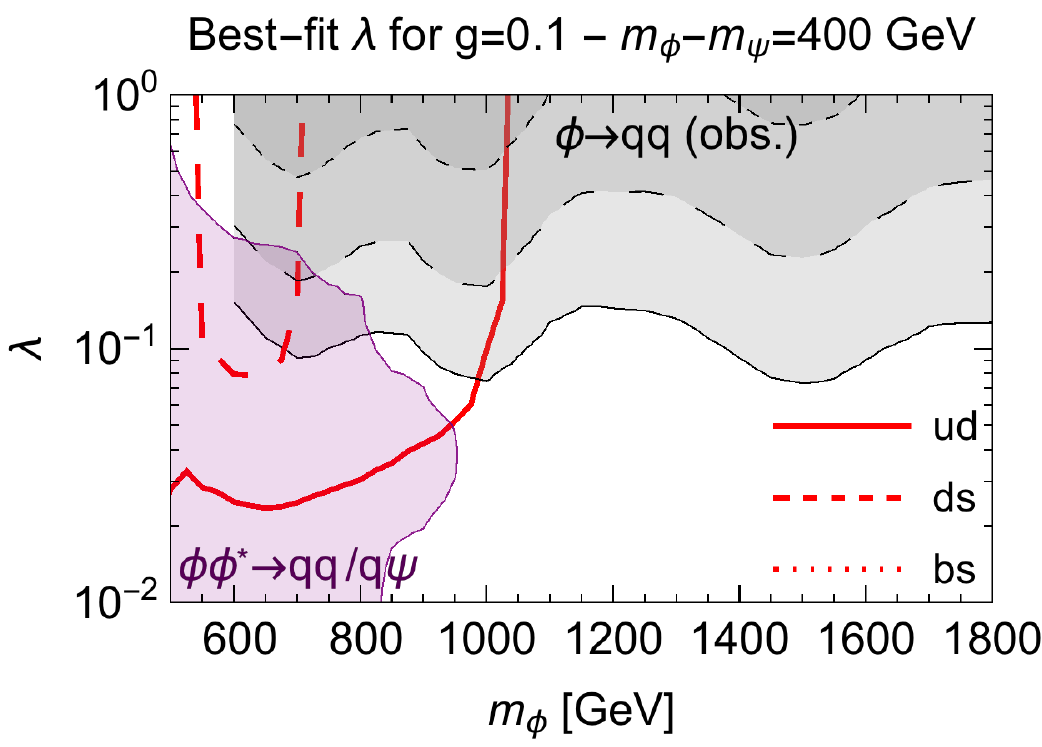}}\qquad%
	\subfigure[]{\includegraphics[width=0.43\textwidth]{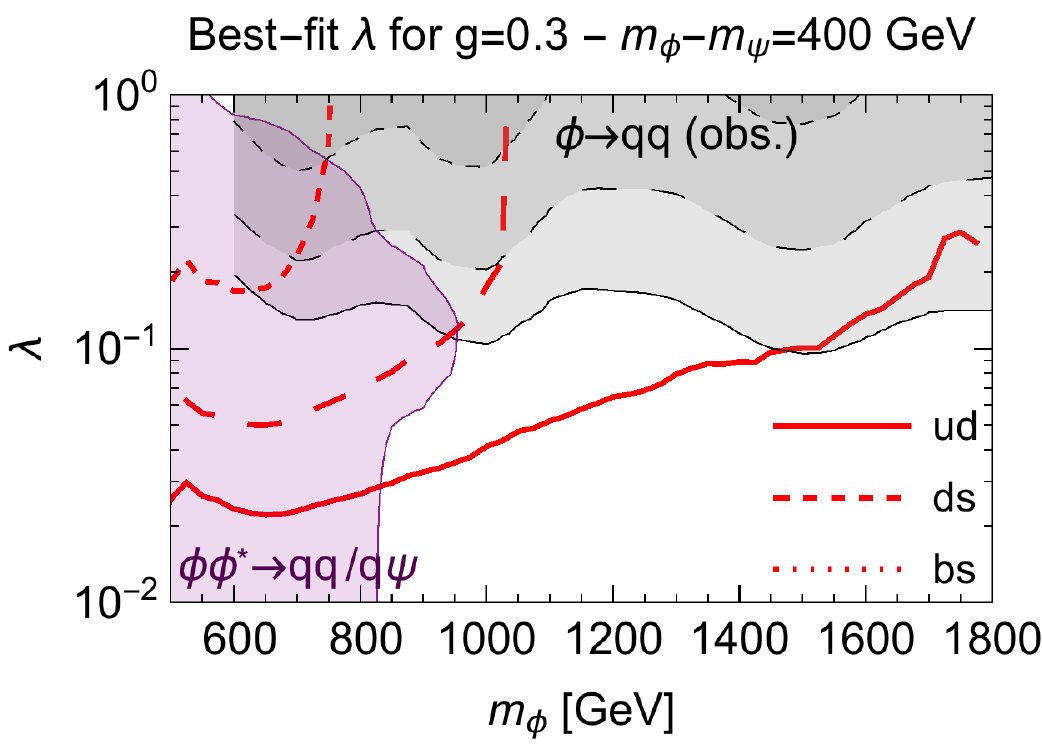}}
	\subfigure[]{\includegraphics[width=0.43\textwidth]{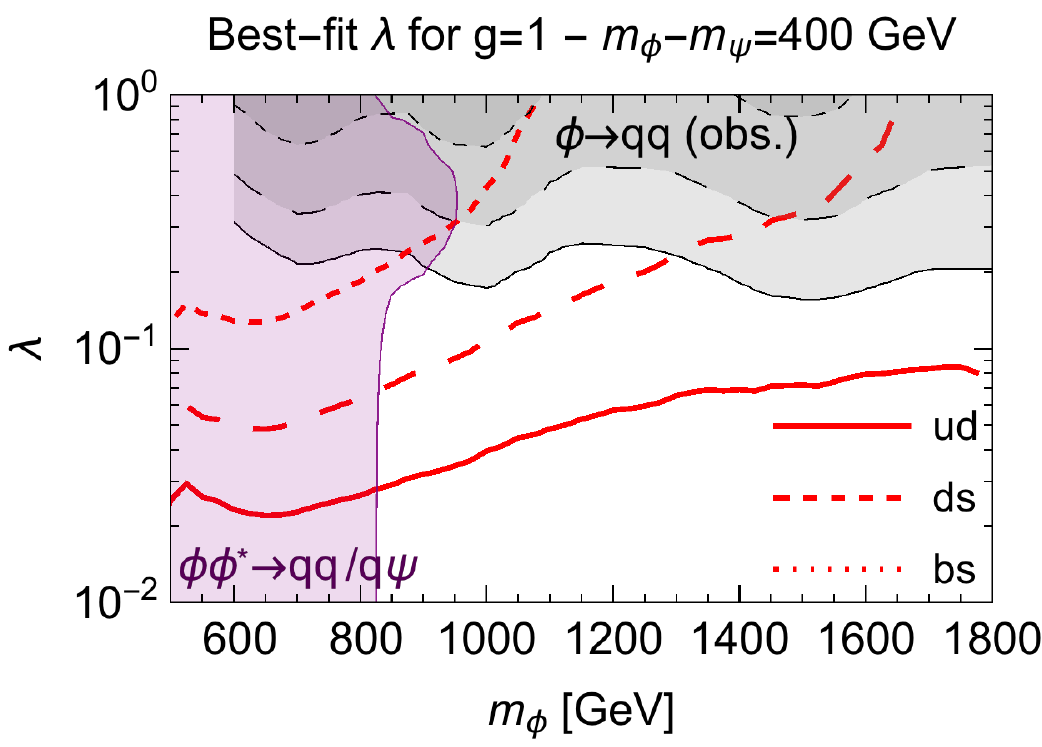}}
\caption{ Best-fit values and constraints for the full model in Eq.~\eqref{eq:model}. In the different plots, we set  $g=0.1,0.3$ and $1$ respectively, while each plot shows in red the value of $\lambda$ needed to reproduce the best-fit cross section as a function of the $\phi$ mass, having fixed $m_\phi-m_\psi=400$~GeV, for different partons in the initial state: $ud$ (solid), $ds$ (dashed) and $bs$ (dotted). The same line styles are used to denote regions excluded by dijet resonance searches \cite{cms-dijet} (which depend on the initial state), while the purple shaded area shows limits set by CMS033 \cite{CMS33} on pair-produced $\phi$'s with mixed decays. }
\label{fig:monophi-lg}
\end{center}
\end{figure}

The production cross section is set by the parton luminosity of the initial state flavors (we use the MSTW2008lo PDF set \cite{Martin:2009iq}) and the coupling $\lambda$, while the branching ratio into the excess, $Br(\phi\to q\psi)$, also depends on the coupling $g$. In Figure~\ref{fig:monophi-lg}, we illustrate the dependence on the production mode and the couplings, by showing the best-fit value of $\lambda$ as a function of the $\phi$ mass, again fixing the mass splitting $m_\phi-m_\psi=400\gev$ to reach the highest significance in CMS036. In each plot, the value of $g$ is fixed at reference values of 0.1 (a), 0.3 (b) and 1 (c). We then vary the initial state from $ud$ (solid lines), $ds$ (dashed) and $bs$ (dotted), with the red line being the value of $\lambda$ that reproduces the best-fit cross section.%
\footnote{In total there are nine possible $q_iq_j$ initial state combinations: for the most part, the parton luminosities  depend on the number of valence vs. sea quarks in the initial state, so that the results for $ds$ apply to $us,ub,cd,db$, while the results for $bs$ apply also for $cs,cb$ initial states. Also note that while for simplicity we refer to $q_iq_j$ initial states, we are including both $q_iq_j\to \phi^*$ and $\bar q_i\bar q_j\to \phi$ processes.}
 At each point on the mass plane, the dijet cross section is fixed and it can be seen if it is allowed or excluded by the dijet search \cite{cms-dijet}: we show limits from dijet resonances on different initial states with the same line-style as for the best-fit $\lambda$ (for example, dijet limits exclude the best-fit cross section for $ds$ initial state when the black solid line is above the colored dashed line).  As mentioned earlier, near $m_\phi=1.2\tev$ the CMS dijet limits show a 2$\sigma$ fluctuation, which could fit naturally in this model. 
 In purple, we show limits on pair-produced $\phi$'s decaying to $qq$ or $q\chi$ depending on the given branching ratios at each point. 

In the limit $\lambda\gg g$, $\sigma \times Br(\phi\to q\psi)\to \text{const.}$, which is the reason for some of the lines to quickly get out of the plot range: in that limit, above a certain mass {\it no coupling} can reproduce the excess, as the desired cross section is too high.

We note that this model is very similar to a previously proposed model of baryogenesis in the context of Twin Higgs, dubbed ``Twin baryogenesis'' \cite{Farina:2016ndq}. There, the hidden sector was formed of twin quarks and $\psi$ decays resulted in the same particle-antiparticle asymmetry in both sectors (due to the Dirac nature of $\psi$), therefore explaining the baryon asymmetry as well as the coincidence between matter and dark matter densities.

Another implementation would be a non-minimal version of RPV SUSY, where $\phi$ can be identified with a right-handed squark and $\psi$ with a bino. Because the model needs Dirac neutralinos as well as a hidden sector, we do not try to pursue a full SUSY implementation, but mention that anomaly-mediated contributions to the neutralino Majorana mass bring back baryon number violation to an unacceptable level \cite{Beauchesne:2017jou}, so that a SUSY model faces many obstacles.

To conclude, we find that for this model to reproduce the excess, it must have $g>0.1$ for any initial state, $g>0.3$ for $ds$ initial state and $g\gtrsim1$ for $bs$. The typical values of the $\lambda$ couplings are $0.05-1$ depending on the initial states. If we require the absence of Landau poles at nearby scales, we predict that $\phi$ couples preferentially to (at least one of the) light quarks, namely to either $ud, us, ub, cd$ or $db$. With such large couplings, one could have new diagrams contributing to $\phi$ pair-production (via a $t$-channel quark and two $\lambda$ insertions, or a $t$-channel $\psi$ and two $g$ insertions), and to associated $\phi-\psi$ production (as in Fig.~\ref{fig:diagramsA}).\footnote{We thank Jared Evans for pointing those out to us.} We neglect those, as even for $O(1)$ couplings, the cross sections only go up to $O(10)$ fb, and will not significantly affect our limits, as well as our best-fit estimates. Finally, we mention that in a complete model, we would expect $\phi$ to couple to all three quark generations, possibly with flavor-dependent couplings, in which case flavor-changing neutral currents could become a strong constraint. We leave this aspect to future work.

\section{Comments on the Look-Elsewhere Effect \label{sec:lookelsewhere}}

In the previous section, we have attempted to fit the  ``mono-jet excess" in the CMS jets+$\slashed{E}_T$ searches to a model consisting of resonant color-triplet production. We saw that the best fit point not excluded by other searches rose to $\sim 3\sigma$ local significance. However, as with any excess, we should also be interested in the global significance of the observed statistical fluctuation. What are the odds of seeing an excess of this size {\em anywhere} in the data set from the Standard Model alone? That is, what is the statistical significance after the look-elsewhere-effect (LEE) is applied?

This question is especially important given the novel method we have used to identify the excess: the rectangular aggregation technique. Given the extremely large number of RAs that we have iterated over, it is certainly tempting to believe that the LEE should reduce a $3\sigma$ anomaly to insignificance. After all, if we have scanned over 33,000 rectangles, have we not looked elsewhere 33,000 times?

As a naive upper bound on the LEE, we first calculate the local significance in all the 33,000 RAs of CMS036 with 1,000 pseudo-experiments. In 15\% of these pseudo-experiments, we see at least one RA with local significance above $N_\sigma=3.5$ (the highest local significance in the real data). This is already far less than a trials factor of 33,000 would imply. Obviously, the 33,000 rectangles are not all independent -- as each rectangle overlaps with many others, an excess in one would typically appear in many.

In fact, we expect the true LEE to be much less severe because, as we saw in Section~\ref{sec:statfluct}, many fluctuations have kinematic and topological characteristics  that make them unlikely to be well-fit by any plausible new physics model. Quantifying this rigorously without resorting to a specific model would require formulating a complete set of signal templates, which is beyond the scope of this work. Instead, we will limit ourselves to demonstrating in the rest of this section that, in the context of the mono-$\phi$ model, the LEE reduces the $3\sigma$ local significance to $2\sigma$ global, which is in line with expectations for the LEE in a traditional ``bump-hunt'' type anomaly \cite{Gross:2010}.\footnote{Note that restricting to a specific model further mitigates the LEE, since not all of the rectangles will tend to be populated by that model. For instance, in the mono-$\phi$ model, only those rectangles dominated by $N_j\lesssim 3$, $N_b=0$ bins will have a chance of showing an excess, no matter where we are in the parameter space.}

To calculate the LEE for the mono-$\phi$ model, we first generate 10,000 pseudo-experiments for CMS036, taking into account the full covariance matrix, assuming only SM contributions. (Note that we cannot quantify the likelihood of the same pseudo-experiment to also give a fluctuation in CMS033 without actually generating Monte Carlo background events, as the searches are largely overlapping.)
Using the number of generated ``observed'' events in each SR, we calculate the statistical preference for the mono-$\phi$ model anywhere in the full parameter space of the model (i.e.\ $(m_\phi,m_\psi,g,\lambda)$ plus the choice of initial state). In practice this amounts to allowing the cross section $\hat\sigma$ to be a free parameter, and fitting in the mass plane. 
We then ask how many pseudo-experiments contain a statistical deviation from the background-only hypothesis at least as significant as the excess seen in the real data ($N_\sigma \approx 3$).

\begin{figure}
\includegraphics[scale=0.8]{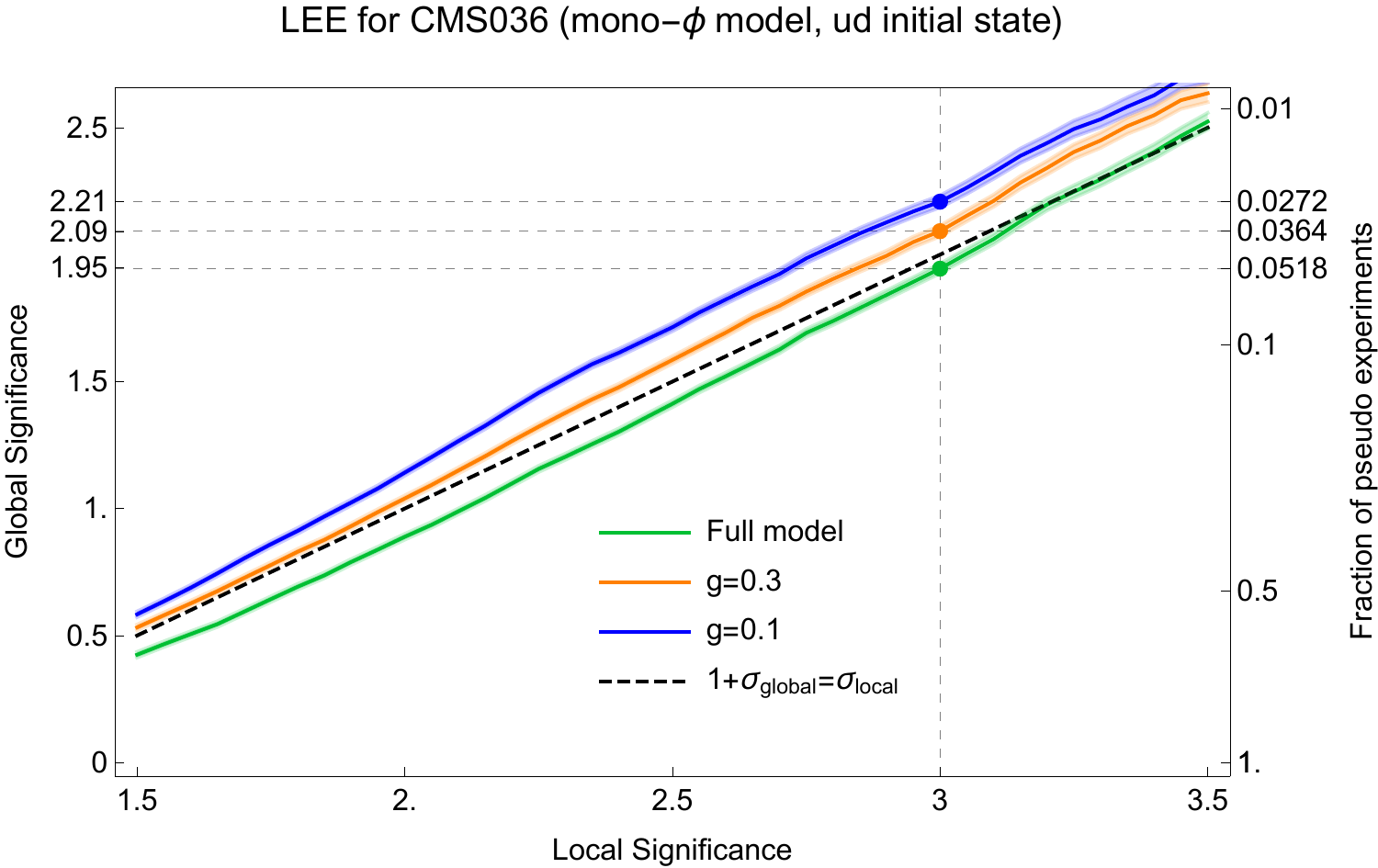}
\caption{Fraction of CMS036 pseudo-experiments (generated from background counts reported in \cite{CMS36}) with a global significance (anywhere on the mass grid of the mono-$\phi$ model) above a specified local significance threshold. The axis on the right shows this fraction, while the axis on the left indicates its equivalent standard normal distribution significance. The green line denotes the fraction of 10,000 pseudo-experiments that have a given local significance in the full parameter space of the model, 
while the orange and blue lines represent the global significance for particular choices of $g$.
The excess in the real data lies on the vertical line at $\sigma_{\rm local} \approx 3$. 
}
\label{fig:LEE}
\end{figure}

Figure \ref{fig:LEE} summarizes the look-elsewhere effect for the model in Section~\ref{sec:model}. It illustrates the fraction of the 10,000 pseudo-experiments that have a local excess above a certain local significance threshold. We also indicate the global $p$-value where the parameter $g$ has been restricted to a couple of special values, $g=0.1,0.3$, with initial state $ud$. 
The choice of this parameter might be motivated by particular frameworks such as SUSY.
We find that there are $518$ pseudo-experiments with an excess at least as significant as the excess in the real data. Of these, 364 (272) have an achievable cross section and are not ruled out by CMS dijets limits \cite{cms-dijet} with $g=0.3$ ($g=0.1$). This corresponds to a $p$-value of $5.2\%$ (or $3.6\%$ and $2.7\%$ for the fixed values of $g=0.3$ or $0.1$) from which an equivalent $1.95$ Gaussian $N_\sigma$ is inferred ($2.1$ and $2.2$ for $g = 0.3$, $0.1$ respectively). Thus, the LEE removes $\sim 1\sigma$ from the $3\sigma$ local excess.

As expected, the look-elsewhere effect, when applied to a specific model, has brought down the significance of the anomaly;
but no more than is typically seen for any experimental excess. Even with that taken into account, we are still left with a non-trivial deviation from the background-only predictions, over $2\sigma$ globally. If this excess is actually a window to new physics, its significance will go up as the LHC dataset increases.

\section{Conclusions \label{sec:conclusion}}
The LHC continues to provide vast amounts of high-quality data across many distinct final states. The sheer volume of information makes it difficult to assess the presence of potential deviations from the Standard Model background predictions. With so many signal regions, statistical fluctuations in individual bins are expected, and some method must be used to combine the signal regions to identify which excesses are interesting potential signals of new physics.

The experimental and phenomenology communities have traditionally addressed this problem by using pre-defined signal templates, often cast in the language of supersymmetry or simplified models. However, this approach is very limiting, and provides a view of the data biased by pre-LHC theoretical assumptions. These expectations are not the only forms new physics can take, and given the lack of evidence for theories such as minimal supersymmetry in the data so far, a more flexible approach is needed. 

Our method of rectangular aggregations provides a systematic approach to the data that allows us to identify a list of interesting excesses, with the only theoretical prior being that new physics should populate a compact set of kinematic and topological variables. As we have demonstrated using the CMS jets+$\slashed{E}_T$ searches, anomalies at the $3\sigma$ level ($2\sigma$ including the look-elsewhere effect) exist in these searches. We have pursued new physics explanations of just one of the identified excesses in this paper, which appears to be shared by both CMS jets+$\slashed{E}_T$ searches as well as possibly the ATLAS jets+$\slashed{E}_T$ search. However the other excesses we identified may also yield interesting results. 

Applying rectangular aggregations to the rest of the current LHC analyses is also an immediate and obvious follow-up to this work.  This requires the collaborations to construct their searches in terms of non-overlapping signal regions covering as much of the kinematic space as possible, and make public the background correlation matrices of these regions -- as CMS has already done for many (but not all) searches. We strongly encourage the ATLAS collaboration to consider this approach as well.

Identifying and categorizing the statistical deviations in the current data provides a first step in the long-term project to monitor the LHC data for interesting anomalies. Early identification of these regions allows for theoretical work (model-building) that can point to better-optimized search strategies. For example,  the mono-$\phi$ model in the monojet channel has distinctive jet kinematics (as one jet comes from the decay of a heavy resonance), which the current jets+$\slashed{E}_T$ searches are not optimizing for. Also, a given model will tend to predict correlated signatures (such as dijets, paired dijet resonances, and $2j/3j$+$\slashed{E}_T$ signatures in the mono-$\phi$ model), whose presence in the data would give much greater confidence in the new physics interpretation. 

Early identification of anomalies also allows the experimental collaborations to freeze the relevant selection criteria. This  point is especially important: as the rate of LHC data collection continues to increase, there will be pressure to raise trigger and selection thresholds. However, this runs the risk of blinding the experiments to interesting physics -- identifying potentially interesting regions will provide another piece of evidence to consider in this process. Already, the high thresholds used in some of the experimental analyses can make new physics searches at low particle masses difficult, and these thresholds should be lowered  where possible.

Our rectangular aggregation method can potentially be improved by refining the templates used to aggregate signal regions. In this paper, we used simple rectangles, which are certainly model-independent, but are insensitive to new physics which populate signal regions in more complicated patterns. For example, if two kinematic variables are highly correlated, the correct aggregation template would be one that moves along a diagonal. The signal templates could possibly be further improved by incorporating information about the effects of initial- and final-state radiation on the topology and kinematics of the hard event.

Though a great deal of discovery potential remains at the LHC, we are entering a slower phase of progress -- at least as measured in absolute increase in the mass of particles. For example, it will take some 20 years to achieve the ultimate reach of $\sim 3\tev$ for gluinos (see \cite{Buckley:2016kvr}).
Given this phase of steady data acquisition, we believe that now is the time to chase ambulances -- as the data comes in at a dependable rate and at fixed energy, it is less productive to just wait around for an excess to grow or shrink. It becomes more and more well motivated to fully explore the data and attempt to fit new physics models to it. We believe that the prevailing view of the LHC data as containing no interesting anomalies is, at best, premature, and a great deal of work remains to fully explore the data set.

\section*{Acknowledgements}

\noindent We are grateful to John Paul Chou, Kyle Cranmer, Yuri Gershtein,
Marumi Kado, Greg Landsberg, Amit Lath, Mariangela Lisanti, Mario Masciovecchio, Kevin Pedro, David Sheffield, Torbjorn Sjostrand, Peter Skands,  and especially Claudio Campagnari for helpful discussions. We thank Jared Evans and Yevgeny Kats for useful comments on the draft. DS is grateful to the Fermilab LPC and the 2017 CERN-CKC Workshop where he presented preliminary versions of this work and received much useful feedback. AM is grateful to the Center for Particle Physics of Marseilles and the IFAC theory group at L2C-LUPM for hospitality and feedback during the completion of this work.
The work of AM, DS was supported by DOE grant DE-SC0013678. The work of PA is supported by DOE grant DE-SC0003883.

\appendix

\section{Statistics}
\label{app:stat}

In this Appendix, we review the profile likelihood ratio methods used to quantify the significance of an excess or to set exclusions on a model. We do not aim to give a complete analysis, for which we refer the reader to \cite{Cranmer1007.1727,SimplifiedLL}, but simply go through the essential concepts.

The likelihood ratio test compares two competing hypotheses, usually referred to as the null hypothesis $H_0$, and the alternative hypothesis $H_1$, and can be related to a $p$-value, that is, the probability of finding a greater or equal test statistic than the observed one if the null hypothesis is true. In a typical LHC search, the data is separated into multiple SRs or bins, 
If there were no theoretical or experimental uncertainties on the predicted backgrounds, the probability of observing $n_i$ events in the $i$-th bin, given a SM expectation $b_i$ and a BSM signal $s_i$, would be given by the Poisson distribution:
\begin{equation}
p(n_i|\mu s_i+b_i) = \frac{(\mu s_i + b_i)^{n_i}e^{-(\mu s_i + b_i)}}{n_i!},
\label{eq:poiss}
\end{equation}
where $\mu$ is a signal strength modifier \cite{SimplifiedLL}: $\mu=0$ stands for no signal beyond the SM and $\mu=1$ refers to the fiducial signal. It is useful to keep explicit the dependence on $\mu$ in order to be able to test how well the data is described by the particular BSM topology under study with other values of $\mu$, which for example can be achieved by changing the branching ratios or the particle multiplicity.

In real experiments, both the SM backgrounds and the BSM signal have systematic uncertainties arising from many sources (theory errors, MC and control region statistics, jet energy scale, fake rates, etc.) In general, the uncertainties are treated as {\it nuisance parameters}, and as outlined in \cite{SimplifiedLL}, they can be well-approximated in many instances by zero-mean Gaussian variables $\theta_i$,  which are added to the background, $b_i\to b_i+\theta_i$, together with a covariance matrix $V$. The likelihood function for all bins is then defined as:
\begin{equation}
{\cal L} (\mu,\theta) = \prod_i  \frac{(\mu s_i + b_i+\theta_i)^{n_i}e^{-(\mu s_i + b_i+\theta_i)}}{n_i!}  \exp \left(	-\frac{1}{2} { \theta}^T  V^{-1}  { \theta} \right),
\label{eq:simplL}
\end{equation}
We then minimize ({\it profile}) the likelihood function with respect to the nuisance parameters $\theta$ (and the signal strength $\mu$), and define a likelihood ratio as:

\beq
\tilde{\lambda}(\mu) \equiv \left\lbrace \begin{array}{ll}
\frac{{\cal L} (\mu,\hat{\theta}_\mu)}{{\cal L}(\hat{\mu},\hat{\hat{\theta}})} & \hat{\mu} \geqslant 0 \\
\frac{{\cal L} (\mu,\hat{\theta}_\mu)}{{\cal L}(0,\hat{\theta}_0)} & \hat{\mu} < 0
\end{array}
\right. ,
\label{eq:lambda}
\eeq
where $\hat{\theta}_\mu$ is a $\theta$ vector that maximizes the likelihood in Eq.~\eqref{eq:simplL} for a given $\mu$ and $(\hat{\mu},\hat{\hat{\theta}})$ are the $\mu$ and the $\theta$ vector that globally maximize the likelihood. $\tilde{\lambda} (\mu)$ is a measure of how far away a given signal ($\mu$) is from being the best model to explain the observed data. Larger values of $\tilde{\lambda} (\mu)$ (notice $0 \leqslant \tilde{\lambda} \leqslant 1$) imply a better compatibility between the signal and the observed data \cite{Cranmer1007.1727}. We assume a signal only increases the event count in each signal region (therefore neglecting cases where interference effects would be important). Thus, $\hat{\mu}<0$ implies that $\mu=0$ has the best agreement with the data while still being a physical value for $\mu$ (this is reflected in the second line in Eq.~\eqref{eq:lambda}).

In the limit of large sample size, it can be found \cite{Wald, Wilks} that $-2\ln\tilde{\lambda}(\mu)$ follows a chi-square distribution with one degree of freedom. For smaller sample sizes, one can either find its distribution by generating toy experiments, or by using the asymptotic formulae in \cite{Cranmer1007.1727}.

We now define two test statistics suitable for our studies:
\bitem 
\item The test statistic for discovery of a positive signal: 
\beq
q_0 \equiv -2 \ln \tilde{\lambda} (0) .
\label{eq:q0}
\eeq
In this case, the goal is to rule out the Standard Model (the null hypothesis), while the alternative hypothesis is the positive BSM signal. While the presence of underfluctuations, where one would find $\hat\mu<0$ in Eq.~\eqref{eq:lambda}, means that the SM is not a good fit, it should not be automatically taken as a sign of new physics but rather point to possible errors in the SM background estimation. In this case we have $q_0=0$.

In the large $N$ limit, one can find the $p$-value $p_0$, and the equivalent Gaussian significance $Z_0$, as:
\beq
p_0=1-\Phi(\sqrt{q_0}), \qquad Z_0=\Phi^{-1}(1-p_0)=\sqrt{q_0}
\eeq
where $\Phi(x)=\frac12 {\rm erfc}(-x/\sqrt2)$ is the cumulative Gaussian distribution.

\item The test statistic for setting upper limits:
\beq
\tilde{q}_\mu \equiv \left\lbrace \begin{array}{ll}
-2 \ln \tilde{\lambda} (\mu) & \hat{\mu} \leqslant \mu \\
0 & \hat{\mu} >\mu
\end{array}
\right. .
\label{eq:qmu}
\eeq
Here we are testing the compatibility between the data and the BSM signal with a signal strength $\mu$, and a larger $\tilde{q}_\mu$ representing increasing incompatibility. The null hypothesis we aim to reject is therefore the BSM signal. In the case that $\hat\mu>\mu$, the best-fit signal contribution is larger than the signal strength we are testing, and we should not reject the signal in favor of the SM (the alternative hypothesis) by setting an upper limit; therefore, $\tilde{q}_\mu$ is set to zero in that range.

In the large $N$ limit, the $p$-value $p_\mu$ and the Gaussian significance $Z_\mu$ are simply given by:
\beq
p_\mu = 1-\Phi(\sqrt{\tilde{q}_\mu}), \qquad Z_\mu=\Phi^{-1}(1-p_\mu)=\sqrt{\tilde{q}_\mu}
\eeq
In particular,  when the $p$-value is below a certain threshold $\alpha$ we say that the signal is excluded at a confidence level of $1-\alpha$. Results are usually quoted at the 95\% confidence level, corresponding to $\alpha=0.05$ or $Z=1.96$. We find this value numerically by varying $\mu$ until we find $\tilde{q}_\mu=4$ (we here gloss over the small difference between $2\sigma$ exclusions ($Z=2$) and $95\% $ C.L. exclusions).
\eitem

The quantities $q_0$ and $\tilde{q}_\mu$ can be calculated for either the full set of SRs of each search, or for the reduced set of SRs found after aggregations, Eq.~\eqref{eq:eventR}. In Section~\ref{sec:jetsMET}, we calculate $q_0$ with the input signal populating {\it only that aggregated region} (which is treated as a single new bin). We are then quantifying how excluded the background-only hypothesis is compared to a hypothetical BSM model that only populates that rectangular aggregation. This number is reported for the RAs in Tables \ref{tab:excess36}-\ref{tab:excess33}. Since we are only studying the fluctuations localized to the aggregated bin, the significance we obtain is local. 

Because multiple searches involve hundreds of exclusive signal regions, each with its nuisance parameter $\theta_i$, the definition of $\tilde{\lambda}(\mu)$  in Eq.~\eqref{eq:lambda} involves  maximizing  the likelihood function ${\cal L}(\mu,\theta)$ with respect to hundreds of variables. While in the absence of correlations, ${\cal L}$ is simply a sum of terms that can be individually maximized, in general this is not an easy task: in this work, we use the powerful \textsc{Minuit} routines \cite{James:1975dr}, interfaced to Python via the \texttt{iminuit} package.

\section{Recasting Pipeline and Validation\label{app:val}}

In this work, we have reinterpreted several ATLAS and CMS searches. In this section, we validate each search by reproducing the exclusion plots on simplified models present in each experimental paper. 

We generate hard events in \textsc{MadGraph5\_aMC@NLO 2.5.3} \cite{Alwall:2014hca}, with additional hard jets in the events if needed. For the SUSY simplified models used for validations, we use the MSSM module included in Madgraph, while for the dark matter simplified models used in \cite{CMS48} we use the \texttt{DMsimp} UFO model \cite{1508.05327}. For the mono-$\phi$ model discussed in Section~\ref{sec:monojet}, we use the \texttt{RPVMSSM} UFO model \cite{1202.4769}, which was generated with \textsc{FeynRules}. While for the validation plots we do not find it necessary to generate more than 10,000 MC events, for the significance plots shown in Section~\ref{sec:monojet} we generate 100,000 MC events to avoid statistical fluctuations in low-efficiency bins. We use the leading-order cross sections as calculated by \textsc{MadGraph}.

Parton-level events are showered and hadronized with \textsc{Pythia8.219} \cite{Sjostrand:2014zea}. If necessary, we match the matrix-element and parton shower events with the MLM technique \cite{MLM}. The resulting particles are reconstructed in a ATLAS- or CMS-like simulated detector using \textsc{Delphes3.4} \cite{deFavereau:2013fsa}, depending on the search, with efficiencies for particle reconstruction taken from the experimental papers. Jets are reconstructed with the \textsc{FastJet} package \cite{Cacciari:2011ma}, using the anti-$k_t$ algorithm \cite{Cacciari:2008gp}; to validate \cite{CMS48}, we also use pruning techniques \cite{Ellis:2009me} on large-R jets, on which $n$-subjettiness variables \cite{Thaler:2010tr} are also calculated.
Finally, cuts and SR definitions in each experimental search are simulated with \textsc{pyROOT}.
For the CMS searches, efficiencies from all the signal regions are used to compute the likelihood, while for ATLAS we use the SR with the best-expected exclusion to set limits (or the best-expected discovery reach for positive significance).

\begin{figure}[h!]
\begin{center}
\includegraphics[width=0.38\columnwidth]{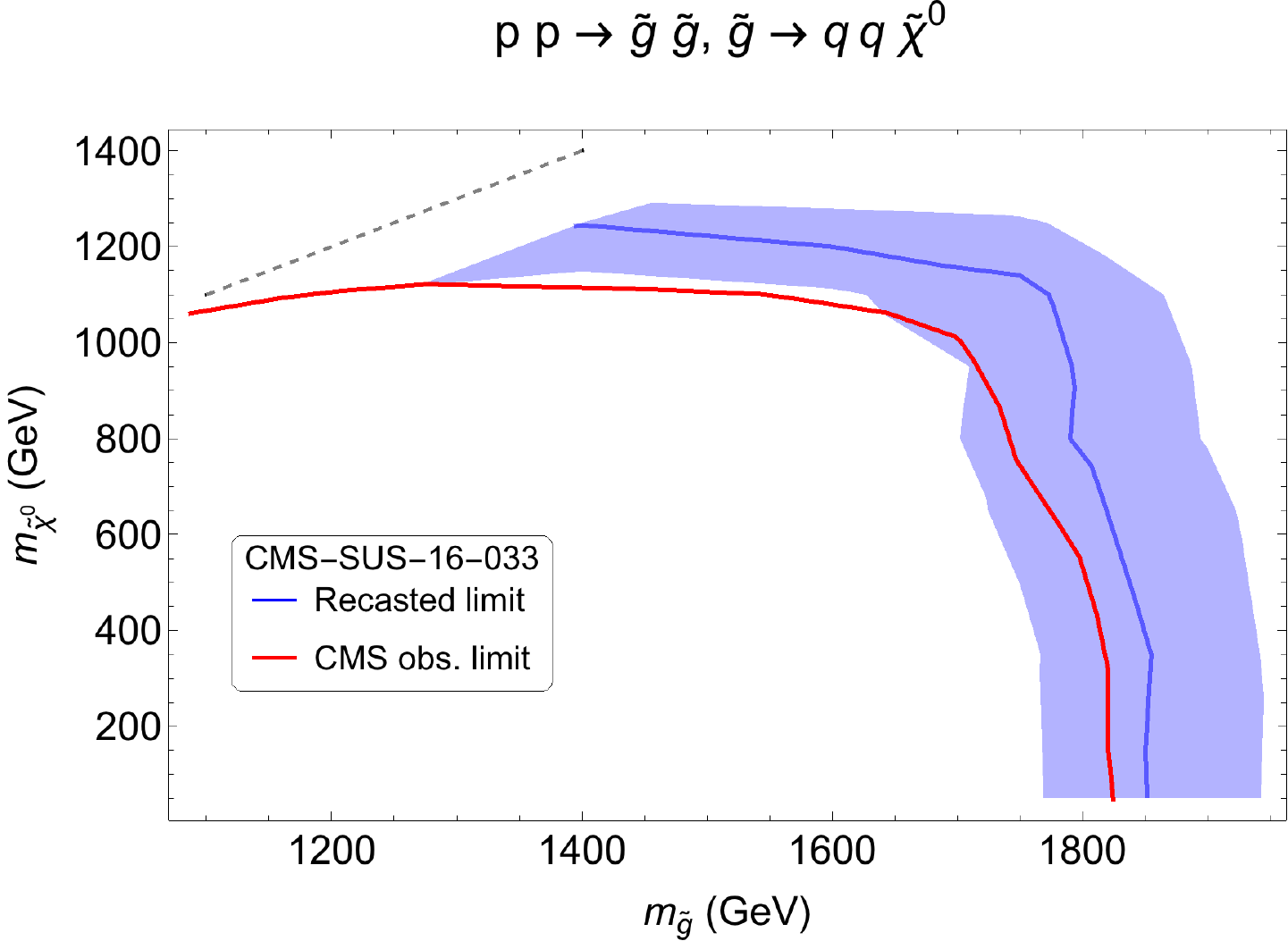}\qquad
\includegraphics[width=0.38\columnwidth]{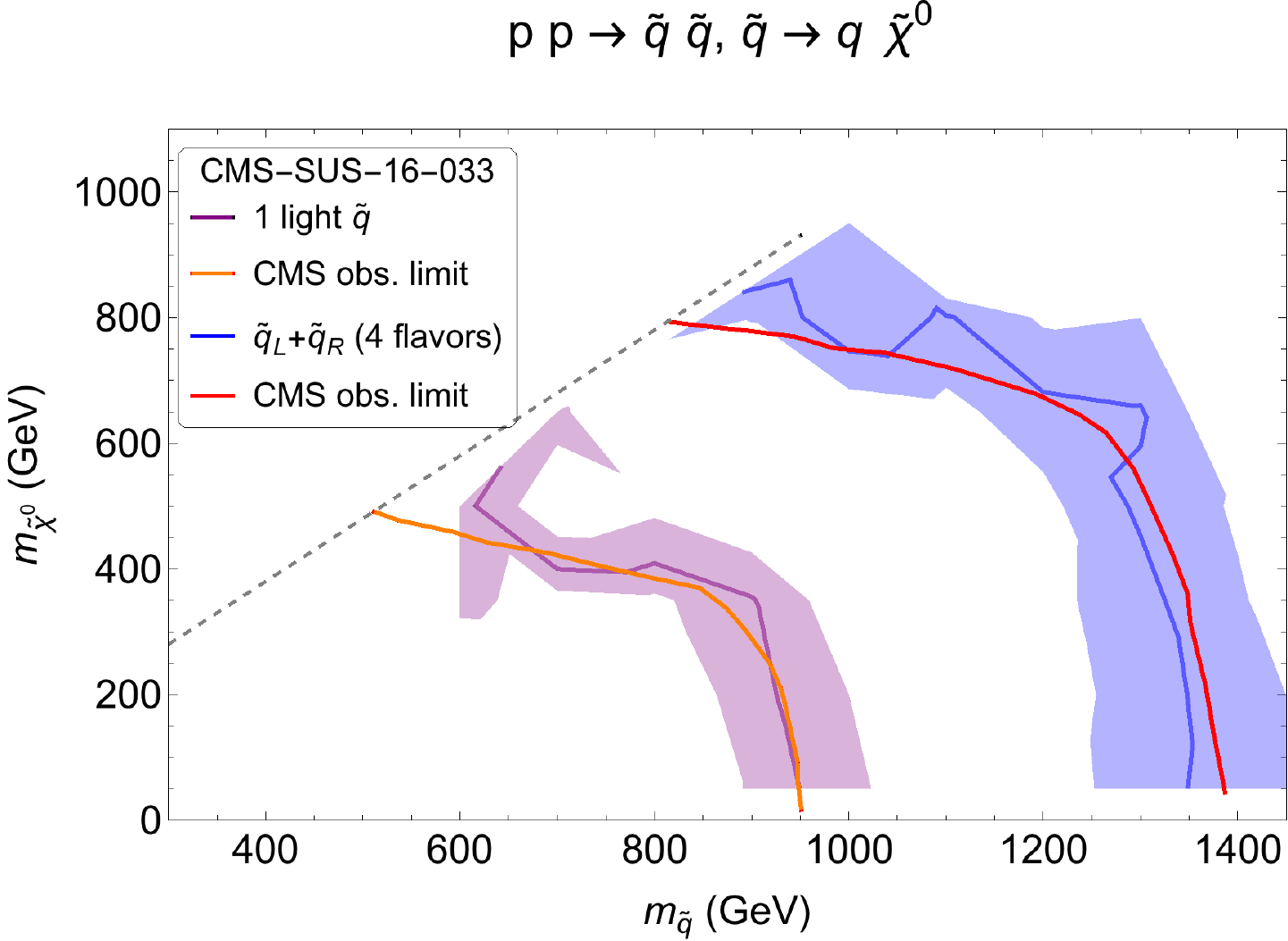}\\
\includegraphics[width=0.38\columnwidth]{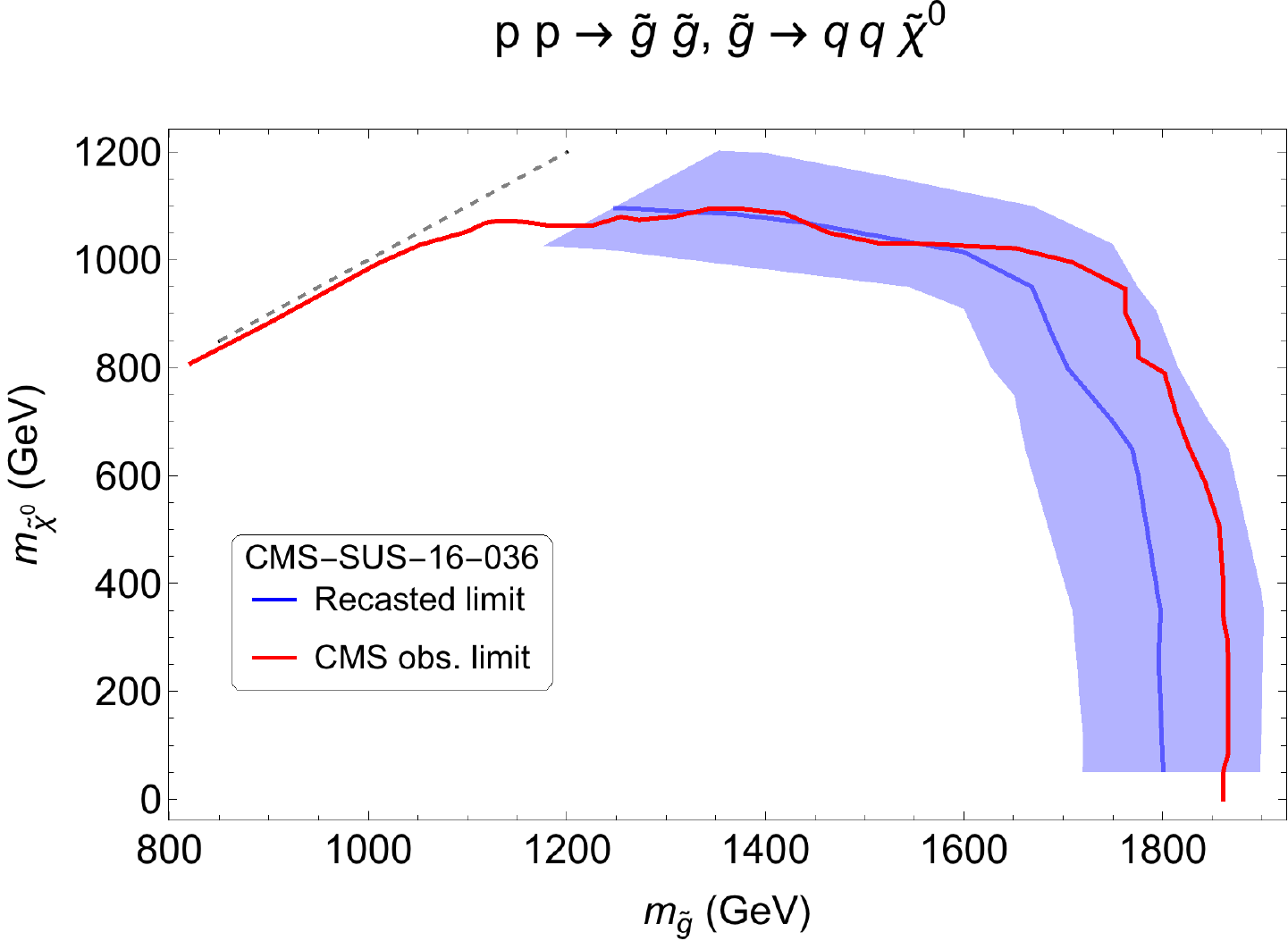}\qquad
\includegraphics[width=0.38\columnwidth]{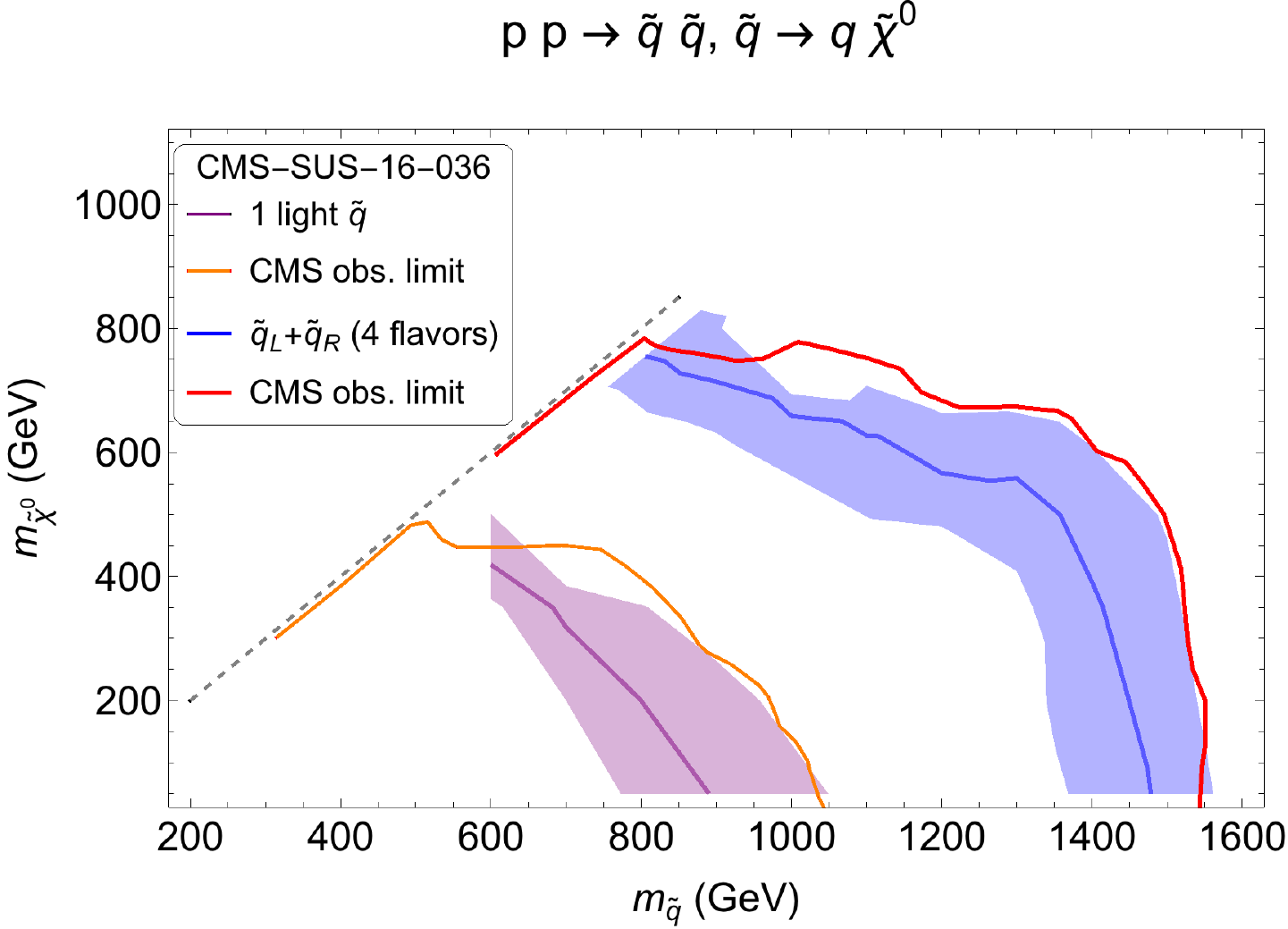}\\
\includegraphics[width=0.38\columnwidth]{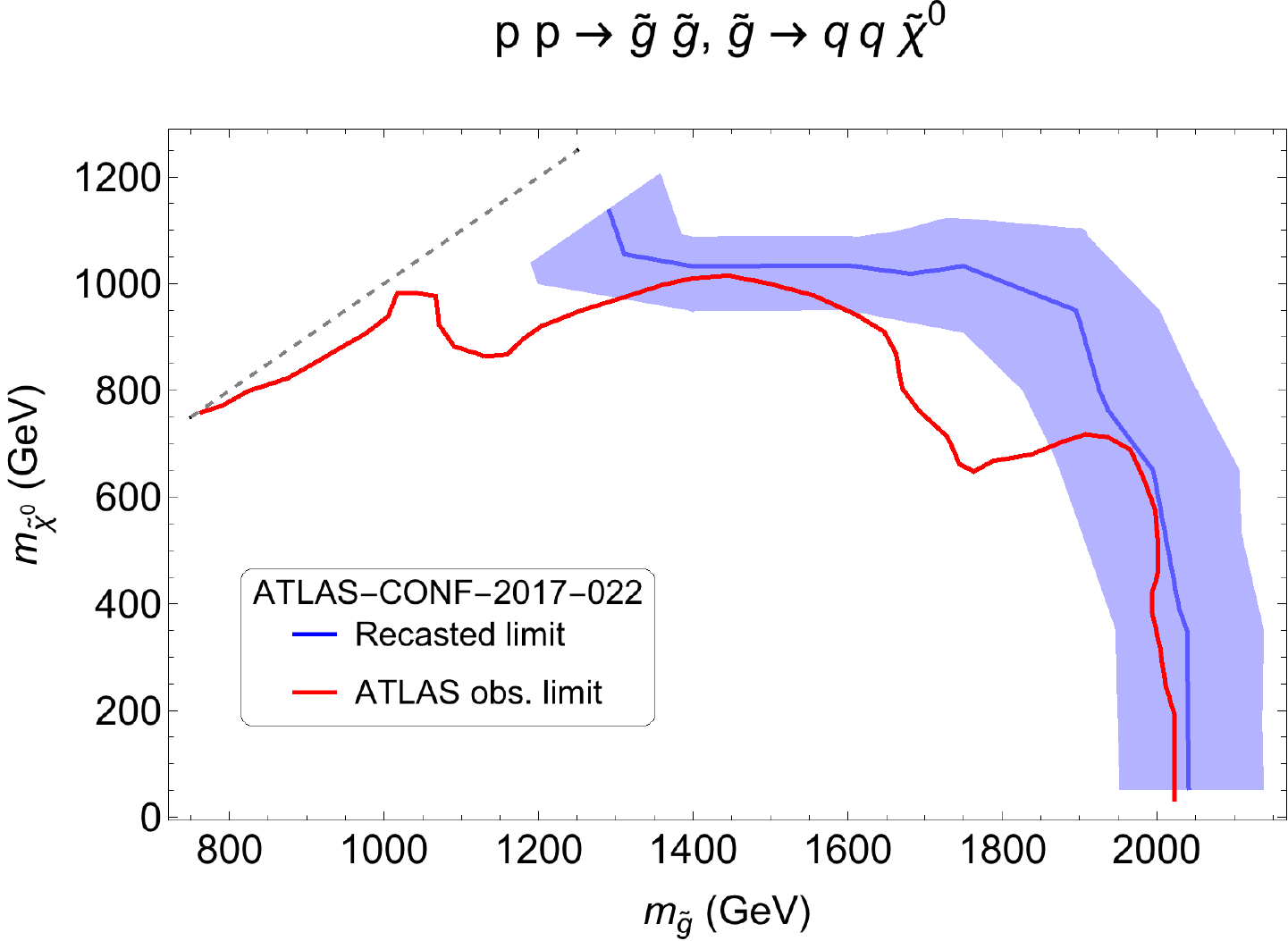}\qquad
\includegraphics[width=0.38\columnwidth]{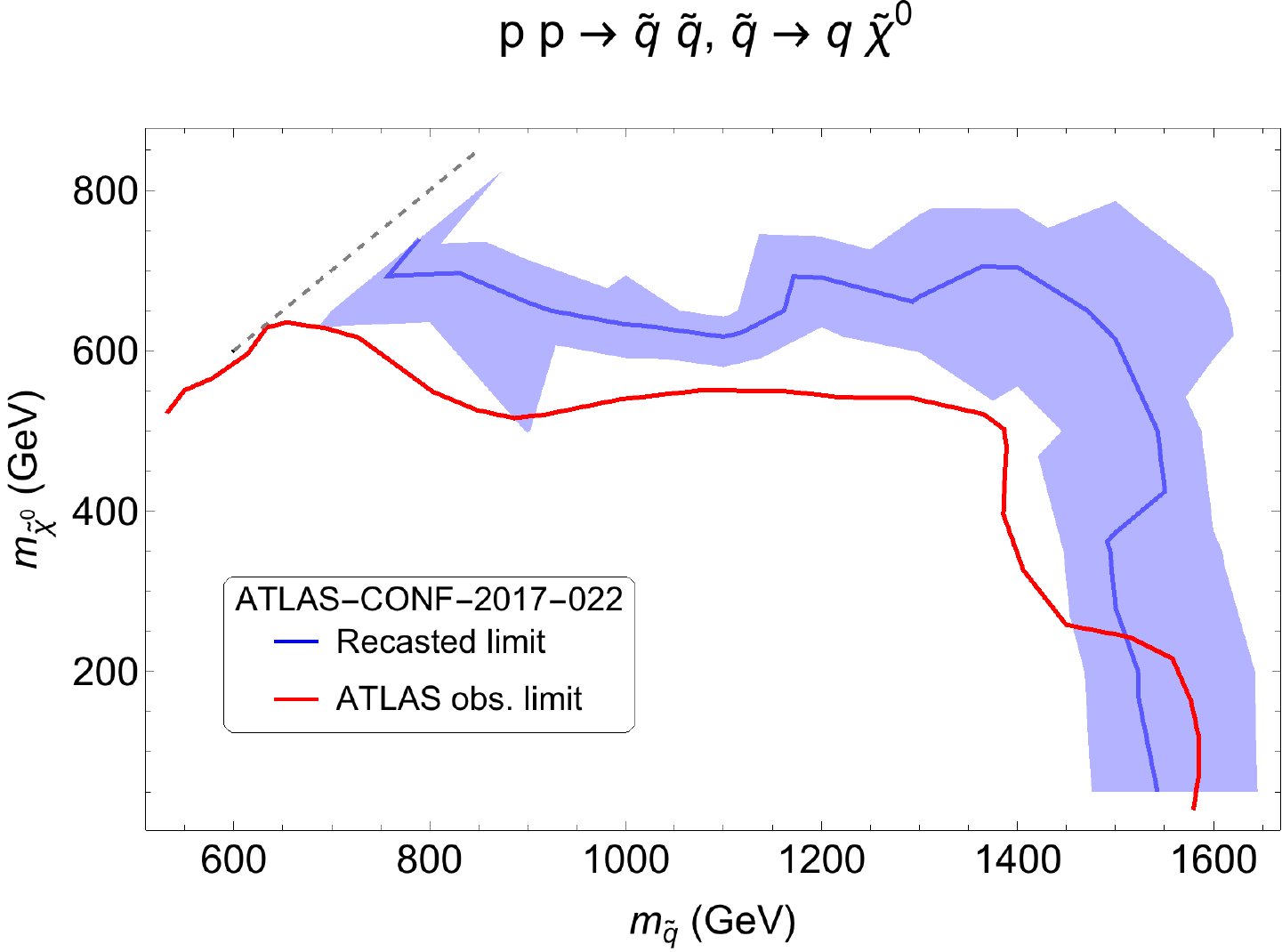}\\

\caption{The validation plots for \cite{CMS33} (first row), \cite{CMS36} (second row) and \cite{ATLAS22} (third row), for the $\tilde g\to qq\tilde\chi^0$ (left) and $\tilde q\to q\chi^0$ (right) simplified models. The blue and purple lines denote the 95$\%$ C.L. limit calculated using the likelihood analysis described in appendix \ref{app:stat}. The shaded regions denote the same limit as the solid line with 50$\%$ error included in our signal strength in each direction (to take the possible recasting errors into account). The red and orange lines are the official observed limits. 
}
\label{fig:33validate}
\end{center}
\end{figure}

\begin{figure}[h]
\begin{center}
\includegraphics[width=0.38\columnwidth]{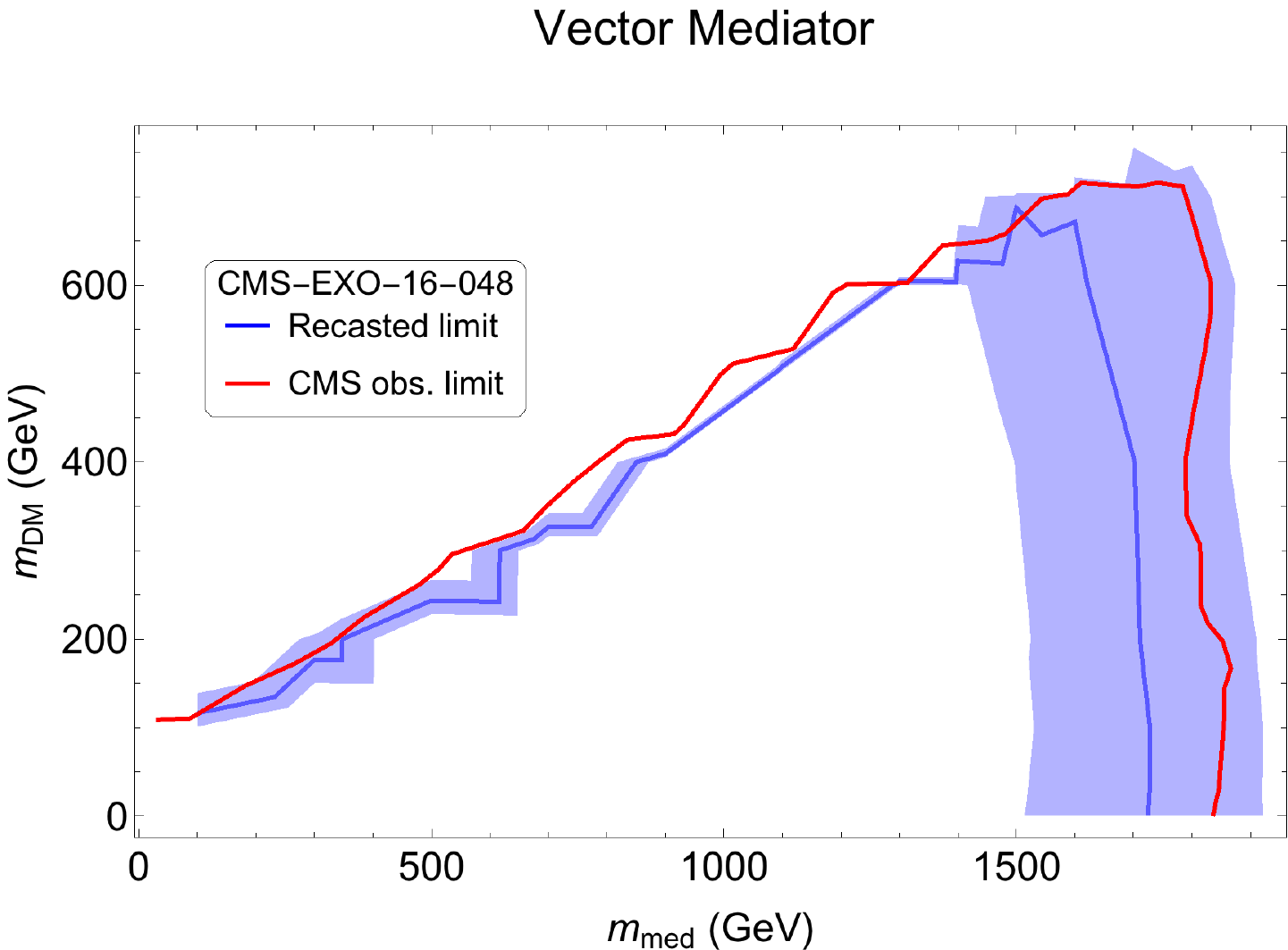}\qquad
\includegraphics[width=0.38\columnwidth]{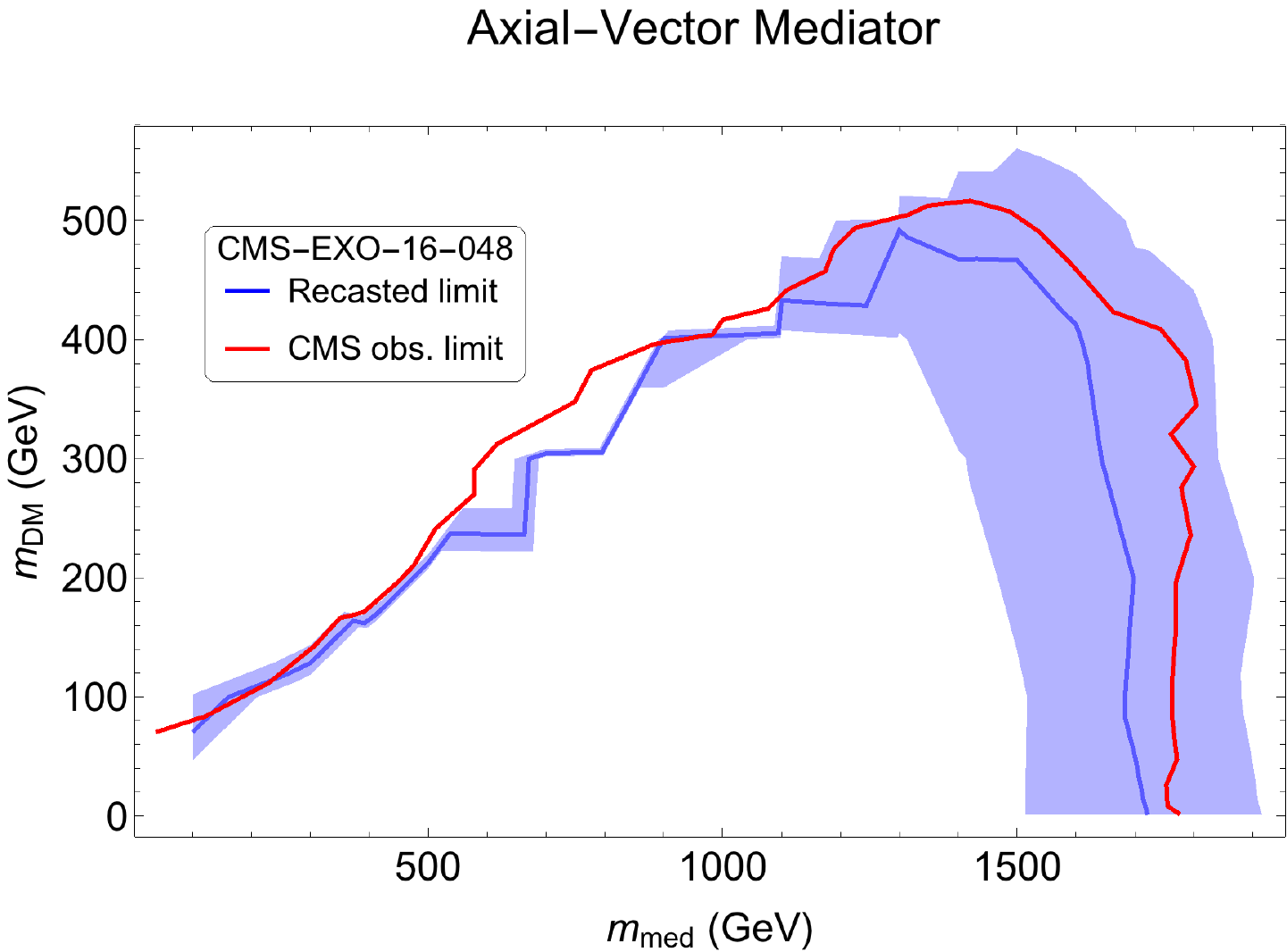}\\
\caption{The validation plots for \cite{CMS48}. The color-coding is the same as in Figure~\ref{fig:33validate}.}
\label{fig:48validate}
\end{center}
\end{figure}

In order to make our work most useful to the community, we also release auxiliary material containing our source code for each analysis as well as the ATLAS and CMS detector cards used in \textsc{Delphes}. With this material, our results can be reproduced as well as extended to different scenarios. We stress that our aim is not to release a public recasting tool, but to make it easier for existing public codes to incorporate the searches that we used.

In the following, we show our validation plots, comparing the experimental exclusion region with the exclusion region we find from our pipeline. In order to show what the uncertainties are in our pipeline, we also show a shaded band around our recasted limits, which are obtained by multiplying or dividing our efficiencies by a factor of $1.5$.  Using the likelihood analysis described above, and the data (observed event counts, expected backgrounds with relative uncertainties and covariance matrices when available) from each search, we compute the test statistics $\tilde{q}_\mu$ as described in Eq.~\eqref{eq:qmu}, and exclude a point when $\tilde{q}_{\mu=1}\geq4$, that is when the nominal cross section is excluded at the 95\% C.L.
Of all the simplified models studied in \cite{CMS33,CMS36,ATLAS22} we only show validation plots for the ones most similar to the topologies studied in Section~\ref{sec:monojet}, namely $pp\to \tilde g\tilde g, \tilde g\to q \bar q\tilde \chi^0_1$ (left column) and $pp\to \tilde q\tilde q^*, \tilde q\to q\tilde \chi^0_1$ (right column) with either one or all eight first and second generation squarks in the spectrum. For \cite{CMS48}, we show the simplified dark matter models with either a vector or an axial-vector mediator.

\section{Identified Excesses Inconsistent with New Physics \label{app:badexcess}}

We here discuss the excesses in Tables~\ref{tab:excess36} and \ref{tab:excess33} that we think are more likely due to statistical fluctuations of the background, according to the criteria outlined in Section~\ref{sec:statfluct}. To map a CMS036 excess into corresponding CMS033 bins, we recall that $\slashed E_T\ge M_{T2}$.

\begin{figure}[t]
\begin{center}
	\subfigure[]{\includegraphics[width=0.45\textwidth]{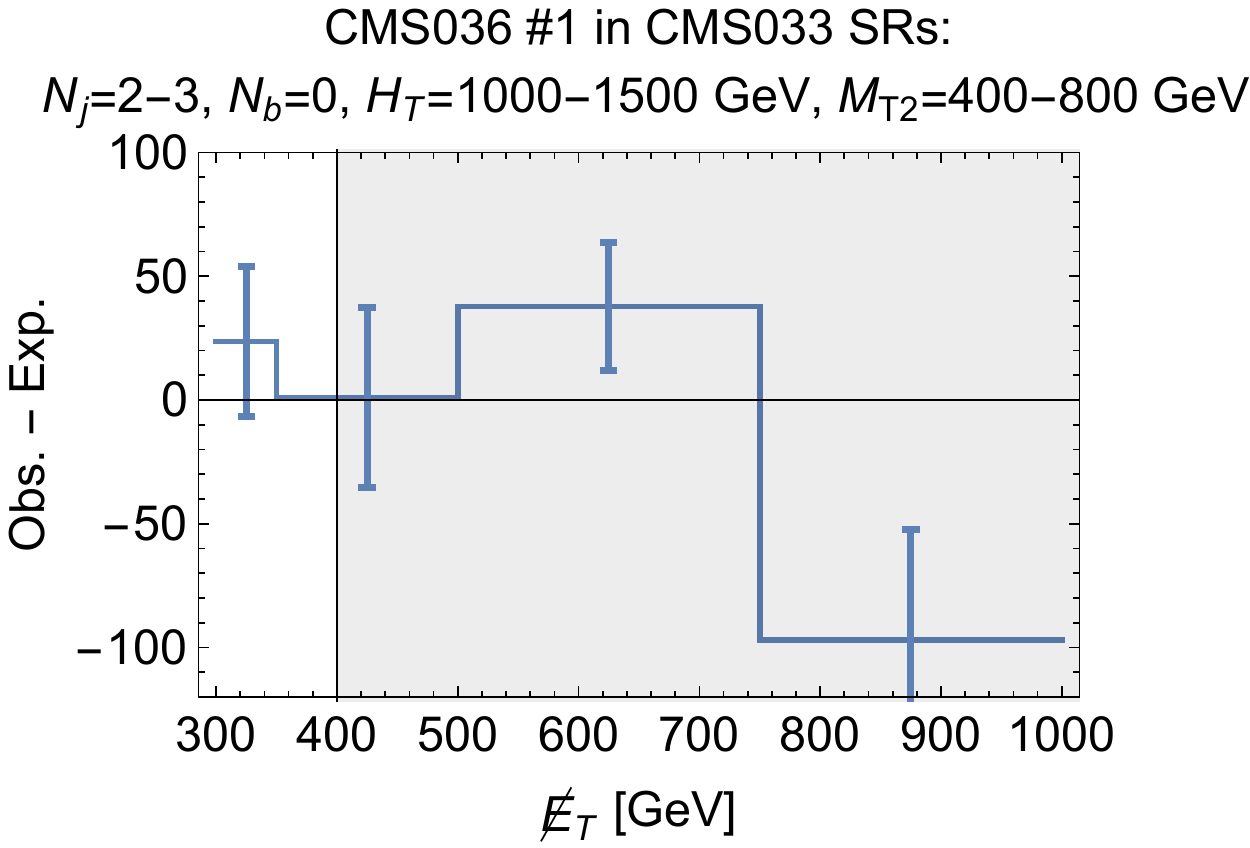}\label{fig:SRhistos36a}}\qquad
	\subfigure[]{\includegraphics[width=0.45\textwidth]{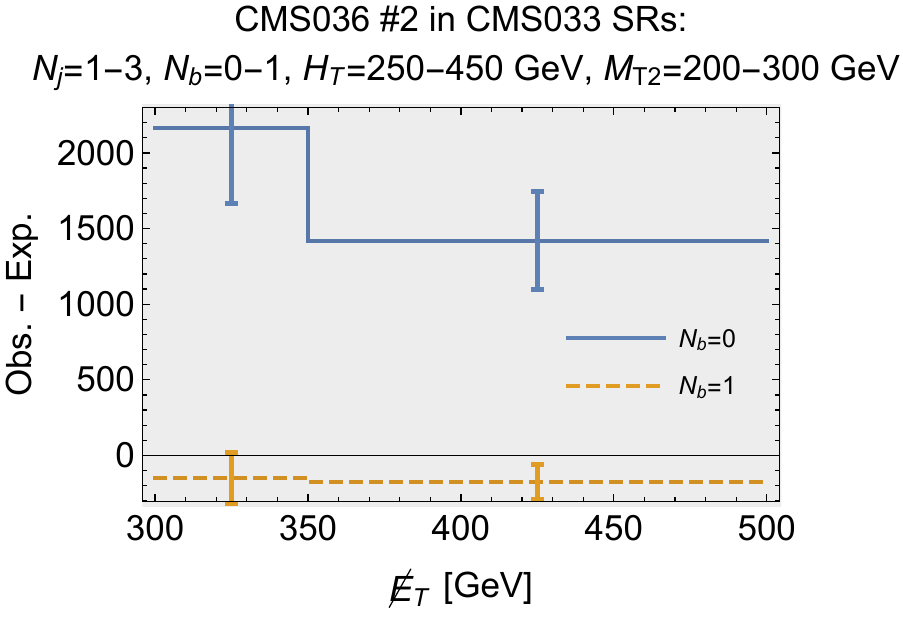}\label{fig:SRhistos36b}}
	\subfigure[]{\includegraphics[width=0.4\textwidth]{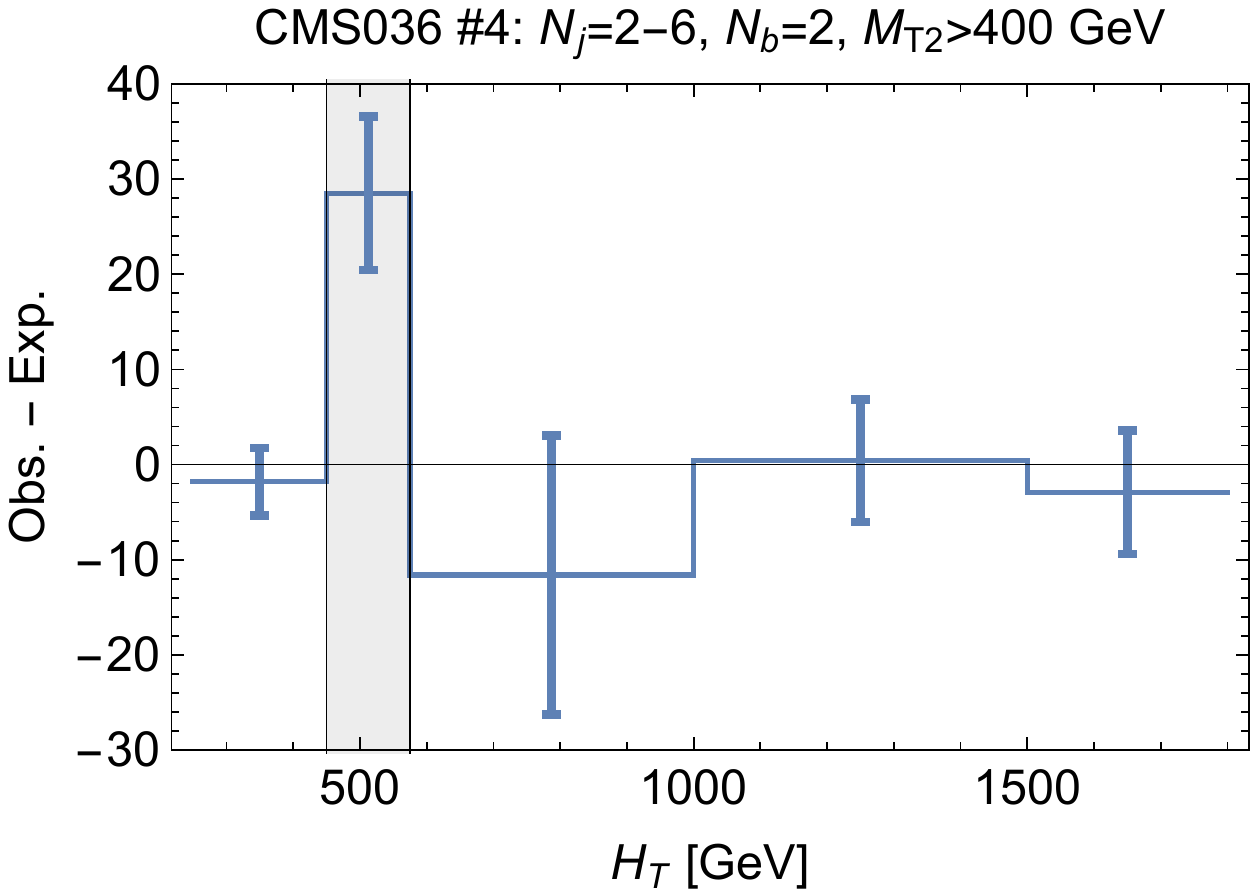}\label{fig:SRhistos36c}}\qquad%
	\subfigure[]{\includegraphics[width=0.4\textwidth]{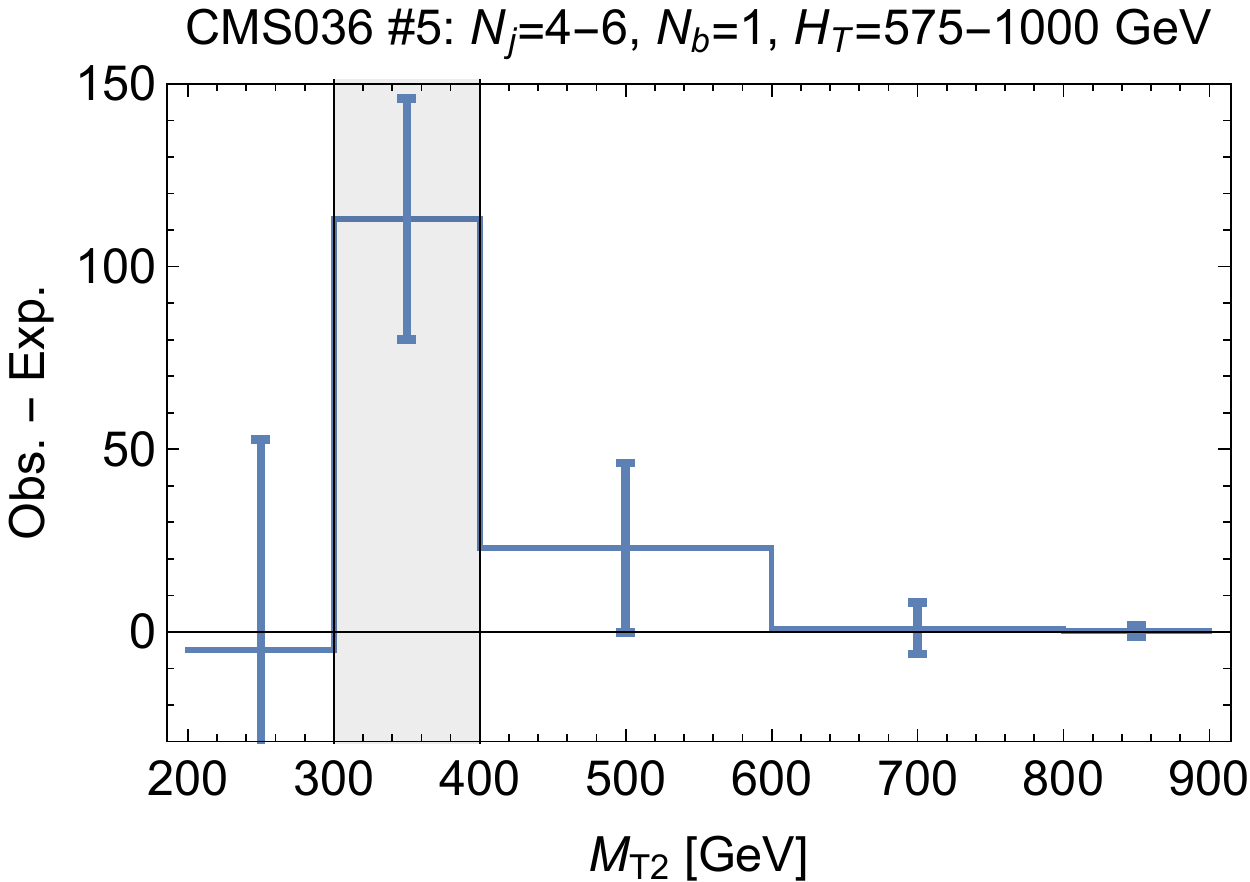}\label{fig:SRhistos36d}}
\caption{Kinematic distributions which suggest various CMS036 excesses (see Table~\ref{tab:excess36}) are unlikely to be new physics. The residuals (observed minus expected) are shown in blue, (with error bars denoting the error). In each plot, the shaded grey region denotes the region with an observed excess. See text for details.\label{fig:SRhistos36}}
\end{center}
\end{figure}

\bitem
\item ROI \#1 of CMS036   is in tension with the lack of any excesses in the corresponding SRs of CMS033. Because the core of the excess is from bins 126, 127 (which by themselves provide 3.1$\sigma$), we will focus on these.  In Figure~\ref{fig:SRhistos36a}, we show the residuals of the CMS033 \MET~distribution for the bins corresponding to the $N_j,N_b,H_T$ ranges of this excess in CMS036 (shaded grey region). Because the $N_j$ bins do not align in the two searches, we include $N_j=3-4$ of CMS033.
The best-fit value for the number of signal events in the CMS036 excess is 71 events: from the \MET~distribution in the corresponding CMS033 bins, we see that the only place for these extra events is the bin $500<\slashed E_T < 750$, which has an excess of $\sim40$ events. Therefore, one would have needed a $\sim1\sigma$ downward fluctuation of the backgrounds in that bin of CMS033 to accomodate the full CMS036 excess. On the other hand, if we reduce the signal strength to $\sim 70\%$ of the best-fit value, the CMS036 significance is reduced from $3.1\sigma$ down to $2.9\sigma$, but the events could be responsible for the small excess in CMS033. In either case, the signal should have a highly peaked $\slashed{E}_T$ distribution, and \MET$\approx$\MTT, which seems implausible. 

\item In the same spirit, in Figure~\ref{fig:SRhistos36b}, we show the \MET~distribution of the CMS033 events corresponding to the CMS036 ROI \#2. As the excess has $200<M_{T2}<300\gev$, any missing energy is in principle allowed. The excess features events with both $0$ and $1$ $b$-tagged jets: we could conjecture a signal with one true $b$ quark in the hard process, in which case we would expect additional ISR jets to populate $N_j\geq2$ bins: in that case, the CMS036 excess should often be seen in the $N_j=2$ bins of CMS033. While there is a large excess in the $N_b=0$ bins (corresponding to CMS033 excesses \#2a,c,d), there is a deficit in the $N_b=1$ bins, so that we deem the $N_b=1$ RAs of this ROI to be inconsistent with the corresponding CMS033 bins. On the other hand, the aggregation with the $N_b=0$ bins is highly significant and mirrors an excess in CMS033, which we have investigated in Sec.~\ref{sec:monojet}.

\item In Figure~\ref{fig:SRhistos36c}, we show the $H_T$ distribution of CMS036 ROI \#4: the nearby bins in $H_T$ have deficits, so that a putative signal would have to have an extremely narrow $H_T$ distribution, which is unlikely for events with up to six jets.

\item In Figure~\ref{fig:SRhistos36d}, we study CMS036 ROI \#5 and show the $M_{T2}$ distribution of  the neighboring bins. For $N_j>2$, the $M_{T2}$ distributions are generally much wider than 100 GeV, so that a signal would contaminate nearby bins, all of which are consistent with backgrounds, especially at lower \MTT.

\begin{figure}[t]
\begin{center}
	\subfigure[]{\includegraphics[height=0.3\textwidth]{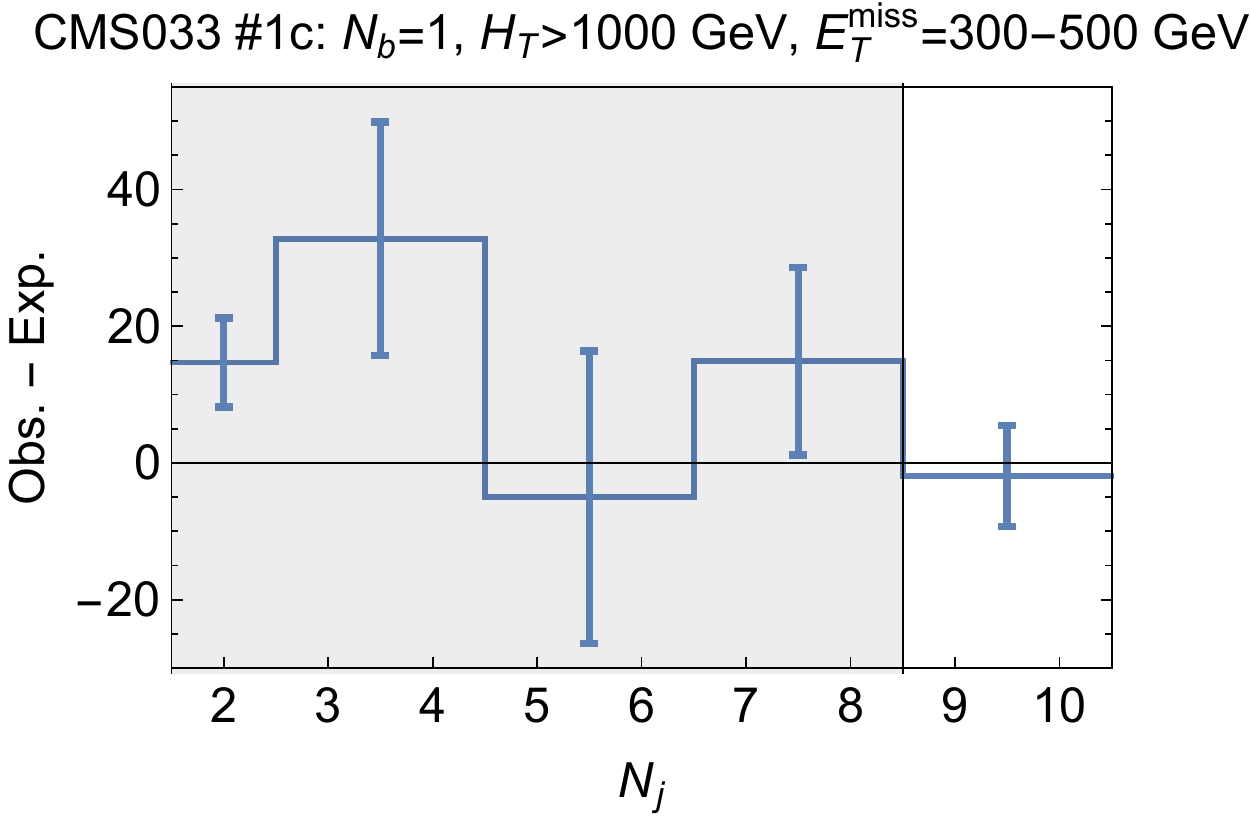}\label{fig:SRhistos33a}}\qquad
	\subfigure[]{\includegraphics[height=0.3\textwidth]{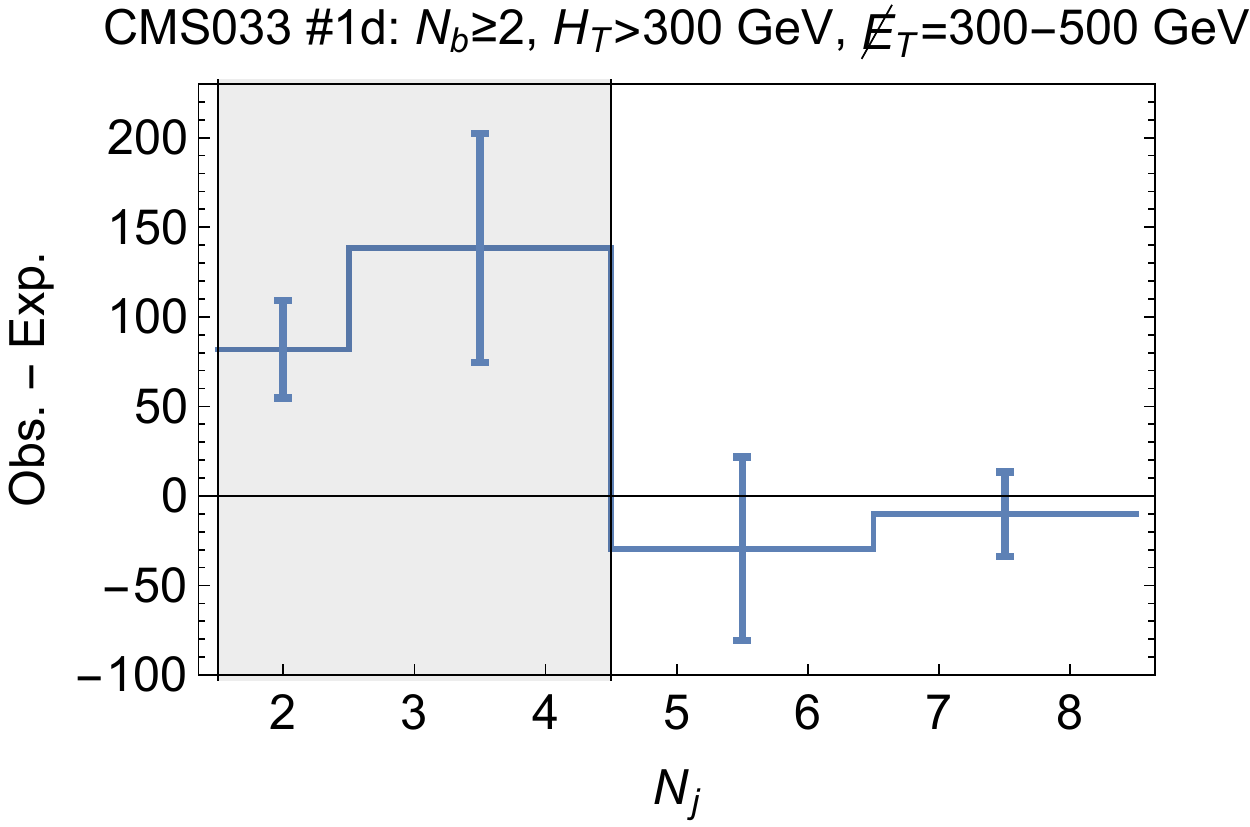}\label{fig:SRhistos33b}}

	\subfigure[]{\includegraphics[height=0.3\textwidth]{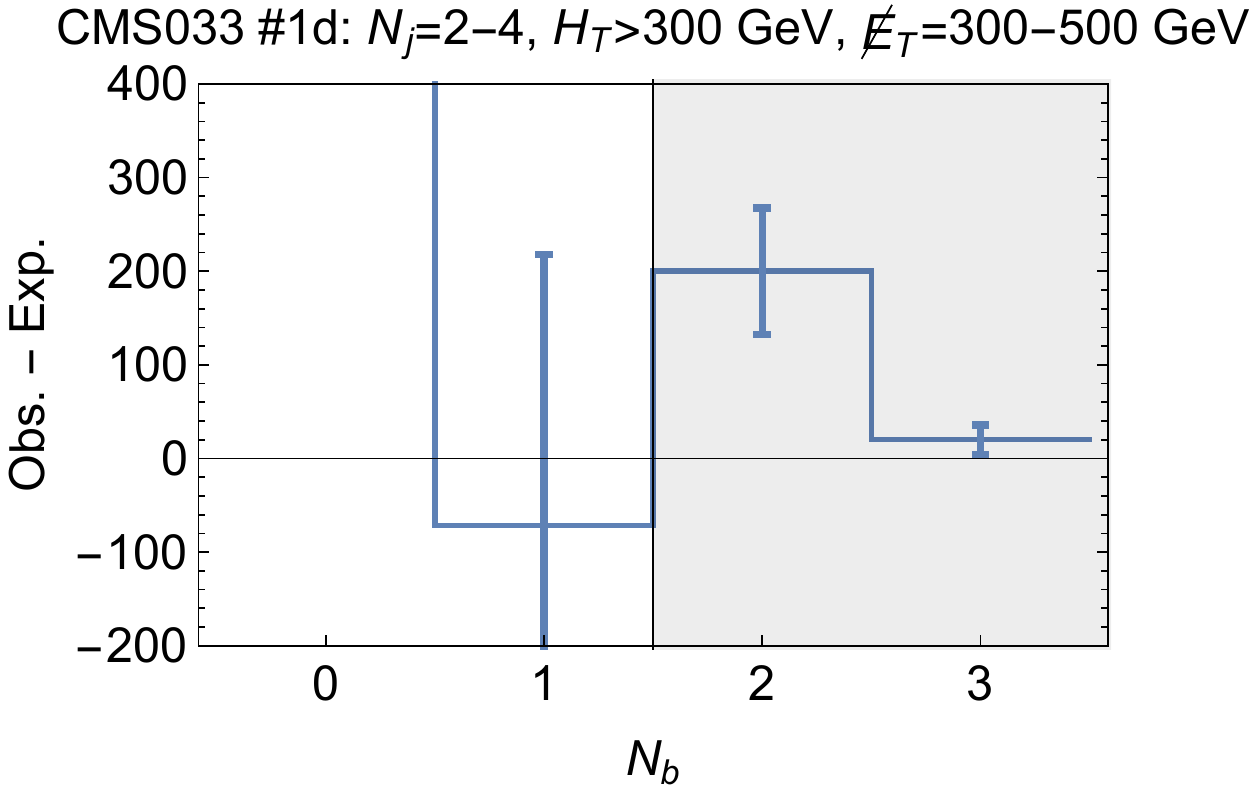}\label{fig:SRhistos33c}}\qquad
	\subfigure[]{\includegraphics[height=0.3\textwidth]{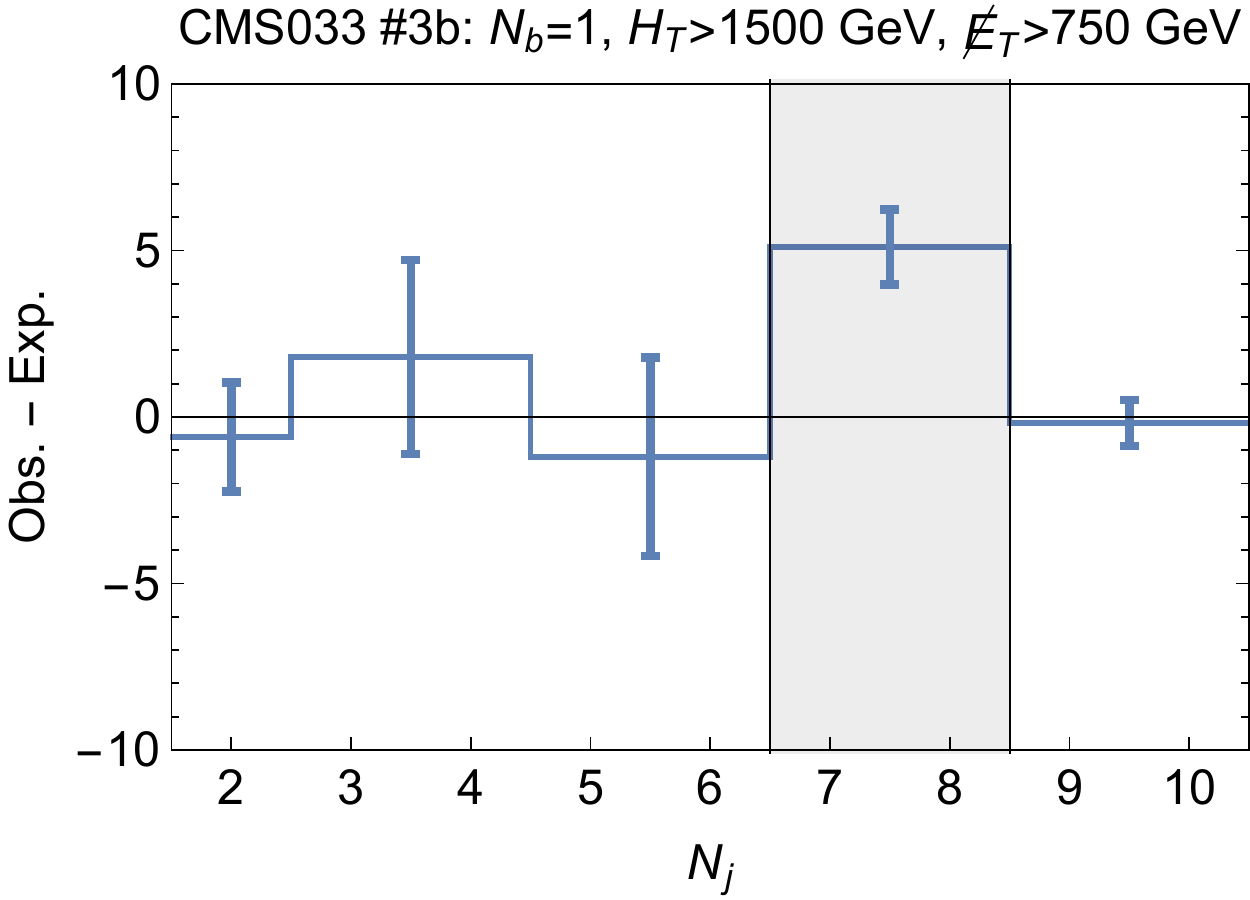}\label{fig:SRhistos33d}}
\caption{Kinematic distributions which suggest various CMS033 excesses (see Table~\ref{tab:excess33}) are unlikely to be new physics. The residuals (observed minus expected) are shown in blue, (with error bars denoting the error). In each plot, the shaded grey region denotes the region with an observed excess. See text for details.}
\label{fig:SRhistos33}
\end{center}
\end{figure}

\item  Now, we turn to the excesses in CMS033, Table~\ref{tab:excess33}: first, aggregation \#1a includes the peculiar combination of $N_j=2$ and $N_b=3$ bins, which we deem unlikely to come from a model with a set decay topology. On the other hand, turning to smaller sub-aggregations results in a viable excess (\#1b, as shown in Section~\ref{sec:statfluct}), as well as less viable ones which we discuss next.

\item in Figure~\ref{fig:SRhistos33a}, we show the $N_j$ distribution of CMS033 aggregation \#1c: the excess spans a large range of $N_j$, but it can be seen that the residual distribution is approximately flat. A signal with a large number of jets in the hard process will be peaked around that number, with tails given by ISR or jets overlapping with each other, while a signal with low jet multiplicity should be peaked at low $N_j$ and have tails from ISR. As the residuals do not look like any of these cases, we think that the $N_j$ distribution of this excess is only compatible with background fluctuations.

\item In Figures~\ref{fig:SRhistos33b}-\ref{fig:SRhistos33c}, we show the $N_j$ and $N_b$ distributions of the CMS033 aggregation \#1d. While the $N_j$ distribution should be sharply peaked at $N_j\leq 4$, the aggregation requires $N_b=2,3$. As $b$-jets from ISR are rare, adding ISR to the $b$-jets would likely overpopulate the high-multiplicity $N_j\geq4$ bins. This excess is incompatible with surrounding bins. Note that the $N_b=0$ bin is out of scale in this plot and overlaps with the viable aggregation \#2c.

\item In Figure~\ref{fig:SRhistos33d}, we show the $N_j$ distribution of CMS033 ROI \#3. We here only plot the distribution of bin 126 which is the core of the excess. It has $N_j=7,8$ and high $H_T$, but nearby bins at both higher and lower jet multiplicities do not show any deviation. Such high jet multiplicities require at least $4-5$ jets at the parton level with the rest coming from ISR/FSR, resulting in wide jet distributions which would populate the nearby bins. We exclude this aggregation due to the $N_j$ distribution.

\eitem

\afterpage{\clearpage}

\bibliographystyle{utphys}
\bibliography{moriond_search}

\end{document}